\documentclass[%
preprint,
 amsmath,amssymb,
 aps,
 prfluids
 , 
]{revtex4-2}

\usepackage{comment}
\usepackage{graphicx}
\usepackage{dcolumn}
\usepackage{bm}
\usepackage{subcaption}
\usepackage{tikz}
\usepackage{hyperref}
\usepackage[mathlines]{lineno}

\begin{document}


\title{Turbulence Modelling of Mixing Layers under Anisotropic Strain}

\author{Bradley Pascoe}
\email{bradley.pascoe@sydney.edu.au}
\affiliation{%
 School of Aerospace, Mechanical and Mechatronic Engineering, University of Sydney,\\
 Sydney, New South Wales 2006, Australia
}%

\author{Michael Groom}
\affiliation{Geophysical Fluids Team, CSIRO Environment, Eveleigh NSW 2015, Australia}%

\author{Ben Thornber}
\affiliation{
 School of Mechanical and Aerospace Engineering, Queen's University Belfast,\\
 Belfast BT9 5AH, Northern Ireland, United Kingdom}%
 
\date{\today}

\begin{abstract}

The development of turbulent mixing layers can be altered by the application of anisotropic strain rates, potentially arising from radial motion in convergent geometry or movement through non-uniform geometry. 
Previous closure models and calibrations of compressible turbulence models tend to focus on incompressible flows or isotropic strain cases, which is in contrast to many real flow conditions.
The treatment of bulk compression under anisotropic strain is investigated using the K-L turbulence model, a two-equation Reynolds-Averaged Navier-Stokes (RANS) model that is commonly used for simulating interfacial instabilities. 
One-dimensional simulations of shock-induced turbulent mixing layers under applied axial or transverse strain rates are performed using three different closures for the bulk compression of the turbulent length scale. 
The default closure method using the mean isotropic strain rate is able to reasonably predict the integral width and turbulent kinetic energy of the mixing layer under the applied strain rates. 
However, the K-L model's performance is improved with the transverse strain closure, while the axial strain closure worsens the model.
The effects of this new closure are investigated for the buoyancy-drag model, showing that a three-equation model which evolves the integral width and turbulent length scale separately is most effective for modelling anisotropic strain. 
Through the equivalence of two-equation RANS models, the modification of the bulk compression closure for the turbulent length scale also suggests an alteration to the K-$\epsilon$ and K-$\omega$ models.

 
\end{abstract}

\maketitle


\section{Introduction}
\label{sec:Introduction}

Strain rates in fluid flow are a key driver of turbulence, responsible for the shear production of turbulent kinetic energy and the vortex stretching of eddies.
Large scale mean strain rates are observed under a variety of scenarios, with shear strain occurring in boundary layers and normal strain occurring at geometric deformations such as nozzles.
Shear driven mixing layers, such as those induced by the Kelvin-Helmholtz instability, are more well studied than the evolution of turbulent mixing layers under mean normal strain. 
In the high strain rate limit, where the strain timescale is much smaller than the turbulence timescale, the evolution of turbulence is described by rapid-distortion theory 
\cite{Hunt_1990_RapidDistortionTheory,Durbin_1992_RapidDistortionTheory,Cambon_1993_RapidDistortionAnalysis,Blaisdell_1996_RapidDistortionTheory}.
These studies provide useful tools for understanding the pressure-dilatation tensor, however many are limited by the assumption of homogenous turbulence. This is in contrast to realistic applications where turbulence is inhomogeneous and may not satisfy the high strain limit. Likewise, the analysis of incompressible fluid flow is simpler than the analysis of compressible flow, with many compressible models relying upon incompressible closures.

Two prominent interfacial instabilities that give rise to turbulent mixing layers are the Rayleigh-Taylor instability (RTI) and the Richtmyer-Meshkov instability (RMI).
These instabilities arise from the deposition of vorticity from baroclinic production at the interfaces between fluids due to the misalignment of pressure and density gradients.
RTI refers to the case of a persistent pressure gradient, causing one fluid to accelerate into the other, however RTI is only unstable for the light fluid accelerated toward the heavy, and stable for the reverse arrangement.
RMI refers to the case of an impulsive acceleration, such as occurs for a shock-wave traversing through the interface.
The deposition of vorticity occurs over a far shorter timespan, with RMI being unstable for both heavy-to-light and light-to-heavy shock passages.
The linear regime of these instabilities is well understood, growing exponentially and linearly for RTI and RMI respectively, while the interface amplitude is much smaller than the wavelength perturbation.
In convergent geometry, alterations to the amplitude growth rate are described by the Bell-Plesset effects, which introduce dependencies on the fluid compression rate and convergence rate of the interface radius \cite{Bell_1951_TaylorInstabilityCylinders, Plesset_1954_StabilityFluidFlows, Penney_1942_ChangingFormNearly,Epstein_2004_BellPlessetEffects}.
The aforementioned compression rate and convergence rate can be rewritten in terms of the radial/axial and circumferential/transverse strain rates \cite{Pascoe_2024_ImpactAxialStrain,Pascoe_2025_ImpactTransverseStrain}, and so the effects of anisotropic strain rates on the linear regime is within the scope of analytical evaluation. 
The late-time behaviour of the mixing layer is non-linear, as the interface amplitudes begin to saturate and roll up due to the Kelvin-Helmholtz instability. The instabilities transition to a turbulent mixing layer, with widths growing quadratically ($\sim t^2$) and sub-linearly ($t^\theta$) for RTI and RMI respectively.
These mixing layers are anisotropic and inhomogeneous, however for high-Reynolds numbers they are able to achieve self-similar growth/decay states. 
A thorough review of the RMI and RTI can be found in Refs. \cite{Zhou_2017_ReviewA,Zhou_2017_ReviewB,Zhou_2021_JourneyThroughScales}.

In spherical configurations, implosions and explosions are examples of bulk fluid moving to inner and outer radii, respectively. A turbulent mixing layer present in an implosion or explosion will experience mean strain rates due to the non-zero radial velocity profile. The mean radial strain rate ($\bar{S}_{rr}$) and circumferential strain rate ($\bar{S}_{\theta \theta}$) for a spherically symmetric flow are given by
\begin{align}
    \bar{S}_{rr} = \frac{\bar{\partial u_r}}{\partial r}, \\
    \bar{S}_{\theta \theta} = \frac{\bar{u}_r}{r}.
\end{align}
These strain rates depend upon the mean radial velocity ($\bar{u}_r$) and the radius ($r$).

\begin{figure}[ht]
    \centering
    \hfill
    \begin{subfigure}{0.4\textwidth}
    \begin{tikzpicture}
        \node[anchor=north west] (image) at (0,0) {\includegraphics[width=\textwidth]{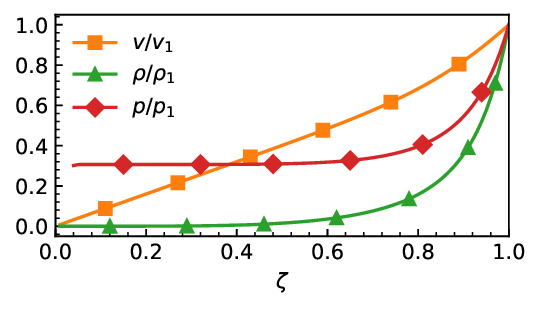}};
        \node[anchor=north east] at (0.2,0) {(\textit{a})};
    \end{tikzpicture}
    \end{subfigure}
    \hfill
    \begin{subfigure}{0.4\textwidth}
    \begin{tikzpicture}
        \node[anchor=north west] (image) at (0,0) {\includegraphics[width=\textwidth]{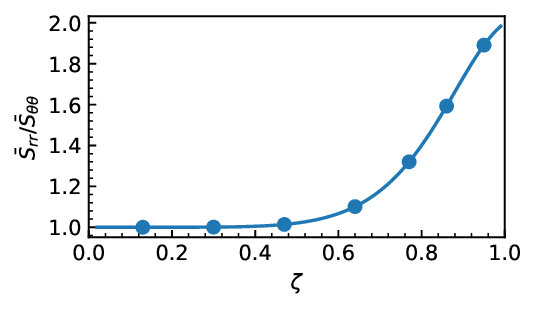}};
        \node[anchor=north east] at (0.2,0) {(\textit{b})};
    \end{tikzpicture}
    \end{subfigure}
    \caption{\label{fig:TaylorSedov}Self similar profiles as a function of relative radius, $\zeta=r/R(t)$, for the Taylor-von Neumann-Sedov blastwave. (\textit{a}) Velocity, density, and pressure normalised to values immediately behind the shock front; (\textit{b}) Strain rate anisotropy.}
\end{figure}

Hydrodynamic instabilities and the advection of fluids play an important role in the evolution of stellar bodies \cite{Arnett_2000_RoleMixingAstrophysics}. 
For example, core-collapse supernovae are some of the largest explosions in the universe, with mixing occurring due to an expansive blast-wave from the core that induces both RMI and RTI \cite{Miles_2009_BlastWaveDrivenInstabilityVehicle}.
An idealised version of this problem is given by the Taylor-von Neumann-Sedov blast-wave, which provides a self-similar analytical solution for a spherically expanding blast-wave \cite{Sedov_1946_PropagationStrongShock,Taylor_1950_FormationBlastWave,Taylor_1950_FormationBlastWavea,Goldstine_1955_BlastWaveCalculation}.
The self-similar profiles are functions of the radius relative to the time-varying radial position of the shock front of the blast-wave, $\zeta=r/R(t)$. The velocity profile scales proportionally to the self-similar function $V(\zeta)$, which is typically solved for numerically. The velocity profile, 
\begin{align}
    v(r,t) = \frac{2r V(\zeta)}{5t}
\end{align}
describes a self-similar profile as shown in Fig.~\ref{fig:TaylorSedov}, where the velocity is normalised by the velocity of the fluid immediately behind the shock-wave, $v_1$.
The velocity profile can be used to obtain the radial and circumferential strain rates.
The anisotropy of the strain rates is also self-similar as a function of $\zeta$:
\begin{align}
    \bar{S}_{rr} / \bar{S}_{\theta \theta} = 1 + \frac{\zeta}{V(\zeta)} \frac{dV(\zeta)}{d\zeta}.
\end{align}
This equation is plotted in Fig. \ref{fig:TaylorSedov} and shows a profile that deviates away from the isotropic value of unity.
Immediately behind the shock-front the strain rates are anisotropic, with the radial strain rate up to twice as large as the circumferential strain rate. 
Whilst this model does not explicitly consider the interaction of the blast-wave with an interface and the resulting wave transmission and reflection, inferences can still be obtained about the evolution of the mixing layer.
The interaction between the blast-wave and an interface between different density fluids will induce RMI due to the blast-wave's impulsive acceleration.
Following the blast-wave interaction, the mixing layer's relative radius ($\zeta$) will decrease from $\zeta=1$, and experience anisotropic strain rates, stretching more in the radial direction than the circumferential direction.
As time continues, the strain rates experienced by the mixing layer will tend toward isotropy.
Whilst there is also a pressure gradient following the blast-wave which can induce RTI, the effects of the buoyancy production under strain rates is not the focus of this work. 

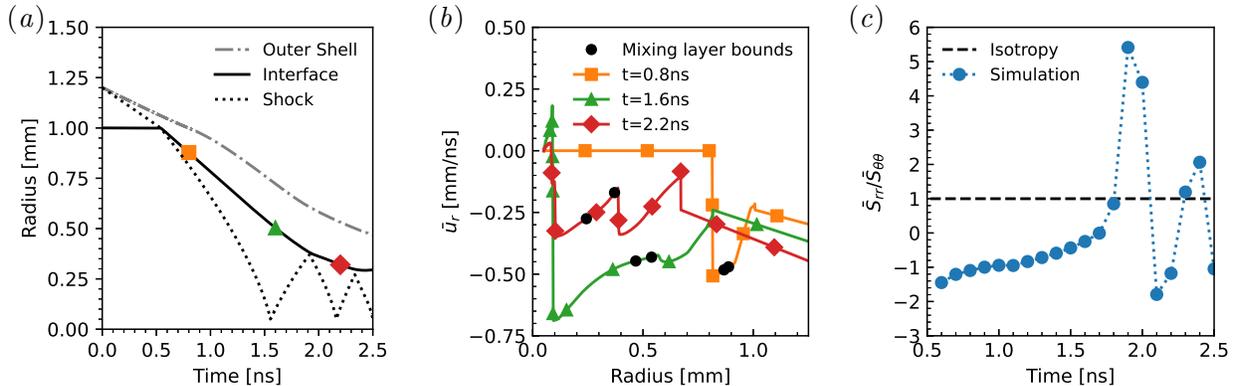
\begin{figure}
    \centering
    \begin{subfigure}{0.32\textwidth}
    \begin{tikzpicture}
       \node[anchor=north west] (image) at (0,0) {\includegraphics[width=\textwidth]{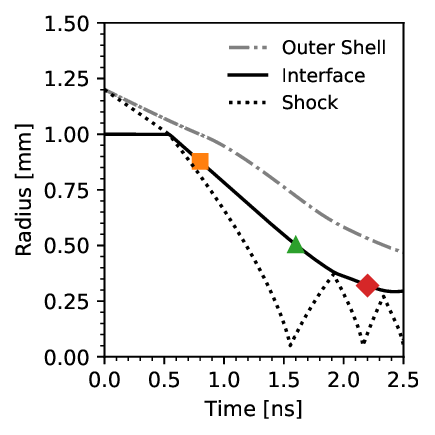}};
       \node[anchor=north west] at (0.,0) {(\textit{a})};
    \end{tikzpicture}
    \end{subfigure}
    \hfill
    \begin{subfigure}{0.32\textwidth}
    \begin{tikzpicture}
       \node[anchor=north west] (image) at (0,0) {\includegraphics[width=\textwidth]{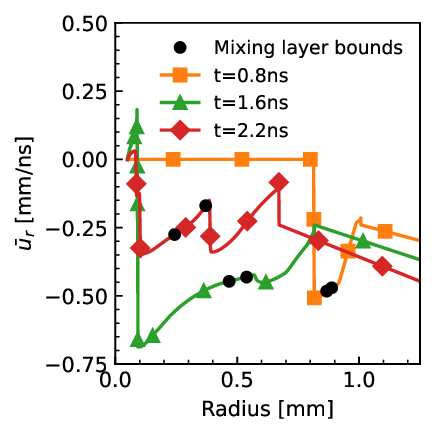}};
       \node[anchor=north west] at (0.,0) {(\textit{b})};
    \end{tikzpicture}
    \end{subfigure}
    \hfill
    \begin{subfigure}{0.32\textwidth}
    \begin{tikzpicture}
       \node[anchor=north west] (image) at (0,0) {\includegraphics[width=\textwidth]{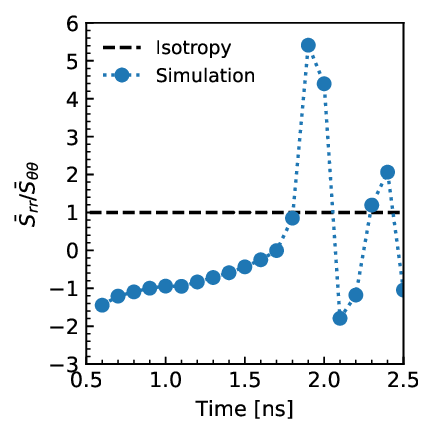}};
       \node[anchor=north west] at (0.,0) {(\textit{c})};
    \end{tikzpicture}
    \end{subfigure}
    
    \caption{Characteristics of idealised implosion simulation \cite{ElRafei_2024_TurbulenceStatisticsTransport}. (\textit{a}) Implosion profile, (\textit{b}) mean radial velocity profile at several time-steps, (\textit{c}) strain rate anisotropy across the mixing layer.}
    \label{fig:Implosion}
\end{figure}

Inertial confinement fusion is an example of an implosion profile, where small capsules containing fuel for nuclear fusion are compressed \cite{Nuckolls_1972_LaserCompressionMatter}, driven directly by lasers, or indirectly by X-rays \cite{Betti_2016_InertialconfinementFusionLasers}.
To achieve the pressures and temperatures necessary for fusion to occur, a series of pulses are used to launch shock waves into the capsule, causing the capsule to compress.
Perturbations in the interfaces between the capsule's layers can arise from limitations in machining, or from asymmetries in the driving on the capsule surface.
These perturbations make the system susceptible to hydrodynamic instabilities, either from shock interaction (RMI) or interface accelerations (RTI). The instabilities are responsible for mixing the cold, outer fuel with the hot fuel at the centre hot-spot, degrading performance and preventing ignition \cite{Lindl_2004_PhysicsBasisIgnition,Lindl_2014_ReviewNationalIgnition}.
A simplified, idealised implosion profile used for modelling the hydrodynamics of inertial confinement fusion was presented in Ref. \cite{Youngs_2008_TurbulentMixingSpherical} and used as a template for many further investigations for both single-mode \cite{Flaig_2018_SinglemodePerturbationGrowth, Heidt_2021_EffectInitialAmplitude} and multi-mode interface perturbations \cite{Joggerst_2014_CrosscodeComparisonsMixing, Boureima_2017_PropertiesTurbulentMixing, ElRafei_2019_ThreedimensionalSimulationsTurbulent, ElRafei_2020_NumericalStudyBuoyancy, ElRafei_2024_TurbulenceStatisticsTransport}.
The mixing layer position and the mean radial velocity profile at several time instants are shown in Fig. \ref{fig:Implosion} for the implosion profile, using the data from Ref. \cite{ElRafei_2024_TurbulenceStatisticsTransport}. 
Evaluating the displayed mean radial velocity profiles, the radial velocity gradient across the mixing layer shows a positive trend indicating expansive radial strain rates. The radial velocity is negative, as expected for an implosion, which correlates to a compressive circumferential strain rate. 
These profiles are taken during the intervals between shock interactions, and suggest that when the mixing layer is evolving between shocks it experiences anisotropic strain rates. 
The anisotropy of the strain rates across the mixing layer is also plotted in Fig. \ref{fig:Implosion}, and shows that the ratio of the strain rates tends towards -1 rather than unity. There are times at which the anisotropy ratio goes above unity, aligning with the times the shock wave is compressing the turbulent mixing layer, such that the radial strain rate is more compressive than the circumferential strain rate. However, for the majority of the simulation the circumferential strain rate is negative (compression), whilst the radial strain rate is positive (expansive). 

Reynolds-Averaged Navier-Stokes (RANS) models are an important tool to used to predict the behaviour of hydrodynamic instabilities at a fraction of the cost of large-eddy simulation (LES) and direct numerical simulation (DNS).
For practical problems such as astrophysical flows and ICF, it can be computationally intractable to completely resolve the complete range of length and time scales involved, driving the necessity for simplified models.
For compressible turbulent mixing layers, Besnard \textit{et al.} \cite{Besnard_1992_TurbulenceTransportEquations} presented a comprehensive derivation of the Favre-averaged equations for variable-density turbulence.
In Ref. \cite{Besnard_1992_TurbulenceTransportEquations}, both a Reynolds stress model and a simpler three equation model were provided, forming the basis of the BHR model.
Both models included an evolution equation for the turbulent kinetic energy dissipation rate ($\epsilon$).
Using the dissipation rate is a popular choice in other fields, given the prominence of the k-$\epsilon$ turbulence model \cite{Jones_1972_PredictionLaminarizationTwoequation,Launder_1974_NumericalComputationTurbulent,Wilcox_2006}, and several other turbulence models designed for compressible turbulent mixing layers have also used the turbulent dissipation rate \cite{Gregoire_2005_SecondorderTurbulenceModel,Moran-Lopez_2014_MulticomponentReynoldsaveragedNavier,Moran-Lopez_2015_ReynoldsAveragedNavierStokes}.
Whilst there is an equivalence between the transport equations, an alternative approach is to transport the turbulent length scale, as recommended in Ref. \cite{Besnard_1992_TurbulenceTransportEquations} and implemented by Dimonte and Tipton \cite{Dimonte_2006_KLTurbulenceModel} for the K-L model.
For interfacial instabilities such as RTI and RMI, the initial turbulent length scale has a greater physical meaning than an initial dissipation rate, and also allows for easier self-similarity analysis at late-time where the turbulent length scale is assumed to be proportional to the mixing layer width.
Improvements have been made to the K-L model to improve performance, through self-similarity analysis \cite{Morgan_2016_LargeeddyUnsteadyRANS,Xiao_2020_UnifiedPredictionReshocked,Xiao_2020_ModelingTurbulentMixing,Zhang_2020_MethodologyDeterminingCoefficients}, improved treatment of buoyancy production for the turbulent kinetic energy \cite{Kokkinakis_2015_TwoequationMultifluidTurbulence}, and realisability of the modelled Reynolds stress \cite{Kokkinakis_2020_TwoequationMultifluidTurbulence,Xiao_2021_KLModelImproved}. Further developments to the BHR model and K-L model have occurred through the introduction of additional transport equations resulting in a multitude of different models \cite{Morgan_2015_ThreeequationModelSelfsimilar,Morgan_2018_LargeeddySimulationReynoldsaveraged,Morgan_2018_TwolengthscaleTurbulenceModel,Morgan_2021_SelfconsistentHighorderSpatial,Morgan_2021_SelfconsistentHighorderSpatial,Morgan_2022_SimulationReynoldsaveragedNavierStokes,Morgan_2023_TwoSelfsimilarReynoldsstress,Banerjee_2010_DevelopmentValidationTurbulentmix,Denissen_2012_ImplementationValidationBHR,Schwarzkopf_2016_TwolengthScaleTurbulence,Braun_2021_MultispeciesTurbulenceModel}.
One of the lesser analysed components of the closure for the RANS models is the effect of bulk compression. The recommended modification by Reynolds \cite{Reynolds_1980_ModelingFluidMotions,Reynolds_1987_FundamentalsOfTurbulence} for the dissipation rate based upon isotropic compression in the rapid distortion regime for homogenous turbulence is equivalent to the term by Dimonte and Tipton \cite{Dimonte_2006_KLTurbulenceModel} for the turbulent length scale derived for conservation of mass of an eddy under isotropic compression \cite{Morel_1982_ModelingTurbulenceInternal}. Investigating isotropic compression of homogenous turbulence for sudden viscous dissipation, Campos and Morgan \cite{Campos_2019_DirectNumericalSimulation} used the same expression for bulk compression but found it necessary to introduce a variable viscosity term in the length scale transport equation to align with DNS results. As noted previously, in practical applications turbulent mixing layers are neither homogenous or experiencing isotropic strain. 

The goal of this paper is to quantify the ability of the K-L turbulence model to capture the characteristics of turbulent mixing layers under anisotropic strain rates and propose an improved model formulation. The work focuses on the bulk compression term for the scaling of the turbulent length scale under strain. Section \ref{sec:ComputationalApproach} introduces the equations of the  K-L model, the method used to apply the uniform strain rates to the simulation domain, and the simulation cases conducted. In Sec. \ref{sec:ResultsDiscussion}, the performance of the K-L model with different scaling approaches is presented, showing results for the integral width and the domain integrated turbulent kinetic energy. Further insight into the K-L model is obtained by performing a self-similar analysis of the K-L model to reduce it down to a buoyancy-drag model. The conclusions are summarised in Sec. \ref{sec:Conclusion}.


\section{Computational Approach}
\label{sec:ComputationalApproach}

\subsection{Strain rates}
\label{subsec:StrainRates}
In order to analyse the influence of the strain rates on the development of the turbulent mixing layer, it is necessary to quantify and control the strain rate. The application of uniform normal strain rates are achieved by applying mean velocity gradients to the simulation domain and moving the domain bounds with the specified strain rates, using boundary conditions which preserve the linear velocity profile. For planar geometry, the strain rates are considered separately in the axial direction ($\bar{S}_A$), representing strain rates in the direction of mixing for the inhomogeneous turbulent mixing layer, and in the transverse direction ($\bar{S}_T$), corresponding to strain rates in the directions of the homogeneous plane.
\begin{eqnarray}
    \bar{S}_A & =&  \frac{\partial \bar{u}_1}{\partial x_1} \\
    \bar{S}_T & =&  \frac{\partial \bar{u}_2}{\partial x_2} = \frac{\partial \bar{u}_3}{\partial x_3}
\end{eqnarray}
A combination of axial and transverse strain is possible, such as for isotropy ($\bar{S}_A=\bar{S}_T$) or for bulk incompressibility with zero divergence ($\bar{S}_A + 2\bar{S}_T=0$).

Two different strain-profiles are utilised. The first is the constant velocity profile, which is the default strain-profile used as it satisfies the momentum equation with uniform pressure. The constant velocity profile corresponds to the case where the domain extent expands/compresses with a constant velocity. The strain rate of this profile varies with time, which can be calculated from the initial strain rate,
\begin{align}
    \bar{S}(t>t_0) = \frac{\bar{S}_0}{1+\bar{S}_0 (t-t_0)}
\end{align}
where $\bar{S}_0$ is the strain rate at an initial time of $t_0$. The second strain profile applies a constant strain rate in time. This is achieved by including a source term to help accelerate the flow to the desired profile, causing the domain extent to expand/compress exponentially.
The source term allows growth without a pressure gradient, thus avoiding the Rayleigh-Taylor instability and isolating the influence of the strain rate on the RMI-induced turbulence.

The strain rates are non-dimensionalised by the inverse of the factor used for the non-dimensionalisation of time:
\begin{align}
    \tau &= \frac{t \dot{W}_0}{\bar{\lambda}} \\
    \hat{S} &= \frac{\hat{S} \bar{\lambda}}{\dot{W}_0}
\end{align}
The non-dimensional time, $\tau$, is calculated by the initial eddy turnover time for the RMI-induced mixing layer which depends on the initial linear growth rate of the integral width, $\dot{W}_0$, and the initial mean wavelength of the mixing layer, $\bar{\lambda}$. This linear transformation preserves the relation $\bar{S} t = \hat{S}\tau$, which is useful for calculating the expansion of the domain as denoted by the expansion factor,
\begin{align}
    \Lambda(t) = \exp\left[\int_0^t\bar{S}(t')dt'\right]=\exp\left[\int_0^\tau\hat{S}(\tau')d\tau'\right] 
\end{align}
which gives the change in length ratio in the direction of the specified strain rate.
\subsection{Governing Equations}
\label{subsec:GoverningEquations}

The simulations conducted use the K-L model, a two-equation RANS model that achieves closure by introducing transport equations for the turbulent length scale  $L$ and turbulent kinetic energy $\tilde{K}=\widetilde{u_i'' u_i''}/2$. The model is derived from the compressible, multi-component Navier-Stokes equations using Favre averages. The Reynolds average decomposition of a scalar is given by,
\begin{align}
    \phi = \bar{\phi} + \phi'
\end{align}
where $\bar{\phi}$ is the Reynolds average and $\phi'$ is the zero-mean fluctuation, $\overline{\phi'} = 0$. The Favre average is a density weighted average defined by,
\begin{align}
    \widetilde{\phi} = \overline{\rho \phi}/\bar{\phi}
\end{align}
that allows a decomposition into the form,
\begin{align}
    \phi = \widetilde{\phi} + \phi''
\end{align}
with a fluctuation term that does not have a zero mean, $\overline{\phi''} \ne 0$, instead it has a zero-mean when density weighted, $\overline{\rho \phi''} = 0$. The governing equations of the K-L model are given by:
\begin{subequations}
     \label{eqn:KLModel}
\begin{eqnarray}
    \frac{\partial \bar{\rho}}{\partial t} + \frac{\partial }{\partial x_j} \left(\bar{\rho}\tilde{u}_j\right) &=& 0\\
    \frac{\partial \bar{\rho}\tilde{u}_i}{\partial t} + \frac{\partial }{\partial x_j} \left( \bar{\rho}\tilde{u}_j\tilde{u}_i + \bar{p}\delta_{ij}\right) &=& \frac{\partial \bar{\tau}_{ij}}{\partial x_j} + \bar{\rho} g_i\\
    \frac{\partial \bar{\rho} \tilde{E}}{\partial t} + \frac{\partial }{\partial x_j} \left(\left(\bar{\rho}\tilde{u}_j + \bar{p}\right)\tilde{u}_j\right)&=& \frac{\partial}{\partial x_j} \left( \frac{\mu_T}{N_H} \frac{\partial \tilde{h}}{\partial x_j} + \frac{\mu_T}{N_K} \frac{\partial \tilde{K}}{\partial x_j} \right) + \frac{\partial \bar{\tau}_{ij} \tilde{u}_i}{\partial x_j} + \bar{\rho} g_i \tilde{u}_i\\
    \frac{\partial \bar{\rho} \tilde{Y}_a}{\partial t} + \frac{\partial }{\partial x_j} \left( \bar{\rho} \tilde{Y}_a \tilde{u}_j \right) &=& \frac{\partial}{\partial x_j} \left(\frac{\mu_T}{N_Y} \frac{\partial \tilde{Y}_a}{\partial x_j} \right)\\
    \frac{\partial \bar{\rho} \tilde{K}}{\partial t} + \frac{\partial}{\partial x_j} \left( \bar{\rho} \tilde{u}_j \tilde{K} \right)  &=& \frac{\partial}{\partial x_j} \left( \frac{\mu_T}{N_K} \frac{\partial \tilde{K}}{\partial x_j} \right) + \bar{\tau}_{ij} \frac{\partial \tilde{u}_i}{\partial x_j} + S_K - C_D \bar{\rho} \frac{\tilde{V}^3}{L} \label{eqn:K_eqn} \\
\frac{\partial \bar{\rho} L}{\partial t} + \frac{\partial}{\partial x_j} \left(\bar{\rho} \tilde{u}_j L \right) &=& \frac{\partial}{\partial x_j} \left( \frac{\mu_T}{N_L} \frac{\partial L}{\partial x_j} \right) + C_L \bar{\rho} V + C_C \bar{\rho} L \frac{\partial \tilde{u}_j}{\partial x_j}
\end{eqnarray}
\end{subequations}

The turbulent velocity is defined from the turbulent kinetic energy, $\tilde{V} = \sqrt{2\tilde{K}}$. The equations have been closed by using a turbulent viscosity with the gradient diffusion approximation,
\begin{align}
    \mu_t = C_\mu \bar{\rho} L \sqrt{2 \tilde{K}}.
\end{align}
The Reynolds stress tensor, $\bar{\tau}_{ij}=-\bar{\rho}\widetilde{u_i'' u_j''}$, is defined with the Boussinesq eddy viscosity assumption, using the mean velocity gradients and turbulent kinetic energy. The coefficient $C_P$ is taken to be $2/3$ to ensure the trace of the Reynolds stress tensor is equal to the turbulent kinetic energy.
\begin{align}
    \bar{\tau}_{ij} = \mu_t \left(\frac{\partial \tilde{u}_i}{\partial x_j} + \frac{\partial \tilde{u}_j}{\partial x_i} - \frac{2}{3} \frac{\partial \tilde{u}_k}{\partial x_k} \delta_{ij} \right) - C_P \bar{\rho} \tilde{K} \delta_{ij} \label{eqn:TauReynolds}
\end{align}
The model differs from the original K-L model by Dimonte and Tipton \cite{Dimonte_2006_KLTurbulenceModel}, instead using the enthalpy diffusion and the equation for total energy conservation given in Ref. \cite{Kokkinakis_2015_TwoequationMultifluidTurbulence}. The source term for the turbulent kinetic energy, $S_K$, is required to differentiate between the different forms of acceleration in RMI and RTI. The method implemented uses the approach by Ref. \cite{Xiao_2020_ModelingTurbulentMixing}, which is based on the work in Ref. \cite{Kokkinakis_2015_TwoequationMultifluidTurbulence}. The source term takes the form,
\begin{align}
    S_K = \left\{ \begin{array}{ll} 
        C_B \bar{\rho} \tilde{V} ~\text{max}\left(A_{L_i} g_i,0 \right), \quad & \Theta_{g_f} \ge \Theta_{g_m} \\ 
        C_B \bar{\rho} \tilde{V} | A_{L_i} g_{L_i} |, \quad & \Theta_{g_f} < \Theta_{g_m}
        \end{array} \right.
\end{align}
where $g_{L_i}=-(1/\bar{\rho})\partial\bar{p}/\partial x_i$ is the local acceleration calculated from the pressure gradient, and $A_{L_i}$ is the local Atwood number to be calculated from the local flow field. The flow is considered RT-like if the turbulent acceleration, $\Theta_{g_f} = \sqrt{\tilde{K}}/\Delta t^*$, is greater than the mean-field acceleration, $\Theta_{g_m} = (1/\bar{\rho}) |\partial p/\partial x|$. The minimum timescale, $\Delta t^*$, is calculated by,
\begin{align}
    \Delta t^* = \min \left( \frac{\min(\Delta x_i) CFL}{|\tilde{u}+\bar{c}|}, \frac{L}{\bar{c}} \right),
\end{align}
using the $CFL$ number, the minimum side length of the cell, $\Delta x_i$, and the local speed of sound, $\bar{c}$. The local Atwood number is calculated from densities reconstructed at the faces using van Leer's monotonicity principle \citep{Kokkinakis_2015_TwoequationMultifluidTurbulence}. 
The second-order slope for the density within each cell is calculated according to
\begin{equation}
    \Delta_i = \left\{\begin{array}{ll}
        \text{sign}(\Delta_{i+\frac{1}{2}})\min(|\Delta_{i-\frac{1}{2}}|,|\Delta_{i+\frac{1}{2}}|), & \text{if } \text{sign}(\Delta_{i-\frac{1}{2}})= \text{sign}(\Delta_{i+\frac{1}{2}})\\
        0, & \text{otherwise }
        \end{array}\right\}
\end{equation}
where $\Delta_{i-\frac{1}{2}}=\bar{\rho}_i-\bar{\rho}_{i-1}$ is the unrestricted difference between the adjacent cells. The reconstructed states at each side of the cells are:
\begin{align}
    \bar{\rho}_i^L = \bar{\rho}_i - \frac{1}{2}\Delta_i, \qquad \bar{\rho}_i^R = \bar{\rho}_i + \frac{1}{2} \Delta_i
\end{align}
Using the reconstructed densities at each face, the values from either side of the face are averaged to give the face density:
\begin{equation}
    \bar{\rho}_{i+\frac{1}{2}} = \frac{1}{2}\left(\bar{\rho}^R_i + \bar{\rho}^L_{i+1}\right)
\end{equation}

The local Atwood number is calculated based upon an initial estimate and a steady-state estimate,
\begin{align}
    A_{L_i} = (1-w_L)A_{0_i} + w_L A_{SS_i}.
\end{align}
 The local initial Atwood number is calculated from the face-averaged density values, which in one-dimension may be written as,
\begin{align}
    A_{0_i} = \left(\frac{\bar{\rho}_{i+\frac{1}{2}} -\bar{\rho}_{i-\frac{1}{2}}}{\bar{\rho}_{i+\frac{1}{2}} + \bar{\rho}_{i-\frac{i}{2}}}\right)_i
\end{align}
The self-similar local Atwood number uses density gradient calculated from the mean density values at the faces. The one dimensional calculation takes the form,
\begin{align}
    \left(\frac{\partial \bar{\rho}} {\partial x}\right)_i &= \frac{\bar{\rho}_{i+\frac{1}{2}} - \bar{\rho}_{i-\frac{1}{2}}}{\Delta x},\\
    A_{SS_i} &= C_A \frac{L}{\bar{\rho} + L | \partial \bar{\rho}/{\partial x} |_i} \left(\frac{\partial \bar{\rho}}{\partial x}\right)_i.
\end{align}

The transition between the two Atwood estimates is controlled by the weighting factor $w_L=\text{min}(L/\Delta x, 1)$. The choice of coefficients will be discussed in Sec. \ref{subsec:ProblemDescription}. 

\subsection{Numerical Implementation}
\label{subsec:NumericalImplementation}

The K-L model is implemented into FLAMENCO, a multi-block structured finite-volume code. The code uses an approach based on the Godunov method \cite{Godunov_1976_NumericalSolutionMultidimensional} to evaluate the inviscid fluxes of the governing equations, corresponding to the spatial fluxes on the left-hand-side of Eq. \ref{eqn:KLModel}. For the inviscid fluxes, the values are reconstructed at the faces using a nominally fifth-order scheme \citep{Kim_2005_AccurateEfficientMonotonic}, and modified with a low-Mach correction to ensure proper dissipation rate scaling as the Mach number decreases \cite{Thornber_2008_EntropyGenerationDissipation,Thornber_2008_ImprovedReconstructionMethod}. The flux is calculated using the HLLC Riemann solver \cite{Toro_1994_RestorationContactSurface}. Viscous fluxes are evaluated using second-order centred differences, and the time-stepping is second order using a total variation diminishing Runge-Kutta method \citep{Spiteri_2002_NewClassOptimal}. To apply the strain rates to the simulation domain, the mesh is moved with the applied strain rate, stretching or compressing according to the theoretical profile. To preserve the uniform strain rate in the flow, moving symmetry planes are applied in the direction of the strain rates, allowing the ghost cells to preserve the specified velocity gradient. The flux calculation is altered using the arbitrary Lagragian-Eulerian approach, modifying the velocity component normal to the face to be the velocity relative to the face velocity for the HLLC Riemann Solver \cite{Luo_2004_ComputationMultimaterialFlows}.

\subsection{Problem Description}
\label{subsec:ProblemDescription}

To investigate the capability of the K-L model to capture the effects of the uniform strain rates on the development of the turbulent mixing layer, one-dimensional RANS simulations are conducted with applied mean strain rates.
Previous implicit large eddy simulations have been conducted looking at the effect of the normal strain rates on the development of the Richtmyer-Meshkov induced mixing layer \cite{Pascoe_2024_ImpactAxialStrain,Pascoe_2025_ImpactTransverseStrain}.
These simulations were based upon the quarter-scale $\theta$-group case \cite{ThetaGroup} which simulated a narrowband multi-mode Richtmyer-Meshkov instability with a $Ma=1.8439$ shock-wave initialising the instability in a heavy-to-light configuration, conducted at $At=0.5$ with an ideal gas equation of state for the fluids. 
The strained simulations were conducted separately for axial strain rates \cite{Pascoe_2024_ImpactAxialStrain}, and transverse strain rates \cite{Pascoe_2025_ImpactTransverseStrain}.
At an early, near self-similar time of $\tau =1$, the strain rates are applied by modifying the velocity profile to achieved the desired velocity gradient, and the domain begins to move with the specified profile to preserve the velocity gradients.

The RANS simulation cases are based on the ILES data in Refs. \cite{ThetaGroup,Pascoe_2024_ImpactAxialStrain,Pascoe_2025_ImpactTransverseStrain}. The ILES cases consist of an axial strain subset with eight strain cases, and a transverse strain subset with eight cases. These strain subsets uses the same strain rate and strain profile combinations for the axial and transverse strain applications. Each case is labelled according to the strain direction, strain profile, and relative strain rate: $D$C$X\pm n$, where $D$ indicates direction (A for axial and T for transverse, C$X$ indicates the strain profile (CV for constant velocity and CS for constant strain rate), and $\pm n$ indicates the ranked magnitude of the applied strain rate ($+n$ is expansive, $-n$ is compressive, larger $n$ is larger magnitude strain rate). The specific strain rates for each case and the simulation run time are listed in Table \ref{tab:SimCases}.

\begin{table}
  
  \caption{\label{tab:SimCases}Properties of the implicit large eddy simulation cases. Simulations cases exist for \textit{D}=A for the axial strain cases, and \textit{D}=T for the transverse cases.}
  \begin{ruledtabular}
  \begin{tabular}{llddc}
    & & \multicolumn{2}{c}{\mbox{Strain rate}}\\ 
    Label   & Strain profile & \mbox{$\bar{S}$ $(s^{-1})$}&  \mbox{$\hat{S}$} & Final $\tau$\\ 
    \hline
    S$0$ & Unstrained & 0.0      & 0.0 & 35\\
    \textit{D}CV$-2$ & Constant Velocity & -2.5     & -0.051 & 10\\
    \textit{D}CV$-1$ & Constant Velocity & -0.625   & -0.013 & 35\\
    \textit{D}CV$+1$ & Constant Velocity & 1.25     & 0.025 & 35\\
    \textit{D}CV$+2$ & Constant Velocity & 5.0      & 0.102 & 10\\
    \textit{D}CS$-2$ & Constant Strain   & -4.00    & -0.081 & 10\\
    \textit{D}CS$-1$ & Constant Strain   & -1.00    & -0.020 & 35\\
    \textit{D}CS$+1$ & Constant Strain   & 1.00     & 0.020 & 35\\
    \textit{D}CS$+2$ & Constant Strain   & 4.00     & 0.081 & 10\\
  \end{tabular}
  \end{ruledtabular}
\end{table}

Whilst the ILES cases are three-dimensional, the homogeneity of the $y$--$z$ planes allows the RANS simulations to be conducted as one-dimensional, using a mesh of 720 cells in the $x$-direction. The domain is initialised using the planar-averages of the initial conditions for the quarter-scale $\theta$-group case. The fluids are set up in a heavy-to-light configurations with densities of 3\,kg\,m$^{-3}$ and 1\,kg\,m$^{-3}$ to achieve a 0.5 Atwood number. The shocked heavy fluid is initialised for $x<3.5$\,m using a $Ma=1.8439$ shock for a pressure ratio of four. The interface between the two fluids is located at $x_0=3.5$\,m with a diffuse error function profile. As the mean profile is applied, the diffusive thickness of the initial interface is taken to use the standard deviation of the ILES perturbation spectrum as opposed to the original diffuse thickness. The volume fraction profile used to initialise the mass fraction profile is
\begin{align}
    \bar{f}_1(x) = \frac{1}{2} \left(1 - \text{erf}\left( \frac{x-x_0}{\sqrt{2} \sigma_0}\right) \right).
\end{align}
The standard deviation $\sigma_0$ is $0.1\lambda_\text{min}$, where the initial perturbation wavelengths ranged between $2\pi/32$ and $2\pi/16$.
The initial turbulent kinetic energy is set to a nominal value of $\tilde{K}=10^{-40}$m\,s$^{-2}$ throughout the domain, whilst the initial turbulent length scale is calibrated to reproduce the integral width of the unstrained ILES case, i.e. the original quarter-scale $\theta$-group simulation. The details of for the turbulent length scale are further discussed below. 
Both fluids use an ideal gas equation of state with $\gamma=5/3$ and have an unshocked pressure of 100\,kPa. The whole domain has an initial velocity offset of $u_1 = -291.575$\,m\,s$^{-1}$ to counteract the velocity shift from the interface-shock interaction, making the post-shock interface stationary. At the non-dimensional time $\tau=1$ the strain is applied to the simulation domain, just as was performed with the ILES cases.

The K-L model, like most RANS models, is not capable of effectively capturing the growth rate of the mixing layer at all stages of the instability evolution. 
Instead the focus during model calibration is on predicting the late-time growth rate for the mixing layers as opposed to the linear regime growth rate.
This is observed in the simulations conducted, as the K-L model will underestimate the growth during the transitional period.
Using the K-L model coefficients posited in Refs. \cite{Xiao_2020_ModelingTurbulentMixing,Zhang_2020_MethodologyDeterminingCoefficients}, it is possible to capture the integral width of the unstrained case at late time, however the turbulent kinetic energy results are out by an order of magnitude.
To provide a better comparison with the ILES data, the coefficients of the model were re-calculated so that both the integral width and turbulent kinetic energy results are aligned for the unstrained simulations.
The coefficients used for the simulations in the present study are listed in Table \ref{tab:KL_coeff}, which were derived using the same methodology laid out in the aforementioned articles. This includes using $\theta=1/4$, which is lower than the value of $\theta=0.291$ measured in the late-time \cite{ThetaGroup} or the value of $\theta=1/3$ used in the buoyancy-drag model \cite{Youngs_2020_BuoyancyDragModelling,Youngs_2020_EarlyTimeModifications}. The value of $\theta$ is time-dependent for the quarter-scale case, likely reaching a value of $\theta=1/4$ for even longer simulation times.
The calibrating coefficient of $\alpha$, which scales the Rayleigh-Taylor growth, was adjusted from the default value of 0.05 to 0.01 to improve the turbulent kinetic energy calculation.
This is a smaller value than $\alpha=0.025$, which is the observed coefficient for narrowband RTI simulations.
This adjustment allows the K-L model to simultaneously capture the mixing zone growth and the decay of the turbulence, however it causes the model to underestimate the Rayleigh-Taylor growth rate.
The ability of the K-L model to capture the influence of the strain rate on the RMI-induced mixing layer is not affected by this recalibration, as simulations conducted with the original coefficients of Refs. \cite{Xiao_2020_ModelingTurbulentMixing,Zhang_2020_MethodologyDeterminingCoefficients} demonstrated the same trends as observed in the results presented here-within. 
The choice of coefficients employed simply allow the turbulent kinetic energy comparison between the ILES and RANS results to be conducted more easily as the unstrained profiles are calibrated to be aligned. Using the original coefficients and matching the integral width causes the turbulent kinetic energy from the RANS to be around a factor of five smaller than the ILES results.

\begin{table}
  \caption{Coefficients for the K-L turbulence model}
  \begin{ruledtabular}
  \label{tab:KL_coeff}
  \begin{tabular}{lcccccccccc}
    Model & $C_A$ & $C_B$ & $C_D$ & $C_L$ & $C_P$ & $C_\mu$ & $N_H$ & $N_K$ & $N_L$ & $N_Y$ \\
    \hline
    Refs. \cite{Xiao_2020_ModelingTurbulentMixing,Zhang_2020_MethodologyDeterminingCoefficients} & 11.2  & 0.76  & 0.2   & 0.19  & 2/3   & 1.19    & 0.35  & 0.43  & 0.04  & 0.35\\
    Current & 5.01 & 0.34  & 0.2   & 0.19  & 2/3   & 1.19    & 1.74  & 2.14  & 0.19  & 1.74
  \end{tabular}
  \end{ruledtabular}
\end{table}

In the prescribed model, there are two coefficients which are not determined from the self-similarity analysis. The first is $C_P$, which is the contribution of the turbulent kinetic energy to the Reynolds stress components. To ensure that the trace of the Reynolds stress tensor equals the turbulent kinetic energy, the value of $C_P$ is taken to be $2/3$. The second term is the coefficient for the bulk compression of the turbulent length scale, $C_C$. The value of this term is typically $1/3$ under the assumption of conservation of mass in an eddy under isotropic compression \citep{Dimonte_2006_KLTurbulenceModel}. If the mass inside the eddy is given by $\rho L^3$, then for isotropic compression
\begin{align}
    \frac{d L}{dt} = \frac{1}{3} L \frac{\partial u_k}{\partial x_k}.
\end{align}
More generally, this approach will evolve the turbulent length scale according to the mean strain rate of the local flow field.
In contrast, the multi-fluid model of Andrews \cite{Andrews_1992_ExperimentalStudyTurbulent} scales the length scale with the negative of the strain tensor magnitude, whilst the multi-fluid of Youngs \cite{Youngs_1994_NumericalSimulationMixing} scales with the divergence but also considers the direction of the strain, using the strain in the direction of mixing as determined by the gradient of the species' volume or mass fraction.
In the cases where the expansion/compression only occurs in $n$ dimensions, Morel \& Mansour \cite{Morel_1982_ModelingTurbulenceInternal} consider the conservation of $\rho L^n$ instead. The resulting evolution of the length scale is
\begin{align}
    \frac{\text{d}L}{\text{d}t} = \frac{1}{n}L \frac{\partial u_k}{\partial u_k}.
\end{align}
Under anisotropic strain rates, length scales in different directions will vary depending upon the strain rate that is aligned with the length scale.
Considering the axial and transverse strain rates, each strain rate will modify the corresponding length scale.
From this, two alternative closure for the bulk compression term are introduced, which use either the axial strain rate or the transverse strain rate:
\begin{align}
    \frac{d L}{dt} &= L \frac{\partial u_1}{\partial x_1}=L S_A, \\
    \frac{d L}{dt} &= L \frac{1}{2} \left(\frac{\partial u_2}{\partial x_2} + \frac{\partial u_3}{\partial x_3} \right)=L S_T.
\end{align}
Using the approach of Mansel \& Mansour \cite{Morel_1982_ModelingTurbulenceInternal}, the axial closure would be obtained if only axial strain was expected, whilst the transverse closure would be obtained under transverse strain. For implementation purposes, the strain rates used in the bulk compression term are derived from the Favre-averaged strain rates, $\tilde{u}_i$, as these are the quantities readily available from the transport equation. An alternative approach could be to use the Reynolds-averaged velocity field which does not include the density weighting. This Reynolds-average velocity could be calculated more easily for a K-L-a model, using the calculation $\bar{u}_i = \tilde{u}_i-a_i$ for the turbulent mass flux $a_i$. The effects of the choice of formulation could be investigated, however this would be best performed with a case at a higher Atwood number to further exacerbate these effects.

These alternative closures for the bulk compression of the turbulent length scale can also be considered as using different $C_C$ coefficients for the different cases. An equivalent $C_C$ coefficient can be calculated for each type of strain application,
\begin{align}
    \hat{C}_C \frac{\partial u_k}{\partial x_k} = S_\phi
\end{align}
where $S_\phi$ represents the desired closure strain rate ($S_A$, $S_T$, or $S_I=\nabla\cdot u/3$), and $\hat{C}_C$ is the equivalent coefficient.
These coefficients are given in Table \ref{tab:EquivalentCC}.
For isotropic strain rates, all three closures will reduce to the same scaling as the mean strain rate is equivalent to the strain rate in any direction.
Under pure axial strain, the transverse strain rate closure reduces $\hat{C}_C$ to zero as the bulk compression term will not activate, whilst the axial strain closure increases the coefficient to unity.
The equivalent coefficients assume that the term only activates under bulk compression/expansion, however the utilisation of Favre averaged velocities in the divergence means that it is affected by the mixing of fluids at different densities.

\begin{table}
    \caption{\label{tab:EquivalentCC}The $\hat{C}_C$ coefficients for the different bulk compression closures under different applied strain rates, representing the equivalent $C_C$ coefficient/scaling of the velocity divergence for the specified applied strain rates.}
    \begin{ruledtabular}
    \begin{tabular}{lccc}
        &\multicolumn{3}{c}{Applied strain rate} \\
        Length scale closure                    & Axial strain & Isotropic strain & Transverse strain\\
        \hline
        Axial, $S_A$           & 1             & 1/3               & 0 \\
        Isotropic, $S_I$   & 1/3           & 1/3               & 1/3 \\
        Transverse, $S_T$              & 0             & 1/3               & 1/2 \\
    \end{tabular}
    \end{ruledtabular}
\end{table}

Given the the possibility of the axial, isotropic or transverse strain closure, the simulation cases listed in Table \ref{tab:SimCases} were conducted for each closure, providing three sets of RANS data.
The effect of the different approaches manifests in the initialisation of the flow variables in the domain.
A common approach to initialising the interface for RMI is to apply a non-negligible $L$ at the cells adjacent to the interface and zero elsewhere, whilst $\tilde{K}$ is set to a negligible value everywhere.
When the shock interacts with the interface, the buoyancy production term will activate, generating turbulent kinetic energy at the interface.
The amount of turbulent kinetic energy deposited, and the resulting growth rate, will depend on the initial value of $L$.
The bulk compression term will also activate due to the velocity gradient across the shock, and as a result different initial $L$ values are required under the different scaling schemes, as shown in Table \ref{tab:L_Initial}.
As the shock is compressive and acts in the axial direction, the axial strain closure simulations require a larger initial $L$ value to counteract the resulting bulk compression of $L$ and achieve the same post-shock growth rate prior to the application of strain.
The listed values of $L$ are obtained by calibrating the initial $L$ value in the cells on either side of the interface to predict the unstrained growth rate.
Outside of the cells neighbouring the interface, the value of $L_0=1\times 10^{-16}$\,m is used.
\begin{table}
  \caption{Values of $L$ used for simulation initialisation}
  \begin{ruledtabular}
  \label{tab:L_Initial}
  \begin{tabular}{ccc}
    Axial closure & Isotropic closure & Transverse closure\\
    \hline
    0.54m & 0.53m & 0.52m
  \end{tabular}
  \end{ruledtabular}
\end{table}


\section{Results and Discussion}
\label{sec:ResultsDiscussion}

\subsection{Integral Width}
\begin{figure}
    \centering
    \begin{tikzpicture}
        \node[anchor=north west] (image) at (0,0) { \includegraphics[width=0.8\textwidth]{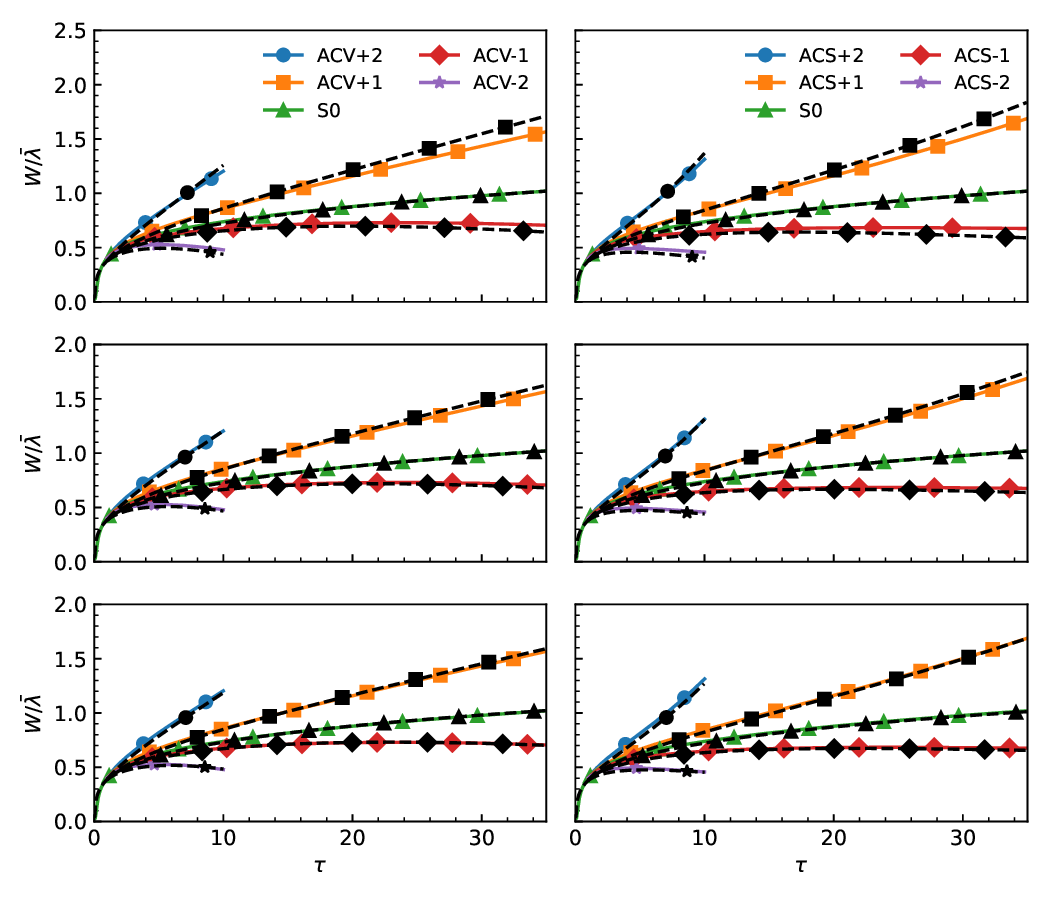}};
        \node[anchor=north west] at (2,0.1) {\large Constant velocity};
        \node[anchor=north west] at (8,0.1) {\large Constant strain rate};
        \node[anchor=north west,text width = 1.8cm,text centered] at (-2,-1.50) {\large \centering Axial Closure};
        \node[anchor=north west,text width = 1.8cm,text centered] at (-2,-5.00) {\large \centering Isotropic Closure};
        \node[anchor=north west,text width = 1.8cm,text centered] at (-2,-8.25) {\large Transverse Closure};
        \node[anchor=north west] at (1.4,-0.6) {\large (\textit{a})};
        \node[anchor=north west] at (7.4,-0.6) {\large (\textit{b})};
        \node[anchor=north west] at (1.4,-4.5) {\large (\textit{c})};
        \node[anchor=north west] at (7.4,-4.5) {\large (\textit{d})};
        \node[anchor=north west] at (1.4,-7.8) {\large (\textit{e})};
        \node[anchor=north west] at (7.4,-7.8) {\large (\textit{f})};
    \end{tikzpicture}
    \caption{Integral width for simulations under the applied axial strain rate with (left) constant velocity profile and (right) constant strain rate profile. Solid lines indicate ILES, dashed lines indicate K-L model with (top) axial closure, (middle) isotropic closure, and (bottom) transverse closure for bulk compression.}
    \label{fig:AxialStrainIWCombined}
\end{figure}
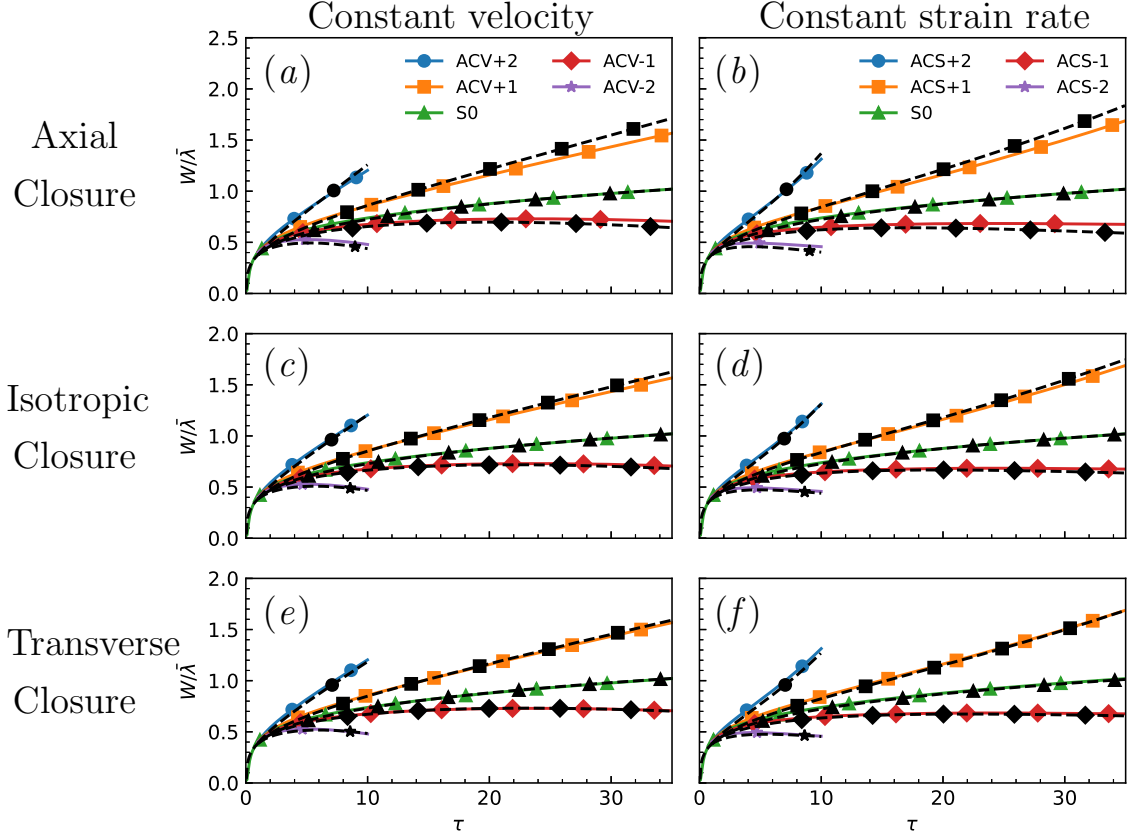

The mixing layer width is measured here using the integral width, which is based upon the mean volume fraction profile,
\begin{align}
    W = \int \bar{f}_1 \bar{f}_2 dx.
\end{align}
The original ILES cases were conducted using the five-equation model of Allaire \textit{et. al} \cite{Allaire_2002_FiveEquationModelSimulation} which includes a transport equation for the volume fraction of each species.
As the K-L model only transports the mean mass fraction, the volume fraction is calculated assuming equal temperature and pressure within the cell, allowing the volume fraction of species 1 to be calculated by
\begin{align}
    \bar{f}_1  = \frac{\tilde{Y}_1/W_1}{\sum_a \tilde{Y}_a/W_a}.
\end{align}
The molecular mass, $W_a$, of each species is $90$\,g~mol$^{-1}$ and $30$\,g~mol$^{-1}$ respectively, inline with the initial density ratio between the fluids.
The integral width results for the simulations with axial strain rates applied are shown in Fig. \ref{fig:AxialStrainIWCombined} for the three K-L models and the ILES results.
The influence of the axial strain rate on the integral width demonstrates an increased integral width for expansive strain rates, as the velocity difference across the mixing layer causes the mixing layer to stretch/compress for positive/negative strain rates \cite{Ge_2022_EvaluatingStretchingCompression}.
The K-L models are unable to capture the early-time growth of the mixing layer as the models are designed for late-time self-similar growth, and as a result are not well designed for the transitional regime.
Despite the underestimation of the mixing layer growth rate in the early time, the K-L models show the same increased growth rate under expansion strain as observed in the ILES.
This stretching/compression effect arises from the velocity difference in the mean flow field which is resolved by the simulation, allowing the simulations to capture this direct effect of the axial strain rate.
The variation between the different bulk compression closure approaches arises from the turbulent growth which is dependent upon the turbulent viscosity, which is proportional to $L$ and $\tilde{K}^{1/2}$.
The value of $L$ is modified by the bulk compression term, with greater variation of $L$ under strain for configurations with larger $\hat{C}_C$ values.
This effect on the mixing layer is evident as the axial strain closure with the applied axial strain ($\hat{C}_C=1$) will have the largest modification of $L$ from the bulk compression term, and is observed to overestimate the influence of the strain rates on the mixing layer.
The model with the best performance, most accurately capturing the turbulent growth, is the transverse strain closure implementation with $\hat{C}_C=0$, effectively not modifying the turbulent length scale for the cases with only axial strain applied.
The default closure with $\hat{C}_C=1/3$ performs quite well for this problem, however the model can be observed to overestimate the influence of the integral width of the mixing layer.
To compare the accuracy of the three implemented models, the mean absolute percentage error (MAPE) between the RANS results and the ILES data is recorded for the low-magnitude strain rate cases (cases ACV$\pm$1 and ACS$\pm1$) at the final time. The high-magnitude strain cases are omitted as their final results are still polluted by the underestimation of the transitional regime by the K-L model. The MAPE of the four cases are recorded in Table \ref{tab:MAPE}, and shows the transverse closure to reduce the MAPE by a factor of four compared to the default isotropic closure, whilst the axial closure performs closer to $2.5\times$ worse.

\begin{table}
  \caption{\label{tab:MAPE} Mean absolute percentage error (MAPE) at the final simulation time for the low-magnitude strain cases. Closure model with lowest MAPE for each strain application are highlighted in bold text.}
  \begin{ruledtabular}
  \begin{tabular}{llcc}
    & & \multicolumn{2}{c}{\mbox{MAPE}}\\ 
    Strain Direction & Closure & \mbox{Integral Width}&  \mbox{TKE} \\ 
    \hline
    Axial       & Axial         & 9.91 & 68.9\\   
    Axial       & Isotropic     & 4.12 & 39.9\\   
    \textbf{Axial}       & \textbf{Transverse}    & \textbf{1.28} & \textbf{25.9}\\   
    Transverse  & Axial         & 9.55 & 47.7\\   
    Transverse  & Isotropic     & 3.77 & 20.4\\   
    \textbf{Transverse}  & \textbf{Transverse}    & \textbf{0.80} & \textbf{~7.2}\\   
  \end{tabular}
  \end{ruledtabular}
\end{table}

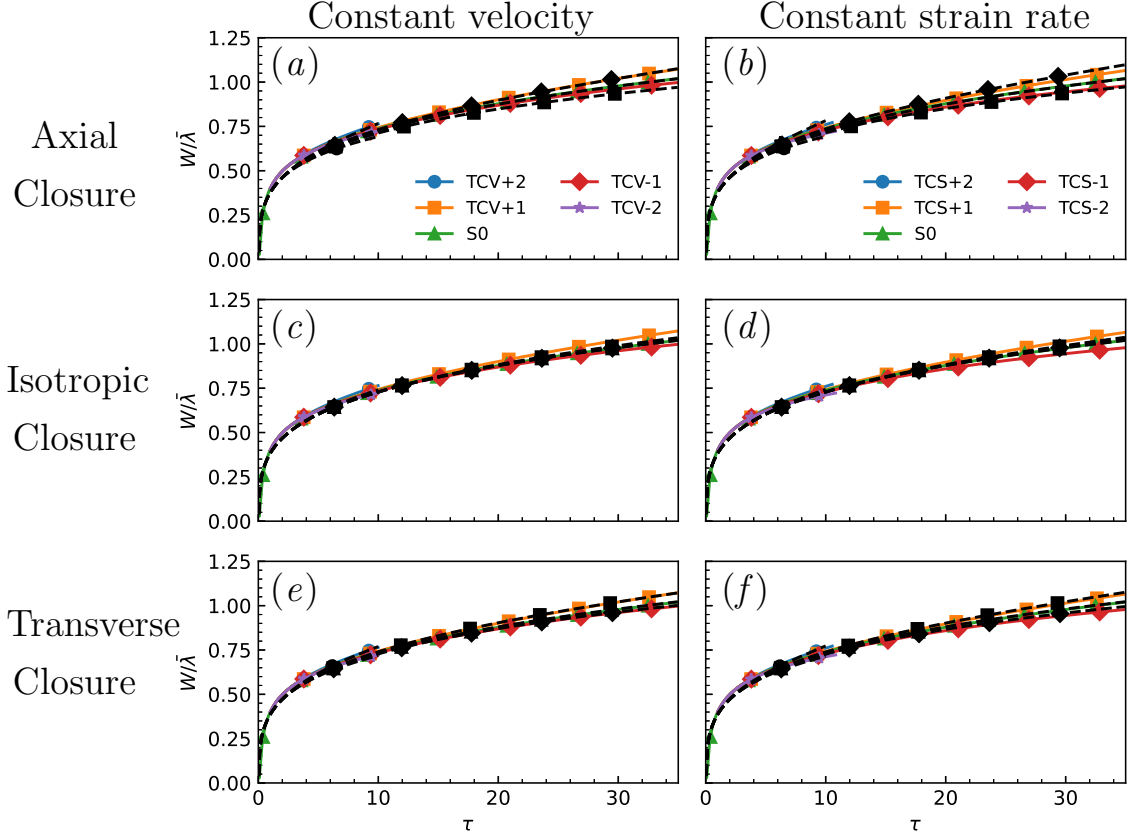
\begin{figure}
    \centering
    \begin{tikzpicture}
        \node[anchor=north west] (image) at (0,0) { \includegraphics[width=0.8\textwidth]{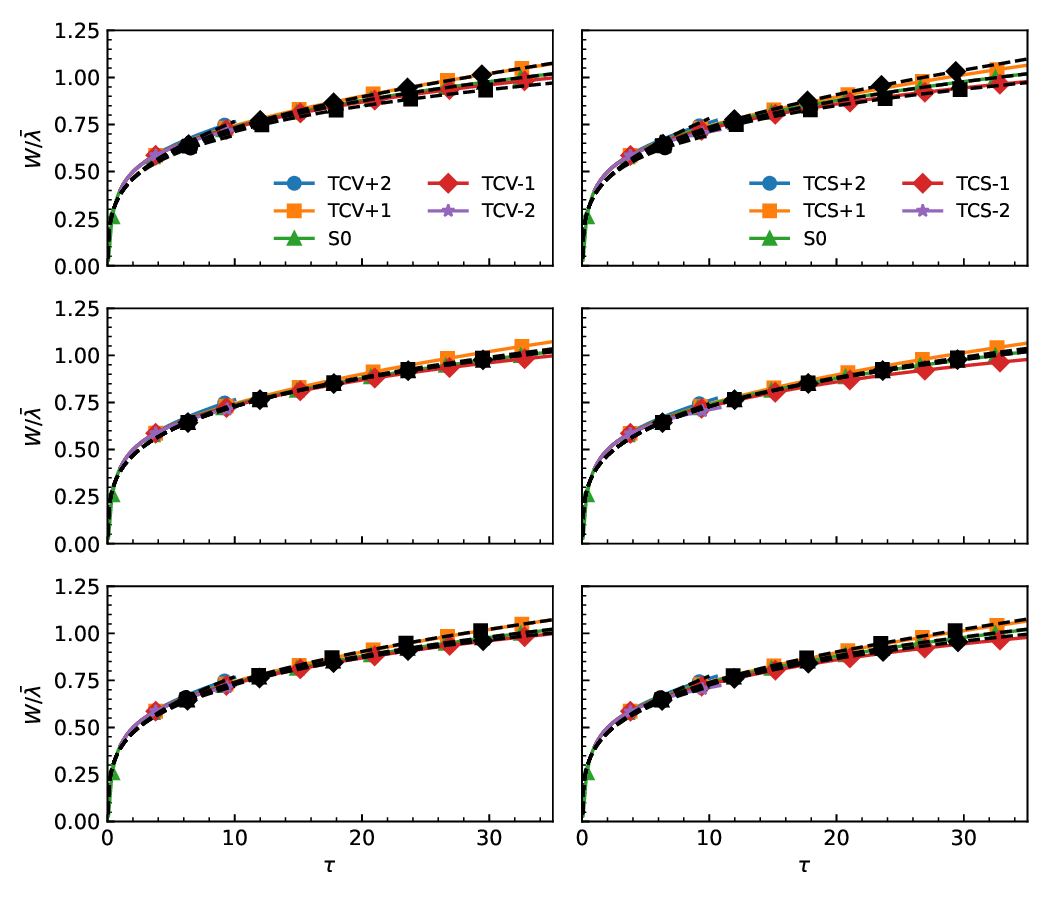}};
        \node[anchor=north west] at (2,0.1) {\large Constant velocity};
        \node[anchor=north west] at (8,0.1) {\large Constant strain rate};
        \node[anchor=north west,text width = 1.8cm,text centered] at (-2,-1.50) {\large \centering Axial Closure};
        \node[anchor=north west,text width = 1.8cm,text centered] at (-2,-4.75) {\large \centering Isotropic Closure};
        \node[anchor=north west,text width = 1.8cm,text centered] at (-2,-8.00) {\large Transverse Closure};
        \node[anchor=north west] at (1.5,-0.5) {\large (\textit{a})};
        \node[anchor=north west] at (7.5,-0.5) {\large (\textit{b})};
        \node[anchor=north west] at (1.5,-4.0) {\large (\textit{c})};
        \node[anchor=north west] at (7.5,-4.0) {\large (\textit{d})};
        \node[anchor=north west] at (1.5,-7.5) {\large (\textit{e})};
        \node[anchor=north west] at (7.5,-7.5) {\large (\textit{f})};
    \end{tikzpicture}
    \caption{Integral width for simulations under the applied transverse strain rate with (left) constant velocity profile and (right) constant strain rate profile. Solid lines indicate ILES, dashed lines indicate K-L model with (top) axial closure, (middle) isotropic closure, and (bottom) transverse closure for bulk compression.}
    \label{fig:TransverseStrainIWCombined}
\end{figure}

Under applied transverse strain rates, the ILES obtains slightly larger growth rates for the expansive strain as compared to the unstrained simulation.
In Ref. \cite{Pascoe_2025_ImpactTransverseStrain}, this trend was considered the result of the turbulence length scale in the simulation stretching under expansion, which in turn effectively decreased the dissipation rate.
A buoyancy-drag model calibrated to the unstrained case \cite{Youngs_2020_EarlyTimeModifications} was modified for the applied transverse strain cases by scaling the unstrained effective drag length scale by the transverse expansion factor, with the resulting model able to accurately predict the integral width \cite{Pascoe_2025_ImpactTransverseStrain}.
For the K-L model simulations the integral width predictions vary, as shown in Fig. \ref{fig:TransverseStrainIWCombined}.
Just as observed for the ILES, the K-L model results show only a small modification to the integral width, as the transverse strain does not directly stretch/compress the mixing layer as the axial strain does.
The isotropic closure shows almost no variation of the integral width, as if the effects of shear production and bulk compression cancel each other out.
Under compression, shear production will deposit turbulent kinetic energy and the bulk compression will reduce the turbulent length scale, which can effectively maintain a steady turbulent viscosity and growth rate.
Using the axial closure, corresponding to $\hat{C}_C=0$, causes the K-L model to incorrectly predict the sign of the change, with the expansion cases growing at a slower rate than the unstrained case.
As there is no bulk compression for this closure and strain combination, the only effect would be shear production, which would decrease the turbulent kinetic energy and turbulent viscosity in turn, producing the observed results for expansive strain.
Using the transverse strain closure improves the performance of the model, aligning to the ILES data more effectively than the isotropic or axial closure.
The transverse closure has a larger effective $C_C$ ($\hat{C}_C=1/2$), causing the turbulent length scale to vary more under the transverse strain.
The MAPE for the low-magnitude strain cases is listed in Table
\ref{tab:MAPE}, and the results confirm that the transverse closure is the most well aligned of the three closure methods.
Similar ratios are observed as for the applied axial strain cases, with the transverse closure of the turbulent length scale attaining a factor of four smaller MAPE than the isotropic closure.

\subsection{Turbulent kinetic energy}

The domain integrated turbulent kinetic energy is calculated by different methods for the ILES and RANS cases. For the ILES, cases the turbulent kinetic energy is evaluated using the Favre velocity fluctuations:
\begin{align}
    TKE = \iiint \frac{1}{2} \rho u_i'' u_i'' dV.
\end{align}
The RANS cases calculate the total turbulent kinetic energy by integrating the transported turbulent kinetic energy variable:
\begin{align}
    TKE = \iiint \bar{\rho} \tilde{K} dV.
\end{align}

\begin{figure}
    \centering
    \begin{tikzpicture}
        \node[anchor=north west] (image) at (0,0) { \includegraphics[width=0.8\textwidth]{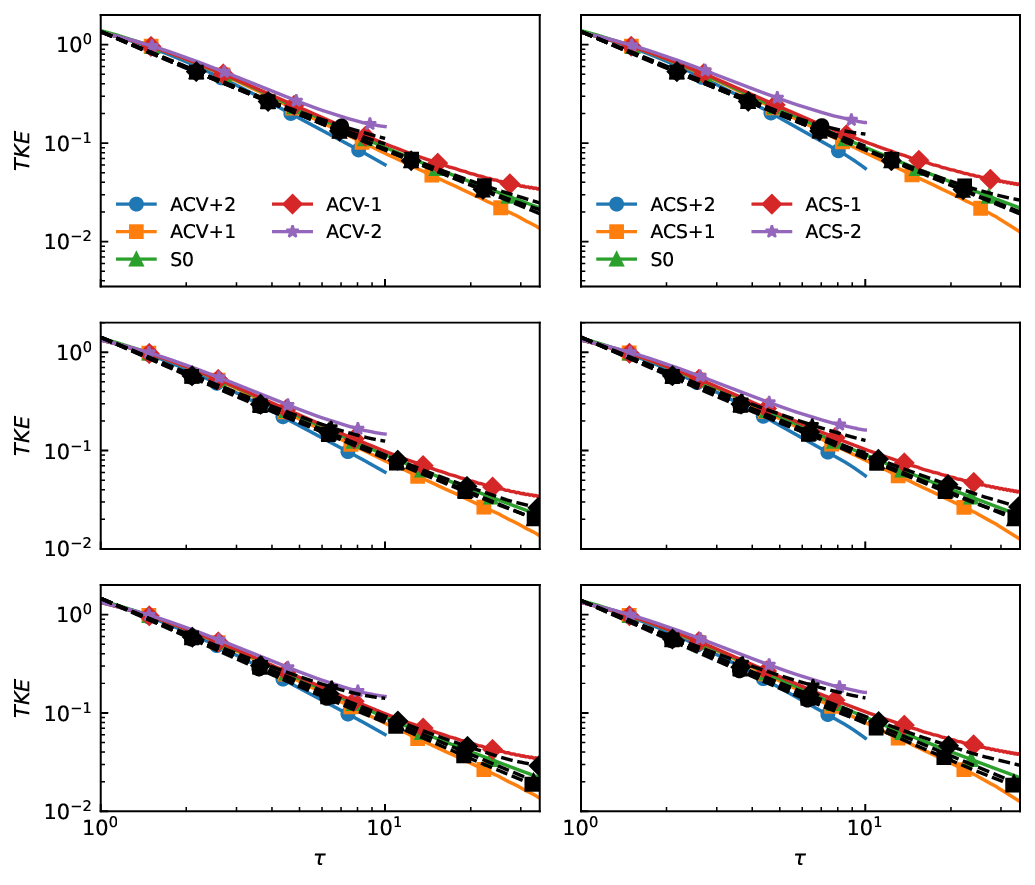}};
        \node[anchor=north west] at (2,0.3) {\large Constant velocity};
        \node[anchor=north west] at (8,0.3) {\large Constant strain rate};
        \node[anchor=north west,text width = 1.8cm,text centered] at (-2,-1.50) {\large \centering Axial Closure};
        \node[anchor=north west,text width = 1.8cm,text centered] at (-2,-5.00) {\large \centering Isotropic Closure};
        \node[anchor=north west,text width = 1.8cm,text centered] at (-2,-8.) {\large Transverse Closure};
        \node[anchor=north west] at (6.0,-0.4) {\large (\textit{a})};
        \node[anchor=north west] at (12.2,-0.4) {\large (\textit{b})};
        \node[anchor=north west] at (6.0,-4.3) {\large (\textit{c})};
        \node[anchor=north west] at (12.2,-4.3) {\large (\textit{d})};
        \node[anchor=north west] at (6.0,-7.6) {\large (\textit{e})};
        \node[anchor=north west] at (12.2,-7.6) {\large (\textit{f})};
    \end{tikzpicture}
    \caption{Turbulent kinetic energy for simulations under the applied axial strain rate with (left) constant velocity profile and (right) constant strain rate profile. Solid lines indicate ILES, dashed lines indicate K-L model with (top) axial closure, (middle) isotropic closure, and (bottom) transverse closure for bulk compression.}
    \label{fig:AxialStrainTKECombined}
\end{figure}
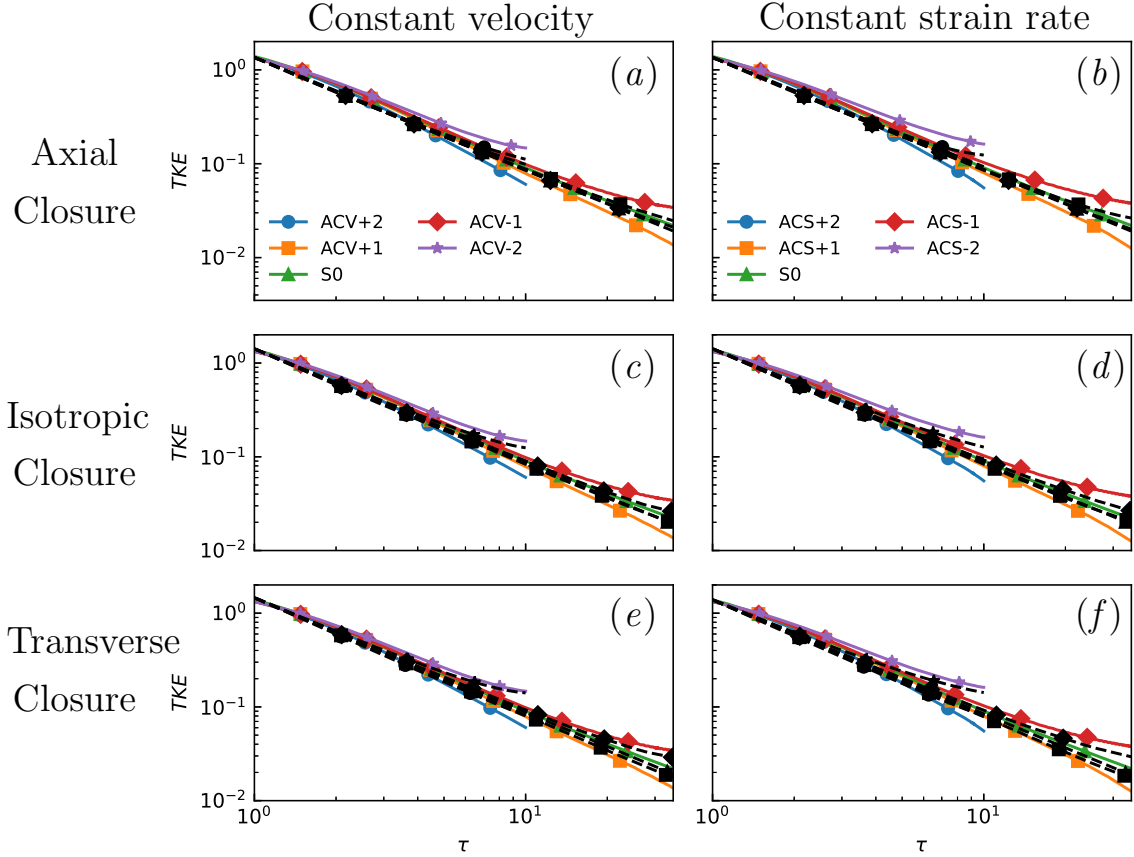

The results for the axially strained cases are plotted in Figure \ref{fig:AxialStrainTKECombined}.
The ILES results show the impact of shear production, with the compressive strain cases obtaining a slight increase in TKE, whilst the compression cases decrease.
The axial closure RANS cases show little variation in the TKE, which can be explained by the change in the turbulent length scale affecting the dissipation rate and counteracting shear production.
In contrast, the transverse closure cases do a better job of capturing the effect of the strain rate on the TKE.
The MAPE for low-magnitude strain cases are calculated in the same manner as was done for the integral width, and is reported in Table \ref{tab:MAPE}.
The results show the same trend as for the integral width, with the transverse closure K-L model performing the best, and the axial closure model performing the worst.
This is not surprising, as for the transverse closure to accurately capture the turbulent growth of the integral width, it would require more accurate representation of the turbulent viscosity and turbulent kinetic energy compared to the other methods.
It is interesting to note that the transverse closure does a better job of capturing the TKE for the compressive cases, as compared to the expansion cases.
As a two-equation model, the turbulent kinetic energy is assumed to be isotropic and is evenly divided across the three normal Reynolds stresses, as indicated by the $-C_P \bar{\rho} \tilde{K} \delta_{ij}$  contribution to $\bar{\tau}_{ij}$ in Eq. \ref{eqn:TauReynolds}.
The expansion cases tend towards isotropy for the TKE as the simulation continues, whilst the compression cases are anisotropic, with the TKE becoming more aligned with the $x$-direction \cite{Pascoe_2024_ImpactAxialStrain}.
It is interesting to note that the model performs better for these anisotropic cases than it does for the isotropic ones. 

\begin{figure}
    \centering
    \begin{tikzpicture}
        \node[anchor=north west] (image) at (0,0) { \includegraphics[width=0.8\textwidth]{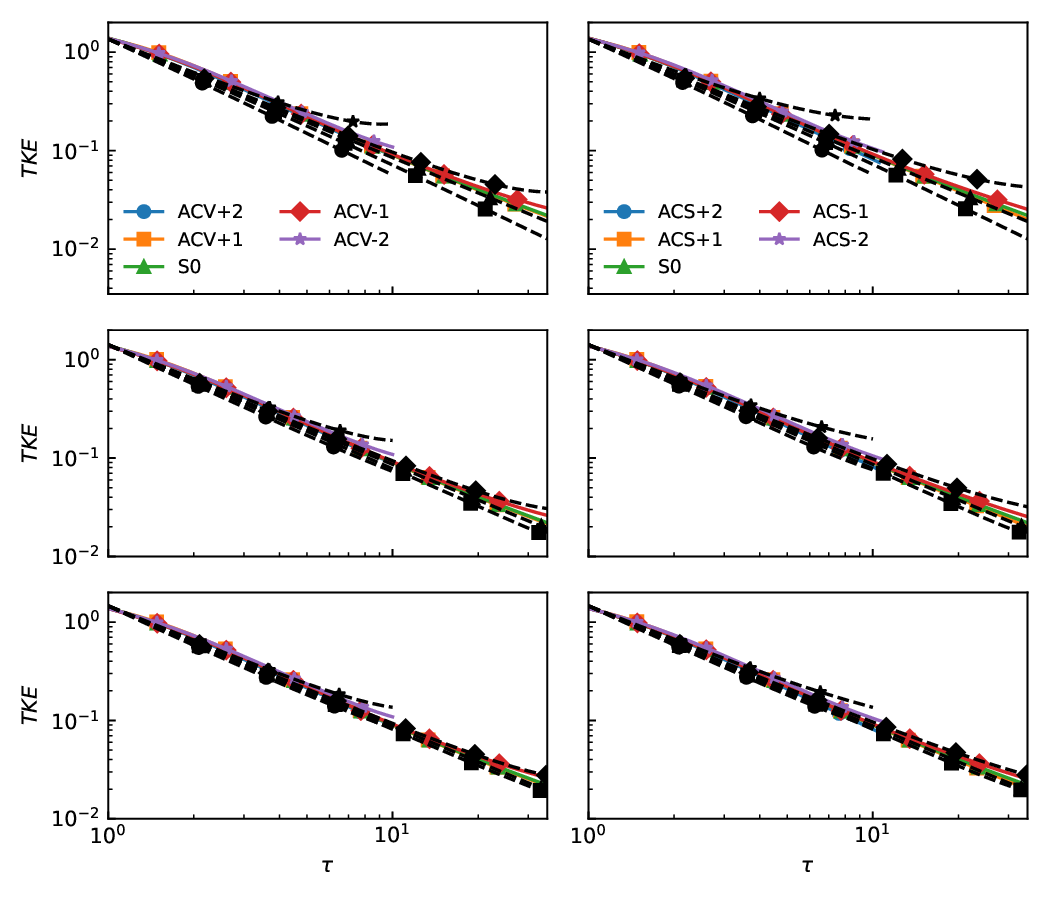}};
        \node[anchor=north west] at (2,0.2) {\large Constant velocity};
        \node[anchor=north west] at (8,0.2) {\large Constant strain rate};
        \node[anchor=north west,text width = 1.8cm,text centered] at (-2,-1.50) {\large \centering Axial Closure};
        \node[anchor=north west,text width = 1.8cm,text centered] at (-2,-5.00) {\large \centering Isotropic Closure};
        \node[anchor=north west,text width = 1.8cm,text centered] at (-2,-8.) {\large Transverse Closure};
        \node[anchor=north west] at (6.0,-0.4) {\large (\textit{a})};
        \node[anchor=north west] at (12.1,-0.4) {\large (\textit{b})};
        \node[anchor=north west] at (6.0,-4.3) {\large (\textit{c})};
        \node[anchor=north west] at (12.1,-4.3) {\large (\textit{d})};
        \node[anchor=north west] at (6.0,-7.6) {\large (\textit{e})};
        \node[anchor=north west] at (12.1,-7.6) {\large (\textit{f})};
    \end{tikzpicture}
    \caption{Turbulent kinetic energy for simulations under the applied transverse strain rate with (left) constant velocity profile and (right) constant strain rate profile. Solid lines indicate ILES, dashed lines indicate K-L model with (top) axial closure, (middle) isotropic closure, and (bottom) transverse closure for bulk compression.}
    \label{fig:TransverseStrainTKECombined}
\end{figure}
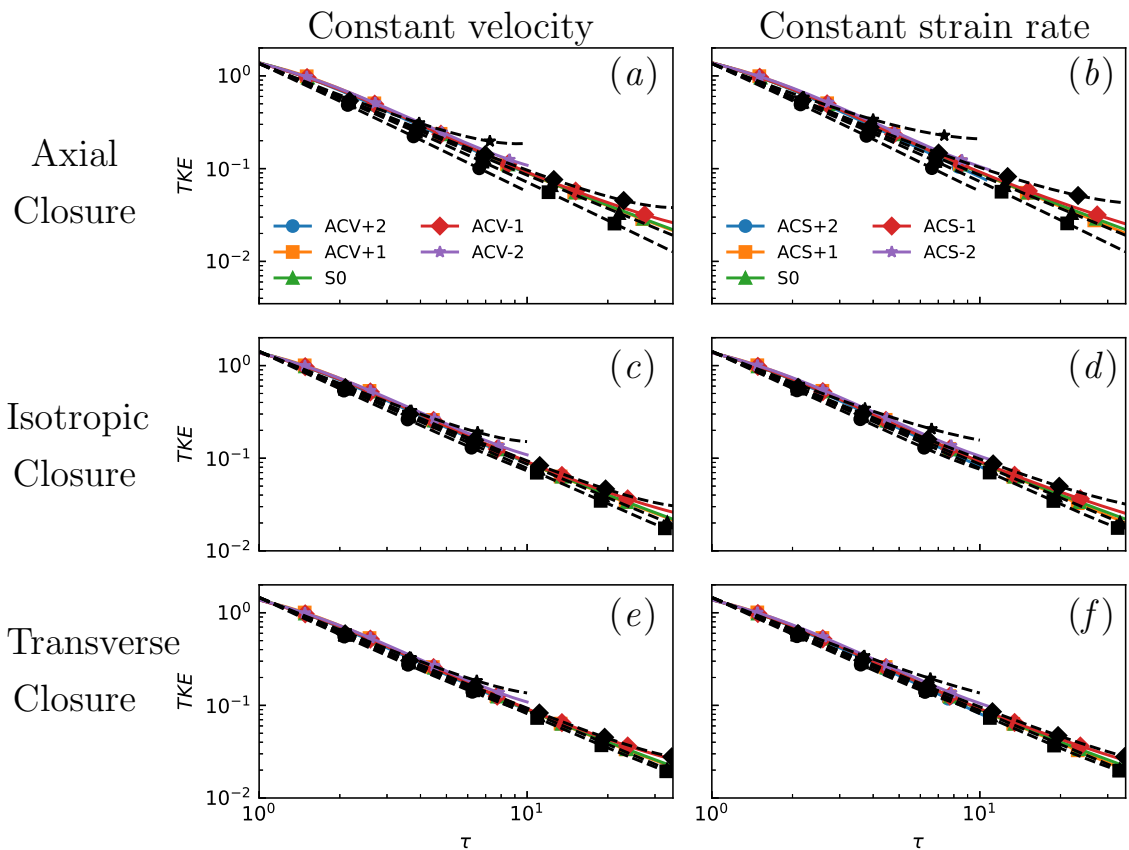

The domain integrated turbulent kinetic energy for the transverse strain cases is shown in Fig. \ref{fig:TransverseStrainTKECombined}. The ILES results show little variation as a result of the change in the turbulent length scale. A simple Reynolds stress model was calibrated to the unstrained case in Ref. \cite{Pascoe_2025_ImpactTransverseStrain}, which showed that the TKE could be accurately estimated if the length scale for the TKE dissipation was made to scale with the geometric mean of the expansion factor in each direction. For compression, this means the length scale decreases and in turn increases the dissipation rate, counteracting shear production that adds energy into the mixing layer from the negative velocity gradients. The counteracting behaviour causes the tight spread observed in the ILES results. Again, the transverse closure appears to do the best job of capturing the tight spread, with the isotropic closure showing a slightly larger variation, and the axial closure overestimating the influence of the strain rate. This is not surprising then, as the axial closure will not modify the turbulent length scale directly ($\hat{C}_C=0$), so there is no counteracting behaviour to shear production. The MAPE is reported in Table \ref{tab:MAPE}, with the transverse closure attaining the lowest error out of the three models.

\subsection{Self-similar analysis of the K-L model}

In order to better understand the K-L model's performance, it is possible to perform a self-similarity analysis of the governing equations.
This will reduce the equations to the form of a buoyancy-drag model, a set of ordinary-differential equations for the evolution of the mixing layer width and mixing layer growth rate.
Following the work in Refs. \cite{Xiao_2020_UnifiedPredictionReshocked,Xiao_2020_ModelingTurbulentMixing,Zhang_2020_MethodologyDeterminingCoefficients}, the profiles of $L$ and $\tilde{K}$ are given by self-similar profiles:
\begin{eqnarray}
    L(x,t) &= L_0(t) X^r(x,t) \label{eqn:L_selfsimilar}\\
    \tilde{K}(x,t) &= \tilde{K}_0(t) X^s(x,t) \label{eqn:K_selfsimilar}
\end{eqnarray}
where the profiles are effectively separated into a magnitude component, and a spatial component using the transformed coordinates:
\begin{eqnarray}
    X(x,t) &= 1 - \chi^2 \\
    \chi(x,t) &= \frac{x}{h(t)}
\end{eqnarray}
where $h$ is half the mixing layer width.
The non-dimensionalised spatial variables are limited to within the mixing layer i.e. $\chi \in [-1,1]$ and $X \in [0,1]$.
Previous works assumed the values for $r=1/2$ and $s=1$ such as in Ref. \cite{Dimonte_2006_KLTurbulenceModel} for the K-L model and Ref. \cite{Morgan_2015_ThreeequationModelSelfsimilar} for the K-L-a model.
Instead, Refs. \cite{Xiao_2020_UnifiedPredictionReshocked,Xiao_2020_ModelingTurbulentMixing,Zhang_2020_MethodologyDeterminingCoefficients} consider arbitrary values of $s$ and $r$, with the constraint of $2r+s=2$, in order to find the optimum fit to the models, ultimately using $s=1.6$.
The generalisation of the exponents meant that the self-similar equations did not separate easily into temporal and spatial contributions, and were solved by integrating over the mixing layer.

For the cases considered, the $\tilde{K}$ and $L$ equations used will assume homogeneity in the transverse directions, and neglect the effects of density by assuming the flow is in low-Atwood number limit.
\begin{subequations}
\begin{eqnarray}
    \bar{\rho} \frac{D \tilde{K}}{D t} &=& \frac{\partial}{\partial x_1} \frac{\mu_T}{N_K} \frac{\partial \tilde{K}}{\partial x_1} +\bar{\tau}_{ij} \frac{\partial \tilde{u}_i}{\partial x_j} - C_D \tilde{V}^3/{L} \\
    \bar{\rho} \frac{D L}{D t} &=& \frac{\partial}{\partial x_1} \frac{\mu_T}{N_L} \frac{\partial L}{\partial x_1} + C_L \bar{\rho} \tilde{V} + \bar{\rho} L S_\phi
\end{eqnarray}
\end{subequations}
The buoyancy production term for $\tilde{K}$ has been neglected, whilst the  Reynolds stress contribution and the $L$ compressibility term are included, two terms which are commonly ignored.  For generality, the $L$ bulk compression term has been represented as scaling with the arbitrary strain rate $S_\phi$, where using the isotropic strain rate closure ($\nabla\cdot \boldsymbol{u}/3$) will provide the original model. To obtain a buoyancy-drag model, the self-similar profiles given in Eqs. (\ref{eqn:L_selfsimilar}) and (\ref{eqn:K_selfsimilar}) were substituted in, a Taylor series about $\chi=0$ was taken to remove the spatial component, and the turbulent kinetic energy was transformed into the turbulent velocity using the relation $\tilde{K}_0 = \tilde{V}_0^2/2$. Assuming only normal velocity gradients are present, the gradients are replaced with the corresponding strain rate.
\begin{subequations}
\begin{eqnarray}
    \frac{d L_0}{dt} &=& \tilde{V}_0 (t) \left(\frac{C_L N_L - 2 C_\mu r \beta^2}{N_L}\right) + L_0 S_\phi\\
    \frac{d \tilde{V}_0}{dt} &=& - \frac{\tilde{V}_0^2(t)}{L_0(t)} \left(\frac{C_D N_K + C_\mu s \beta^2}{N_K} \right) + \frac{4}{3}C_\mu L_0 ({S}_A-{S}_T)^2 - \frac{1}{2} C_P ({S}_A+2 {S}_T) \tilde{V}
\end{eqnarray}
\end{subequations}

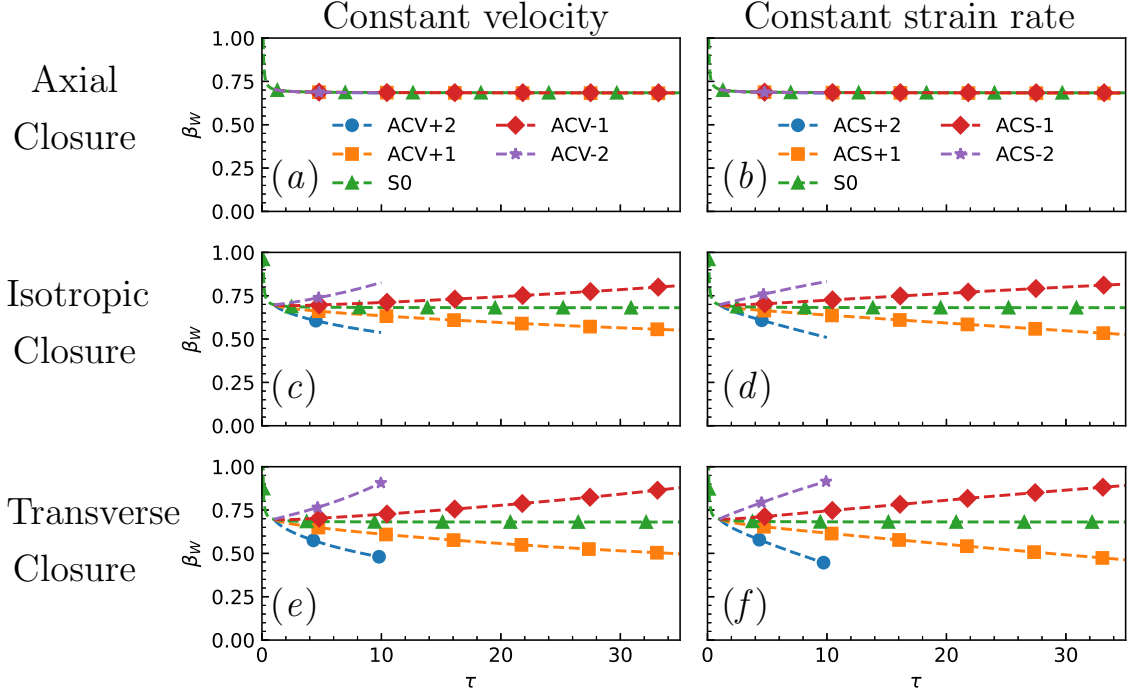
\begin{figure}
    \centering
    \begin{tikzpicture}
        \node[anchor=north west] (image) at (0,0) { \includegraphics[width=0.8\textwidth]{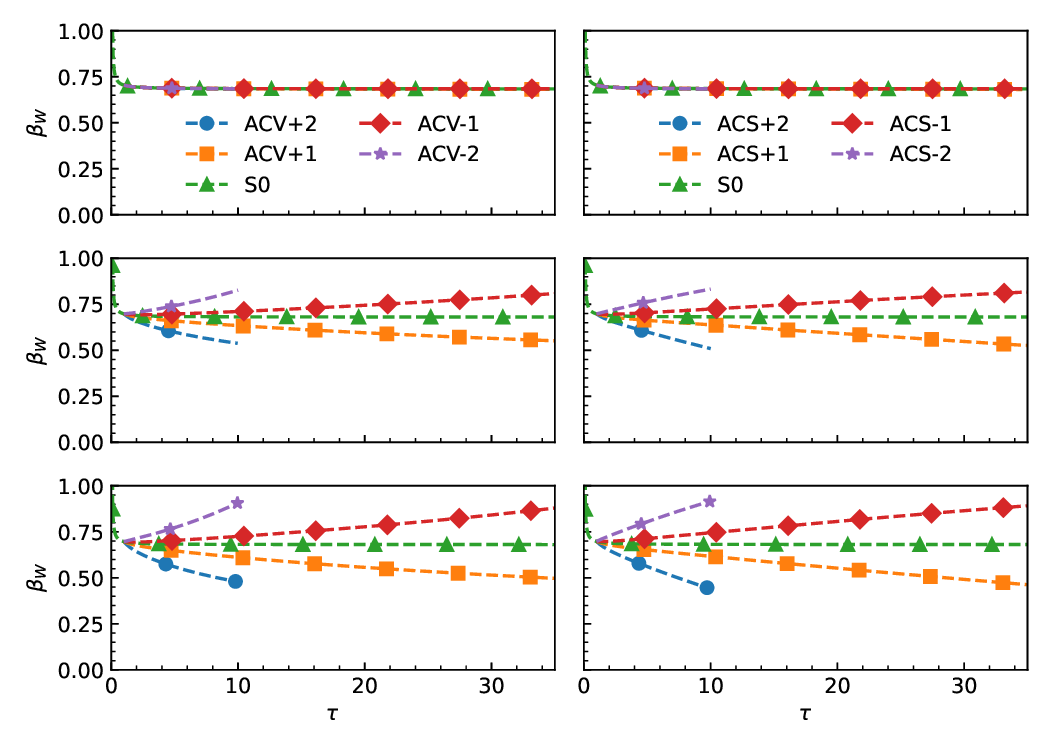}};
        \node[anchor=north west] at (2.2,0.1) {\large Constant velocity};
        \node[anchor=north west] at (7.8,0.1) {\large Constant strain rate};
        \node[anchor=north west,text width = 1.8cm,text centered] at (-2,-0.75) {\large \centering Axial Closure};
        \node[anchor=north west,text width = 1.8cm,text centered] at (-2,-3.6) {\large \centering Isotropic Closure};
        \node[anchor=north west,text width = 1.8cm,text centered] at (-2,-6.5) {\large Transverse Closure};
        \node[anchor=north west] at (1.5,-2.0) {\large (\textit{a})};
        \node[anchor=north west] at (7.5,-2.0) {\large (\textit{b})};
        \node[anchor=north west] at (1.5,-4.8) {\large (\textit{c})};
        \node[anchor=north west] at (7.5,-4.8) {\large (\textit{d})};
        \node[anchor=north west] at (1.5,-7.7) {\large (\textit{e})};
        \node[anchor=north west] at (7.5,-7.7) {\large (\textit{f})};
    \end{tikzpicture}
    \caption{$\beta_W$ for simulations under the applied axial strain rate with (left) constant velocity profile and (right) constant strain rate profile for the K-L model with (top) axial closure, (middle) isotropic closure, and (bottom) transverse closure for bulk compression.}
    \label{fig:AxialStrainBetaCombined}
\end{figure}

\begin{figure}
    \centering
    \begin{tikzpicture}
        \node[anchor=north west] (image) at (0,0) { \includegraphics[width=0.8\textwidth]{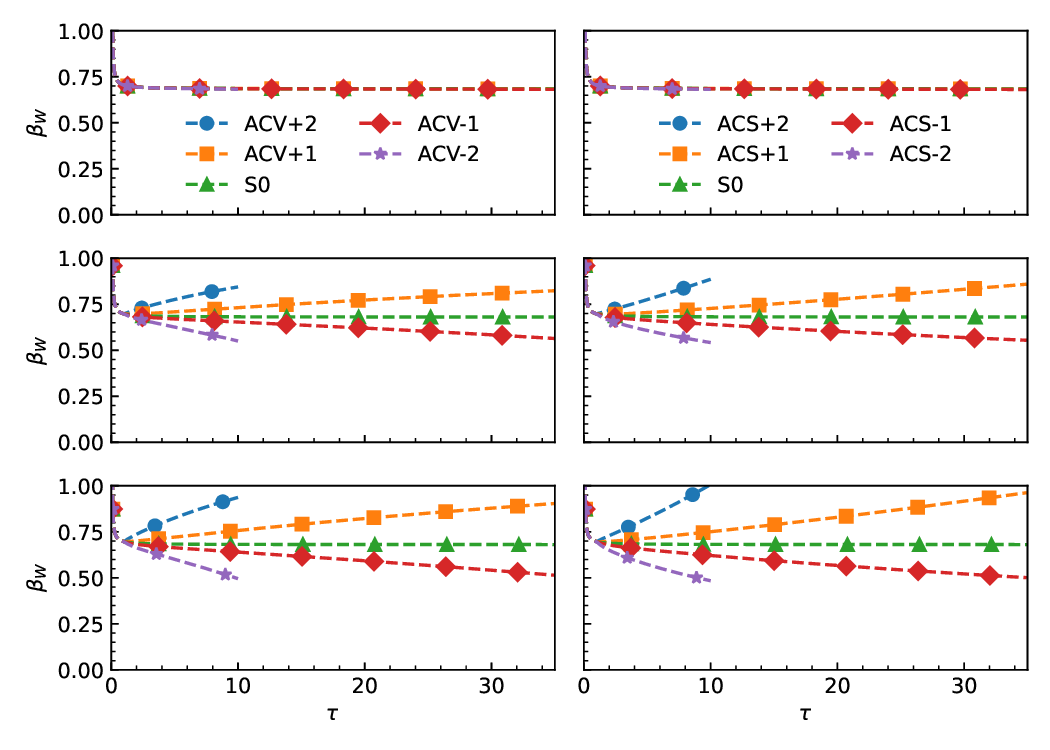}};
        \node[anchor=north west] at (2.2,0.1) {\large Constant velocity};
        \node[anchor=north west] at (7.8,0.1) {\large Constant strain rate};
        \node[anchor=north west,text width = 1.8cm,text centered] at (-2,-0.75) {\large \centering Axial Closure};
        \node[anchor=north west,text width = 1.8cm,text centered] at (-2,-3.6) {\large \centering Isotropic Closure};
        \node[anchor=north west,text width = 1.8cm,text centered] at (-2,-6.5) {\large Transverse Closure};
        \node[anchor=north west] at (1.5,-2.0) {\large (\textit{a})};
        \node[anchor=north west] at (7.5,-2.0) {\large (\textit{b})};
        \node[anchor=north west] at (1.5,-4.8) {\large (\textit{c})};
        \node[anchor=north west] at (7.5,-4.8) {\large (\textit{d})};
        \node[anchor=north west] at (1.5,-7.7) {\large (\textit{e})};
        \node[anchor=north west] at (7.5,-7.7) {\large (\textit{f})};
    \end{tikzpicture}
    \caption{$\beta_W$ for simulations under the applied transverse strain rate with (left) constant velocity profile and (right) constant strain rate profile for the K-L model with (top) axial closure, (middle) isotropic closure, and (bottom) transverse closure for bulk compression.}
    \label{fig:TransverseStrainBetaCombined}
\end{figure}

Typically, the evolution equation of $L_0$ will be transformed into the evolution equation of the mixing layer half width, $h$, as these terms are considered to be proportional by a constant within the regime of self-similarity.
\begin{align}
    \beta = \frac{L_0(t)}{h(t)}
\end{align}
As the integral width has been used as the measure of the mixing layer width in the previous sections, the system will be transformed by $\beta_W = L_0/W$.
However, the assumption of self-similarity is not valid for the strained simulation cases.
Under the application of axial and transverse strain, the turbulent kinetic energy anisotropy and the mixedness of the mixing layer do not approach an asymptotic value, instead deviating away from the unstrained case's self-similar behaviour \cite{Pascoe_2024_ImpactAxialStrain,Pascoe_2025_ImpactTransverseStrain}.
This is further evidenced by the variation of $\beta_W$ for the RANS simulations conducted, shown in Figs. \ref{fig:AxialStrainBetaCombined} and \ref{fig:TransverseStrainBetaCombined}.
The unstrained cases approach a final value of $\beta_W=0.687$, suggesting the unstrained case has entered this regime of self-similar growth for the RANS model.
It is possible to further validate this asymptotic value, as for a linear mean volume fraction profile the integral width should be three times smaller than the half width, giving the relation of $\beta_W = 3\beta$.
The theoretical value of $\beta$ for the present model is calculated according to $\beta=1/C_A$ \cite{Xiao_2020_UnifiedPredictionReshocked,Xiao_2020_ModelingTurbulentMixing,Zhang_2020_MethodologyDeterminingCoefficients}.
The actual relation observed from the simulations is $\beta_W=3.44\beta$, a slightly higher ratio due to the mean volume fraction profile not being linear.
Whilst the unstrained cases plateau, the strained cases show a variation in $\beta_W$, suggesting that the peak value of $L$ in the domain and the integral width do not grow in the same proportion for the isotropic and transverse closure cases.
In contrast, the axial closure cases do maintain the late-time value of $\beta_W$.
Whilst the self-similarity assumption simplifies the buoyancy-drag reduction analysis, it is not an accurate representation of the state of self-similarity for the ILES cases. 
This necessitates the inclusion of time dependence for $\beta$.

Adjusting both equations to use the integral width, the growth rate of the integral width is defined by
\begin{align}
    V_W = \tilde{V}_0 \frac{C_L N_L - 2 C_\mu r \beta^2}{N_L \beta_W}.
\end{align}
Substituting and simplifying gives the new expressions:
\begin{subequations}
\begin{eqnarray}
    \frac{dW}{dt} &=& V_W + \left( S_\phi- \frac{\dot{\beta}_W}{\beta_W} \right)W\\
    \frac{dV_W}{dt} &=& - \frac{V_W^2}{l^\text{eff}} +  C_W ({S}-{S}_T)^2 W - \frac{1}{2} C_P ({S}_A+2{S}_T) V_W  + C_\beta V_W 
\end{eqnarray}
\end{subequations}
with the definitions
\begin{subequations}
\begin{eqnarray}
    l^\text{eff} &=& \frac{N_K \left(C_L N_L - 2C_\mu r \beta^2 \right)}{N_L \left(C_D N_K + C_\mu s \beta^2\right)} W\\ 
    C_W &=&\frac{4 C_\mu (C_L N_L - 2C_\mu r \beta^2)}{3N_L} \\ 
    C_\beta &=& - \frac{\dot{\beta}_W}{\beta_W} - \frac{4C_\mu r\beta}{C_L N_L - 2C_\mu r \beta^2}\dot{\beta}.
\end{eqnarray}
\end{subequations}
This derivation has not assumed a constant $\beta$ or $\beta_W$, causing the derivatives of each to occur in the differential equations. The equations may be simplified to use only a single $\beta$ term as they are proportional for a specified mean volume fraction profile. The model is greatly simplified by using constant $\beta$. It may be possible to empirically determine a relationship for the variation in $\beta$ under strain for the specific bulk compression closure employed. For now, the analysis is conducted with the assumption of self-similarity between the turbulent length scale and the mixing layer width. Using this assumption reduces the model to,
\begin{subequations}
\label{eqn:derived_BD}
\begin{eqnarray}
    \frac{dW}{dt} &=& V_W + S_\phi W \\
    \frac{dV_W}{dt} &=& - \frac{V_W^2}{l^\text{eff}} +  C_W ({S}_A-{S}_T)^2 W - \frac{1}{2} C_P ({S}_A+2{S}_T) V_W. 
\end{eqnarray}
\end{subequations}
In this model the growth of the integral width is dependent upon the velocity as well as a contribution from the background strain rate, $S_\phi W$.
For the case of axial closure ($S_\phi=S_A$), this contribution is identical to the RMI linear regime correction with axial strain, where the amplitude growth rate is the superposition of the amplitude growth rate and the background stretching contribution \cite{Epstein_2004_BellPlessetEffects,Pascoe_2024_ImpactAxialStrain}.
It makes sense then that axial closure preserves the $\beta_W$ ratio, as it allows the turbulent length scale to grow at the same rate as the integral width, where the integral width will naturally vary from the velocity difference across the mixing layer.
Despite this, the transverse closure provides better performance for the model by adjusting the dissipation rate for the transverse strain expansion factor.
Without any strain and with the coefficients provided in Table \ref{tab:KL_coeff}, the effective drag lengthscale is calculated to be $l^\text{eff} = 0.383W$, which for unstrained RMI suggests a power law growth $W=A(t-t_0)^\theta$, where $\theta = 0.277$. This is in between the prescribed coefficient used for the K-L model derivation of $\theta = 0.25$, and the late-time coefficient of $\theta=0.292$ for the quarter-scale $\theta$-group case \cite{ThetaGroup}.

The velocity equation has two additional terms that are dependent upon strain rate.
These two terms are the result of shear production, a product of the Reynolds stress and velocity gradients.
Similar terms were hypothesised and calibrated to improve the performance of a buoyancy-drag model to described the ILES cases under axial strain \cite{Pascoe_2024_ImpactAxialStrain}.
Whilst the $C_W$ term will always act as a positive source for the evolution of the growth rate, the $C_P$ contribution will depend upon the signs of the applied strain rate as expected for shear production contribution.

Given the clear separation in behaviour of the two length scales, the integral width which scales with axial strain and the turbulent length scale which scales with transverse strain, a two length scale approach could be used to generate a buoyancy-drag model. Whilst two length scales are employed in many modern RANS models, those length scales are used to separate the turbulence transport and destruction length scales \cite{Schwarzkopf_2016_TwolengthScaleTurbulence,Morgan_2018_TwolengthscaleTurbulenceModel}. In the derivation of the K-2L-a model, the turbulent length scale equation is considered to grow self-similarly with the mixing layer width, while the ratio between the turbulent transport and destruction length scale is not necessarily constant \cite{Morgan_2018_TwolengthscaleTurbulenceModel}. 

\begin{figure}
    \centering
    \begin{tikzpicture}
        \node[anchor=north west] (image) at (0,0) { \includegraphics[width=0.8\textwidth]{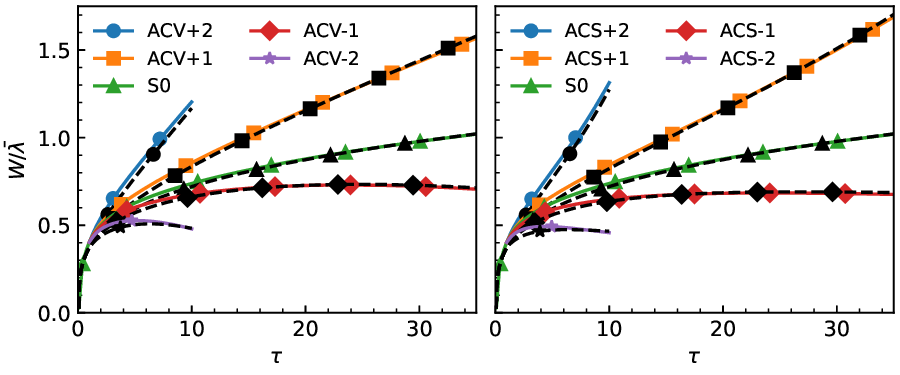}};
        \node[anchor=north west] at (2.2,0.4) {\large Constant velocity};
        \node[anchor=north west] at (8.0,0.4) {\large Constant strain rate};
        \node[anchor=north west] at (6.1,-3.9) {\large (\textit{a})};
        \node[anchor=north west] at (12.3,-3.9) {\large (\textit{b})};
    \end{tikzpicture}
    \caption{Buoyancy-drag model for simulations under the applied axial strain rate with (left) constant velocity profile and (right) constant strain rate. Sold line indicates ILES, dashed line indicates buoyancy-drag model from Eq. \ref{eqn:new_BD}.}
    \label{fig:Axial_BD}
\end{figure}

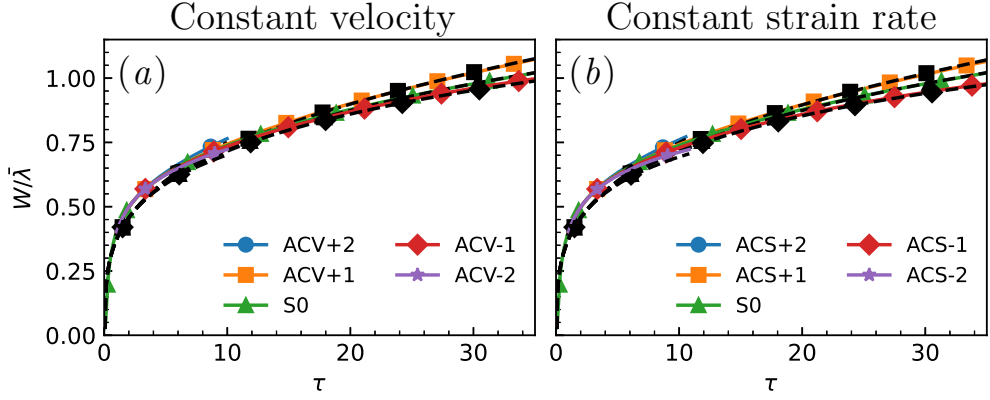
\begin{figure}
    \centering
    \begin{tikzpicture}
        \node[anchor=north west] (image) at (0,0) { \includegraphics[width=0.8\textwidth]{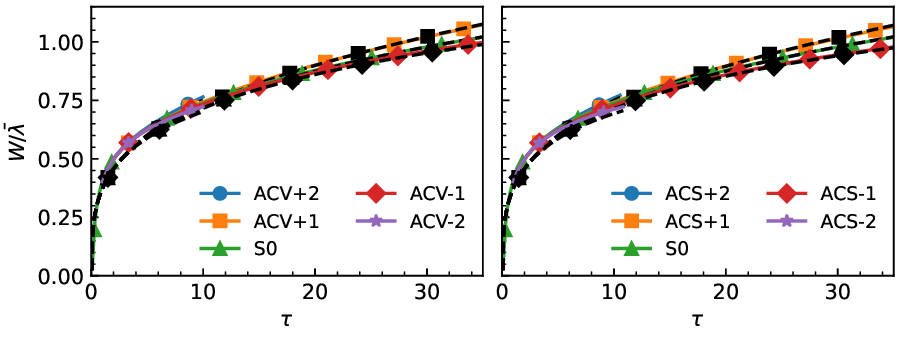}};
        \node[anchor=north west] at (2.2,0.4) {\large Constant velocity};
        \node[anchor=north west] at (8.0,0.4) {\large Constant strain rate};
        \node[anchor=north west] at (1.5,-0.3) {\large (\textit{a})};
        \node[anchor=north west] at (7.5,-0.3) {\large (\textit{b})};
    \end{tikzpicture}
    \caption{Buoyancy-drag model for simulations under the applied transverse strain rate with (left) constant velocity profile and (right) constant strain rate. Solid line indicates ILES, dashed line indicates buoyancy-drag model from Eq. \ref{eqn:new_BD}.}
    \label{fig:Transverse_BD}
\end{figure}

Two further modifications have been made to obtain the buoyancy-drag model in Eq. \ref{eqn:new_BD} from Eq. \ref{eqn:derived_BD}.
Firstly, the transverse strain contributions from shear production have been omitted from the evolution of the turbulent velocity growth to improve agreement with the transverse results.
Whilst turbulent kinetic energy is added under compressive transverse strain, it is not distributed to the axial direction which the isotropic model for $\tilde{K}$ does not account for and so causes the mixing layer growth rate to be incorrectly estimated.
An alternate modification would be to scale the turbulent velocity more strongly with the transverse strain rate to counteract this effect.
Secondly, the magnitude of the coefficient of the $S_A V_W$ term has been increased to improve the alignment with the axial results, effectively increasing the strain drag.
Whilst these modifications are phenomenological, they can be considered corrections to the prior removal of the terms which depended upon the time variation of $\beta$.
\begin{subequations}
\label{eqn:new_BD}
\begin{eqnarray}
    \frac{dW}{dt} &=& V_W + S_A W \\
    \frac{dL}{dt} &=& V_W + S_T L \\
    \frac{dV_W}{dt} &=& - \frac{V_W^2}{l^\text{eff}} +  0.143 S_A^2 L - 0.933 S_A V_W \\
    l^\text{eff} &=& 0.383 L
\end{eqnarray}
\end{subequations}
For simplicity, both length scales are scaled by the turbulent velocity as opposed to being different by a factor of $\beta_W$. 
For a case without applied strain, the proposed model will reduce down to the more typical two-equation model for buoyancy-drag modelling. 
The performance of the model is plotted in Figs. \ref{fig:Axial_BD} and \ref{fig:Transverse_BD}, using initial conditions of [$W_0$,$L_0$,$V_{W0}$]=[0.00638,~0.00638,1679] and an initial time offset of $\tau=0.08$ to better align with the shock transition.
The initial length scale values use the post-shock integral width \cite{ThetaGroup,Youngs_2020_EarlyTimeModifications}, as with the omission of the source term in the analysis the resulting buoyancy-drag is not able to capture the shock transition.
As the model uses an effective drag length scale proportional to the turbulent length scale which is initialised the same as the integral width, it is unable to capture the linear and transitional regimes, just like the RANS model.
The initial drag is very high due to the small turbulent length scale/integral width, requiring a very large initial velocity (36$\times$ larger than the theoretical growth rate\cite{ThetaGroup,Youngs_2020_EarlyTimeModifications}) to counteract the drag. This could be rectified by using a piecewise effective drag length scale as done in the buoyancy-drag models of Youngs and Thornber \cite{Youngs_2020_EarlyTimeModifications,Youngs_2020_BuoyancyDragModelling}.
As the drag term is derived from the diffusion and dissipation terms for the turbulent kinetic energy, a modification to these terms could improve the performance of the K-L model in the early time.
The late time performance of the derived buoyancy-drag model is well aligned with the ILES, and is far less computationally expensive to run than both the ILES and RANS simulations.

\subsection{Equivalence of two-equation models}
The bulk compression closure for the turbulent length scale can be generalised to other two-equation RANS models, such as for the K-$\epsilon$ model which transports the turbulent kinetic energy dissipation rate. Starting from the turbulent kinetic energy and the turbulent length scale, alternate formulations are obtained by using the transformation \cite{Pope_2000_TurbulentFlows,Wilcox_2006}
\begin{align}
    Z = C_Z K^m L^n.
\end{align}
Given the formulation of the dissipation of turbulent kinetic energy in Eq. \ref{eqn:K_eqn}, for $Z=\epsilon$ the values of $m=3/2$, $n=-1$, and $C_Z = 2\sqrt{2}C_D$ are obtained. 
The transport equation for $Z$ is then obtained via
\begin{align}
    \bar{\rho}\frac{\tilde{D}Z}{\tilde{D}t} &= m \frac{Z}{\tilde{K}} \bar{\rho}\frac{\tilde{D}\tilde{K}}{\tilde{D}t} + n \frac{Z}{L} \bar{\rho}\frac{\tilde{D}L}{\tilde{D}t} \\
    \begin{split}
    &= m \frac{Z}{\tilde{K}} \left[ \frac{\partial}{\partial x_j} \left(\frac{\mu_T}{N_K} \frac{\partial K}{\partial x_j} \right) + \bar{\tau}_{ij} \frac{\partial \tilde{u}_i}{\partial x_j} + S_K - \bar{\rho} \epsilon\right] \\&\qquad+ n \frac{Z}{L} \left[ \frac{\partial}{\partial x_j} \left(\frac{\mu_T}{N_L} \frac{\partial L}{\partial x_j} \right)+C_L \bar{\rho}\tilde{V} + \bar{\rho} LS_\phi \right]
    \end{split}
\end{align}
where $\tilde{D}/\tilde{D}t=\partial/\partial t + \tilde{u}_i\partial/\partial x_i$ refers to the total derivative for the Favre-averaged flow field. For the turbulent dissipation rate, the equations evaluate to
\begin{align}
    \begin{split}
    \bar{\rho}\frac{\tilde{D}\epsilon}{\tilde{D}t} = \frac{\partial}{\partial x_j}\left( \mu_T \left(\frac{3}{2N_K}+\frac{1}{N_L} \right) \frac{\partial \epsilon}{\partial x_j}\right)
    +\frac{3}{2}\frac{\epsilon S_K}{K} 
    +\frac{3}{2} \frac{\epsilon}{\tilde{K}}\bar{\tau}^{(d)}_{ij} \frac{\partial \tilde{u}_i}{\partial x_j} 
    \\- \left(\frac{3}{2}+\frac{C_L}{2C_D} \right) \bar{\rho}\frac{\epsilon^2}{\tilde{K}} 
    - \left(\frac{3}{2} C_P\frac{\partial \tilde{u}_i}{\partial x_i} + S_\phi\right) \bar{\rho} \epsilon  +C.D.\label{eqn:eqv_epsilon} 
    \end{split}
\end{align}
On the right-hand side of Eq. \ref{eqn:eqv_epsilon}, the terms represent diffusion, shear production, buoyancy production, dissipation, bulk compression, and cross-diffusion respectively for the turbulent dissipation rate.
The re-arrangement of the diffusion terms for $\tilde{K}$ and $L$ into the diffusion term for $\epsilon$ introduces cross-diffusion terms \cite{Wilcox_2006}, which will depend upon the gradients of $\tilde{K}$, $\epsilon$ and density, but not the strain rates of the flow. For simplicity, these terms have not been included as the focus of this comparison is on the bulk compression. 
The Reynolds stress in the shear production term has been split into the deviatoric component $\bar{\tau}^{(d)}_{ij}$ and the isotropic component $- C_P \bar{\rho}K\delta_{ij}$. The isotropic component of the Reynolds stress combines with the bulk compression term for the turbulent length scale to produce the bulk compression term for the turbulent dissipation rate. For comparison, the transport equation for the turbulent dissipation rate in Refs. \cite{Moran-Lopez_2013_MulticomponentReynoldsaveragedNavier,Moran-Lopez_2014_MulticomponentReynoldsaveragedNavier,Gauthier_1990_KeModelTurbulent,Schilling_2015_ComparativeStudyPredictions,Schilling_2021_SelfsimilarReynoldsaveragedMechanical,Schilling_2024_SelfsimilarReynoldsaveragedMechanical} is
\begin{align}
    \bar{\rho}\frac{\tilde{D}\epsilon}{\tilde{D} t} = 
    \frac{\partial}{\partial x_j} \left[ \left( \bar{\mu} +\frac{\mu_t}{\sigma_\varepsilon} \right) \frac{\partial \varepsilon}{\partial x_j} \right] 
    + C_{\varepsilon 0} \frac{\varepsilon}{K} S_K 
    + C_{\varepsilon 1} \frac{\varepsilon}{K} \tau_{ij}^{(d)} \frac{\partial \tilde{u}_i}{\partial x_j} 
    - C_{\varepsilon 2} \frac{\rho  \epsilon^2}{K}
    - \frac{2}{3} C_{\varepsilon 3} \rho \varepsilon \frac{\partial \tilde{u}_j}{\partial x_j} \label{eqn:epsilon_true}
\end{align}
where the form of $S_K$ varies depending on the model, a physical viscosity $\bar{\mu}$ has been included in the diffusion term, and $\bar{\tau}_{ij}$ has been updated changed to use current notation. The values of $\sigma_\epsilon$ and $C_{\epsilon j}$ are model constants. The last term on the right-hand-side of Eq. \ref{eqn:epsilon_true} is the bulk compression term for the dissipation rate. For incompressible flows it is common to set the corresponding coefficient $C_{\epsilon 3}$ to zero, whilst for compressible flows a value of $C_{\epsilon 3}=2$ is used to achieve the coefficient of $-4/3$ \cite{Reynolds_1987_FundamentalsOfTurbulence,Speziale_1991_SecondorderClosureModels}.  Using the default model coefficients for the K-L model with $C_P=2/3$ and $S_\phi=\nabla\cdot \mathbf{\tilde{u}}/3$, the resulting bulk compression term for the turbulent dissipation rate is
\begin{align}
    - \left(\frac{3}{2} C_P\frac{\partial \tilde{u}_i}{\partial x_i} + S_I\right) \bar{\rho} \epsilon =
    - \frac{4}{3} \frac{\partial \tilde{u}_i}{\partial x_i} \bar{\rho} \epsilon,
\end{align}
giving the same expression as used for K-$\epsilon$ models. The model in Ref. \cite{Xie_2021_PredictingDifferentTurbulent} neglects the contribution of the isotropic production from $K$ in the derivation for the bulk compression closure of $\epsilon$, instead attaining $C_{\epsilon 3}=1/2$. For inhomogeneous turbulence with anisotropic strain, the transverse strain closure model for the bulk compression of the turbulent length scale performed better than the default closure for isotropic strain. Using the transverse strain closure, the expression for the bulk compression of the turbulent dissipation rate becomes 
\begin{align}
    - \left(\frac{3}{2} C_P\frac{\partial \tilde{u}_i}{\partial x_i} + S_T\right) \bar{\rho} \epsilon =
    - \left(\frac{\partial \tilde{u}_i}{\partial x_i}+S_T\right) \bar{\rho} \epsilon.
\end{align}
This expression is very similar, in essence swapping out a mean strain rate contribution (one third of the divergence) for the transverse strain contribution. Under isotropic strain, the expressions are identical in the low-Atwood number limit. The transformation process can be repeated for the K-$\omega$ model, using $m=1/2$, $L=-1$, and leaving $C_\omega$ as an arbitrary constant. The equivalent transport equation for the specific dissipation rate $\omega$ is 
\begin{align}
    \begin{split}
    \bar{\rho}\frac{\tilde{D}\omega}{\tilde{D}t} = \frac{\partial}{\partial x_j}\left( \mu_T \left(\frac{1}{2N_K}+\frac{1}{N_L} \right) \frac{\partial \omega}{\partial x_j}\right)
    +\frac{1}{2}\frac{\omega S_K}{K} 
    +\frac{1}{2} \frac{\omega}{\tilde{K}}\bar{\tau}^{(d)}_{ij} \frac{\partial \tilde{u}_i}{\partial x_j} 
    \\- \left(\frac{C_D^{3/2}2^{1/2}}{C_\omega}+\frac{C_L2^{1/2}}{2C_\omega} \right) \bar{\rho}\frac{\omega^2}{\tilde{K}} 
    - \left(\frac{1}{2} C_P\frac{\partial \tilde{u}_i}{\partial x_i} + S_\phi\right) \bar{\rho} \omega  +C.D.\label{eqn:eqv_omega} 
    \end{split}
\end{align}
The terms on the right-hand-side of Eqn. \ref{eqn:eqv_omega} are presented in the same order as for Eqn. \ref{eqn:eqv_epsilon}. Evaluation of the bulk compression term with the default isotropic closure gives a scaling coefficient of $-2/3$. The 1988 Wilcox K-$\omega$ model \cite{Wilcox_1998_TurbulenceModelingCFD} is given by 
\begin{align}
    \bar{\rho}\frac{\tilde{D}\omega}{\tilde{D}t} = 
    \frac{\partial}{\partial x_j}\left[ \left(\mu +\frac{\mu_T}{\sigma_\omega}\right) \frac{\partial \omega}{\partial x_j}\right]
    +\gamma \frac{\omega}{\tilde{K}}\bar{\tau}^{(d)}_{ij} \frac{\partial \tilde{u}_i}{\partial x_j} 
    - \beta \rho \omega^2 
    - \gamma \frac{2}{3} \bar{\rho} \omega \frac{\partial \tilde{u}_k}{\partial x_k}.
\end{align}
The terms on the right-hand-side are diffusion, shear production, dissipation, and a bulk compression term which has been extracted from the shear production. This model does not have any buoyancy production, and it does not include the cross-diffusion term for simplicity. The coefficient $\gamma=13/25$ is calibrated for wall properties, but is very close to the derived factor of $1/2$ from the model equivalence. The bulk compression term in this K-$\omega$ model will then evaluate to around $-1/3$. Based off Morel \& Mansour's derivation, Refs. \cite{Vuong_1987_ModelingTurbulenceHypersonic,Coakley_1992_TurbulenceModelingHigh} adjust the coefficient by assuming that one-dimensional rapid compression from shock waves is the primary activation of the bulk compression. This results in a scaling of the divergence by $-4/3$ as compared to the isotropic compression closure of $-2/3$. Using the transverse closure will result in the relation
\begin{align}
    - \left(\frac{1}{2} C_P\frac{\partial \tilde{u}_i}{\partial x_i} + S_T\right) \bar{\rho} \omega= 
    - \left(\frac{1}{3} \frac{\partial \tilde{u}_i}{\partial x_i} + S_T\right) \bar{\rho} \omega
\end{align}

An interesting inclusion in the K-$\omega$ model is the 'Pope correction' \cite{Pope_1978_ExplanationTurbulentJet,Wilcox_2006}, which modifies $\beta$ in the dissipation term of $\omega$. This modification was created due to the anomaly of the different spreading rates of the round-jet and plane-jet. The plane-jet is a two-dimensional problem, whilst the round-jet is three-dimensional and will experience vortex stretching. The correction introduces a dependence upon $\chi$, a non-dimensional measure of the vortex stretching.  
Whilst full details can be found in the aforementioned references, the non-dimensional measure for the vortex stretching in the K-$\omega$ model is given by
\begin{align}
    \chi_\omega = \left|\frac{\Omega_{ij} \Omega_{jk} \hat{S}_{ki}}{(\beta^* \omega)^3}\right|
\end{align}
For the strained simulations conducted, this correction would not contribute as $\chi$ would be zero due to the dependence on the rotation tensor of the flow.

The same closure for the bulk compression of the turbulent length scale should be valid for alternative RANS models that include additional equations, such as equations for the turbulent mass flux, scalar variance or density self-correlation. Two points of future investigation would be for the treatment of length scales in two length scale models, where dissipation and transport length scales may be most accurately modelled with different bulk compression closures, and for Reynolds stress models which no longer assume isotropy of the turbulent kinetic energy. An additional complication for this alternate closure can arise for depending upon the specific case configuration. For turbulent mixing layers in standard configurations with clear distinctions between the axial (normal to the interface) and transverse (parallel to interface or plane of homogeneity), identification of the transverse strain rate is a simple endeavour. For more complicated mixing layers and flow structures, the transverse direction may not be uniform in space or time. For example, the tilted rocket rig \cite{Read_1984_ExperimentalInvestigationTurbulent,Youngs_1989_ModellingTurbulentMixing,Andrews_2014_ComputationalStudiesTwoDimensional,Ferguson_2023_MassMomentumTransport}, which experiences anisotropic strain, is a popular validation case for RANS models as it is difficult to simultaneously capture the mixing layer width, turbulent kinetic energy, turbulent mass flux and density-specific volume correlation \cite{Denissen_2014_TiltedRocketRig,Morgan_2023_TwoSelfsimilarReynoldsstress}.


\section{Conclusion\label{sec:Conclusion}}

The ability of the K-L turbulence model to predict the growth of a Richtmyer-Meshkov induced turbulent mixing layer under anisotropic strain was investigated through a series of simulations. Based upon the quarter-scale $\theta$-group case \cite{ThetaGroup}, the strained ILES of Refs. \cite{Pascoe_2024_ImpactAxialStrain,Pascoe_2025_ImpactTransverseStrain} served as the basis for comparison. 
Three sets of simulations were conducted, each utilising a different closure for the the treatment of the turbulent length scale under bulk compression of the fluid, a term that is not theoretically defined for anisotropic strains. 
The three closure approaches correspond to scaling the turbulent length scale with the axial strain rate, the transverse strain rate, and the default methodology of scaling with the mean strain.
Under mean axial strain rates, the K-L model was able to predict the expansion/compression of the mixing layer due to the ability to resolve the velocity difference in the fluid flow which expands/compresses the mixing layer.
The results for each closure differed in their ability to model the turbulent growth and the total turbulent kinetic energy, with the transverse strain closure performing the best of the three options. 
In a similar fashion, the transverse strain closure performed the best of the three methods for both the integral width and turbulent kinetic energy.
These results suggest that for modelling of turbulent mixing layers under anisotropic strain rates, such as observed in implosion and explosion profiles in spherical geometries, the performance of the K-L model can be improved through the modification of the bulk compression term for the turbulent length scale to instead vary with the transverse strain rate. 
A self-similarity analysis of the K-L model produced a buoyancy-drag model for modelling the mixing layer growth.
The development of the model was limited by the assumption of the proportionality of the turbulent length scale and integral width, a common assumption for model calibration in K-L type models. 
To rectify this constraint, a three-equation buoyancy-drag model was developed, which evolves both the integral width and a turbulent length scale.
With this design, the integral width is allowed to scale with the axial strain which stretches/compresses the mixing layer, whilst the turbulent length scale scales with the transverse strain rate.
Through the equivalence of two-equation RANS models, the closure for the bulk compression was also derived for the K-$\epsilon$ and K-$\omega$ models. Further avenues of research should be conducted to determine the proficiency of these modifications in describing other flow features.

\begin{acknowledgments}
The authors would like to thank D.L. Youngs for his helpful advice and input on the project. The authors acknowledge the Sydney Informatics Hub and the use of the University of Sydney’s high performance computing cluster, Artemis. This research was supported by the Australian Government's National Collaborative
Research Infrastructure Strategy (NCRIS), with access to computational resources
provided by the Setonix supercomputer (https://doi.org/10.48569/18sb-8s43) through the National Computational Merit Allocation Scheme. The authors would like to thank EPSRC for the computational time made available on the UK supercomputing facility ARCHER/ARCHER2 via the UK Turbulence Consortium (EP/R029326/1).

\end{acknowledgments}
\appendix


\bibliography{MyLibrary}

\providecommand{\noopsort}[1]{}\providecommand{\singleletter}[1]{#1}%
\begin{thebibliography}{91}%
\makeatletter
\providecommand \@ifxundefined [1]{%
 \@ifx{#1\undefined}
}%
\providecommand \@ifnum [1]{%
 \ifnum #1\expandafter \@firstoftwo
 \else \expandafter \@secondoftwo
 \fi
}%
\providecommand \@ifx [1]{%
 \ifx #1\expandafter \@firstoftwo
 \else \expandafter \@secondoftwo
 \fi
}%
\providecommand \natexlab [1]{#1}%
\providecommand \enquote  [1]{``#1''}%
\providecommand \bibnamefont  [1]{#1}%
\providecommand \bibfnamefont [1]{#1}%
\providecommand \citenamefont [1]{#1}%
\providecommand \href@noop [0]{\@secondoftwo}%
\providecommand \href [0]{\begingroup \@sanitize@url \@href}%
\providecommand \@href[1]{\@@startlink{#1}\@@href}%
\providecommand \@@href[1]{\endgroup#1\@@endlink}%
\providecommand \@sanitize@url [0]{\catcode `\\12\catcode `\$12\catcode `\&12\catcode `\#12\catcode `\^12\catcode `\_12\catcode `\%12\relax}%
\providecommand \@@startlink[1]{}%
\providecommand \@@endlink[0]{}%
\providecommand \url  [0]{\begingroup\@sanitize@url \@url }%
\providecommand \@url [1]{\endgroup\@href {#1}{\urlprefix }}%
\providecommand \urlprefix  [0]{URL }%
\providecommand \Eprint [0]{\href }%
\providecommand \doibase [0]{https://doi.org/}%
\providecommand \selectlanguage [0]{\@gobble}%
\providecommand \bibinfo  [0]{\@secondoftwo}%
\providecommand \bibfield  [0]{\@secondoftwo}%
\providecommand \translation [1]{[#1]}%
\providecommand \BibitemOpen [0]{}%
\providecommand \bibitemStop [0]{}%
\providecommand \bibitemNoStop [0]{.\EOS\space}%
\providecommand \EOS [0]{\spacefactor3000\relax}%
\providecommand \BibitemShut  [1]{\csname bibitem#1\endcsname}%
\let\auto@bib@innerbib\@empty
\bibitem [{\citenamefont {Hunt}\ and\ \citenamefont {Carruthers}(1990)}]{Hunt_1990_RapidDistortionTheory}%
  \BibitemOpen
  \bibfield  {author} {\bibinfo {author} {\bibfnamefont {J.~C.~R.}\ \bibnamefont {Hunt}}\ and\ \bibinfo {author} {\bibfnamefont {D.~J.}\ \bibnamefont {Carruthers}},\ }\bibfield  {title} {\bibinfo {title} {Rapid distortion theory and the `problems' of turbulence},\ }\href@noop {} {\bibfield  {journal} {\bibinfo  {journal} {Journal of Fluid Mechanics}\ }\textbf {\bibinfo {volume} {212}},\ \bibinfo {pages} {497} (\bibinfo {year} {1990})}\BibitemShut {NoStop}%
\bibitem [{\citenamefont {Durbin}\ and\ \citenamefont {Zeman}(1992)}]{Durbin_1992_RapidDistortionTheory}%
  \BibitemOpen
  \bibfield  {author} {\bibinfo {author} {\bibfnamefont {P.~A.}\ \bibnamefont {Durbin}}\ and\ \bibinfo {author} {\bibfnamefont {O.}~\bibnamefont {Zeman}},\ }\bibfield  {title} {\bibinfo {title} {Rapid distortion theory for homogeneous compressed turbulence with application to modelling},\ }\href@noop {} {\bibfield  {journal} {\bibinfo  {journal} {Journal of Fluid Mechanics}\ }\textbf {\bibinfo {volume} {242}},\ \bibinfo {pages} {349} (\bibinfo {year} {1992})}\BibitemShut {NoStop}%
\bibitem [{\citenamefont {Cambon}\ \emph {et~al.}(1993)\citenamefont {Cambon}, \citenamefont {Coleman},\ and\ \citenamefont {Mansour}}]{Cambon_1993_RapidDistortionAnalysis}%
  \BibitemOpen
  \bibfield  {author} {\bibinfo {author} {\bibfnamefont {C.}~\bibnamefont {Cambon}}, \bibinfo {author} {\bibfnamefont {G.~N.}\ \bibnamefont {Coleman}},\ and\ \bibinfo {author} {\bibfnamefont {N.~N.}\ \bibnamefont {Mansour}},\ }\bibfield  {title} {\bibinfo {title} {Rapid distortion analysis and direct simulation of compressible homogeneous turbulence at finite {{Mach}} number},\ }\href@noop {} {\bibfield  {journal} {\bibinfo  {journal} {Journal of Fluid Mechanics}\ }\textbf {\bibinfo {volume} {257}},\ \bibinfo {pages} {641} (\bibinfo {year} {1993})}\BibitemShut {NoStop}%
\bibitem [{\citenamefont {Blaisdell}\ \emph {et~al.}(1996)\citenamefont {Blaisdell}, \citenamefont {Coleman},\ and\ \citenamefont {Mansour}}]{Blaisdell_1996_RapidDistortionTheory}%
  \BibitemOpen
  \bibfield  {author} {\bibinfo {author} {\bibfnamefont {G.~A.}\ \bibnamefont {Blaisdell}}, \bibinfo {author} {\bibfnamefont {G.~N.}\ \bibnamefont {Coleman}},\ and\ \bibinfo {author} {\bibfnamefont {N.~N.}\ \bibnamefont {Mansour}},\ }\bibfield  {title} {\bibinfo {title} {Rapid distortion theory for compressible homogeneous turbulence under isotropic mean strain},\ }\href@noop {} {\bibfield  {journal} {\bibinfo  {journal} {Physics of Fluids}\ }\textbf {\bibinfo {volume} {8}},\ \bibinfo {pages} {2692} (\bibinfo {year} {1996})}\BibitemShut {NoStop}%
\bibitem [{\citenamefont {Bell}(1951)}]{Bell_1951_TaylorInstabilityCylinders}%
  \BibitemOpen
  \bibfield  {author} {\bibinfo {author} {\bibfnamefont {G.~I.}\ \bibnamefont {Bell}},\ }\href@noop {} {\emph {\bibinfo {title} {Taylor Instability on Cylinders and Spheres in the Small Amplitude Approximation}}},\ \bibinfo {type} {Tech. Rep.}\ \bibinfo {number} {LA-1321}\ (\bibinfo  {institution} {Los Alamos Scientific Laboratory},\ \bibinfo {year} {1951})\BibitemShut {NoStop}%
\bibitem [{\citenamefont {Plesset}(1954)}]{Plesset_1954_StabilityFluidFlows}%
  \BibitemOpen
  \bibfield  {author} {\bibinfo {author} {\bibfnamefont {M.~S.}\ \bibnamefont {Plesset}},\ }\bibfield  {title} {\bibinfo {title} {On the {{Stability}} of {{Fluid Flows}} with {{Spherical Symmetry}}},\ }\href@noop {} {\bibfield  {journal} {\bibinfo  {journal} {Journal of Applied Physics}\ }\textbf {\bibinfo {volume} {25}},\ \bibinfo {pages} {96} (\bibinfo {year} {1954})}\BibitemShut {NoStop}%
\bibitem [{\citenamefont {Penney}\ and\ \citenamefont {Price}(1942)}]{Penney_1942_ChangingFormNearly}%
  \BibitemOpen
  \bibfield  {author} {\bibinfo {author} {\bibfnamefont {W.}~\bibnamefont {Penney}}\ and\ \bibinfo {author} {\bibfnamefont {A.}~\bibnamefont {Price}},\ }\bibfield  {title} {\bibinfo {title} {On the changing form of a nearly spherical submarine bubble},\ }\href@noop {} {\bibfield  {journal} {\bibinfo  {journal} {Underwarter Explosion Research}\ }\textbf {\bibinfo {volume} {2}},\ \bibinfo {pages} {145} (\bibinfo {year} {1942})}\BibitemShut {NoStop}%
\bibitem [{\citenamefont {Epstein}(2004)}]{Epstein_2004_BellPlessetEffects}%
  \BibitemOpen
  \bibfield  {author} {\bibinfo {author} {\bibfnamefont {R.}~\bibnamefont {Epstein}},\ }\bibfield  {title} {\bibinfo {title} {On the {{Bell}}--{{Plesset}} effects: {{The}} effects of uniform compression and geometrical convergence on the classical {{Rayleigh}}--{{Taylor}} instability},\ }\href@noop {} {\bibfield  {journal} {\bibinfo  {journal} {Physics of Plasmas}\ }\textbf {\bibinfo {volume} {11}},\ \bibinfo {pages} {5114} (\bibinfo {year} {2004})}\BibitemShut {NoStop}%
\bibitem [{\citenamefont {Pascoe}\ \emph {et~al.}(2024)\citenamefont {Pascoe}, \citenamefont {Groom}, \citenamefont {Youngs},\ and\ \citenamefont {Thornber}}]{Pascoe_2024_ImpactAxialStrain}%
  \BibitemOpen
  \bibfield  {author} {\bibinfo {author} {\bibfnamefont {B.}~\bibnamefont {Pascoe}}, \bibinfo {author} {\bibfnamefont {M.}~\bibnamefont {Groom}}, \bibinfo {author} {\bibfnamefont {D.~L.}\ \bibnamefont {Youngs}},\ and\ \bibinfo {author} {\bibfnamefont {B.}~\bibnamefont {Thornber}},\ }\bibfield  {title} {\bibinfo {title} {Impact of axial strain on linear, transitional and self-similar turbulent mixing layers},\ }\href@noop {} {\bibfield  {journal} {\bibinfo  {journal} {Journal of Fluid Mechanics}\ }\textbf {\bibinfo {volume} {999}},\ \bibinfo {pages} {A5} (\bibinfo {year} {2024})}\BibitemShut {NoStop}%
\bibitem [{\citenamefont {Pascoe}\ \emph {et~al.}(2025)\citenamefont {Pascoe}, \citenamefont {Groom}, \citenamefont {Youngs},\ and\ \citenamefont {Thornber}}]{Pascoe_2025_ImpactTransverseStrain}%
  \BibitemOpen
  \bibfield  {author} {\bibinfo {author} {\bibfnamefont {B.}~\bibnamefont {Pascoe}}, \bibinfo {author} {\bibfnamefont {M.}~\bibnamefont {Groom}}, \bibinfo {author} {\bibfnamefont {D.}~\bibnamefont {Youngs}},\ and\ \bibinfo {author} {\bibfnamefont {B.}~\bibnamefont {Thornber}},\ }\bibfield  {title} {\bibinfo {title} {The impact of transverse strain of linear, transitional, and self-similar turbulent mixing layers},\ }\href@noop {} {\bibfield  {journal} {\bibinfo  {journal} {(Submitted)}\ } (\bibinfo {year} {2025})}\BibitemShut {NoStop}%
\bibitem [{\citenamefont {Zhou}(2017{\natexlab{a}})}]{Zhou_2017_ReviewA}%
  \BibitemOpen
  \bibfield  {author} {\bibinfo {author} {\bibfnamefont {Y.}~\bibnamefont {Zhou}},\ }\bibfield  {title} {\bibinfo {title} {Rayleigh--{{Taylor}} and {{Richtmyer}}--{{Meshkov}} instability induced flow, turbulence, and mixing. {{I}}},\ }\href@noop {} {\bibfield  {journal} {\bibinfo  {journal} {Physics Reports}\ }\bibinfo {series} {Rayleigh-{{Taylor}} and {{Richtmyer-Meshkov}} Instability Induced Flow, Turbulence, and Mixing. {{I}}},\ \textbf {\bibinfo {volume} {720--722}},\ \bibinfo {pages} {1} (\bibinfo {year} {2017}{\natexlab{a}})}\BibitemShut {NoStop}%
\bibitem [{\citenamefont {Zhou}(2017{\natexlab{b}})}]{Zhou_2017_ReviewB}%
  \BibitemOpen
  \bibfield  {author} {\bibinfo {author} {\bibfnamefont {Y.}~\bibnamefont {Zhou}},\ }\bibfield  {title} {\bibinfo {title} {Rayleigh--{{Taylor}} and {{Richtmyer}}--{{Meshkov}} instability induced flow, turbulence, and mixing. {{II}}},\ }\href@noop {} {\bibfield  {journal} {\bibinfo  {journal} {Physics Reports}\ }\bibinfo {series} {Rayleigh--{{Taylor}} and {{Richtmyer}}--{{Meshkov}} Instability Induced Flow, Turbulence, and Mixing. {{II}}},\ \textbf {\bibinfo {volume} {723--725}},\ \bibinfo {pages} {1} (\bibinfo {year} {2017}{\natexlab{b}})}\BibitemShut {NoStop}%
\bibitem [{\citenamefont {Zhou}\ \emph {et~al.}(2021)\citenamefont {Zhou}, \citenamefont {Williams}, \citenamefont {Ramaprabhu}, \citenamefont {Groom}, \citenamefont {Thornber}, \citenamefont {Hillier}, \citenamefont {Mostert}, \citenamefont {Rollin}, \citenamefont {Balachandar}, \citenamefont {Powell}, \citenamefont {Mahalov},\ and\ \citenamefont {Attal}}]{Zhou_2021_JourneyThroughScales}%
  \BibitemOpen
  \bibfield  {author} {\bibinfo {author} {\bibfnamefont {Y.}~\bibnamefont {Zhou}}, \bibinfo {author} {\bibfnamefont {R.~J.~R.}\ \bibnamefont {Williams}}, \bibinfo {author} {\bibfnamefont {P.}~\bibnamefont {Ramaprabhu}}, \bibinfo {author} {\bibfnamefont {M.}~\bibnamefont {Groom}}, \bibinfo {author} {\bibfnamefont {B.}~\bibnamefont {Thornber}}, \bibinfo {author} {\bibfnamefont {A.}~\bibnamefont {Hillier}}, \bibinfo {author} {\bibfnamefont {W.}~\bibnamefont {Mostert}}, \bibinfo {author} {\bibfnamefont {B.}~\bibnamefont {Rollin}}, \bibinfo {author} {\bibfnamefont {S.}~\bibnamefont {Balachandar}}, \bibinfo {author} {\bibfnamefont {P.~D.}\ \bibnamefont {Powell}}, \bibinfo {author} {\bibfnamefont {A.}~\bibnamefont {Mahalov}},\ and\ \bibinfo {author} {\bibfnamefont {N.}~\bibnamefont {Attal}},\ }\bibfield  {title} {\bibinfo {title} {Rayleigh--{{Taylor}} and {{Richtmyer}}--{{Meshkov}} instabilities: {{A}} journey through scales},\ }\href@noop {} {\bibfield  {journal} {\bibinfo  {journal} {Physica D: Nonlinear
  Phenomena}\ }\textbf {\bibinfo {volume} {423}},\ \bibinfo {pages} {132838} (\bibinfo {year} {2021})}\BibitemShut {NoStop}%
\bibitem [{\citenamefont {Arnett}(2000)}]{Arnett_2000_RoleMixingAstrophysics}%
  \BibitemOpen
  \bibfield  {author} {\bibinfo {author} {\bibfnamefont {D.}~\bibnamefont {Arnett}},\ }\bibfield  {title} {\bibinfo {title} {The {{Role}} of {{Mixing}} in {{Astrophysics}}},\ }\href@noop {} {\bibfield  {journal} {\bibinfo  {journal} {The Astrophysical Journal Supplement Series}\ }\textbf {\bibinfo {volume} {127}},\ \bibinfo {pages} {213} (\bibinfo {year} {2000})}\BibitemShut {NoStop}%
\bibitem [{\citenamefont {Miles}(2009)}]{Miles_2009_BlastWaveDrivenInstabilityVehicle}%
  \BibitemOpen
  \bibfield  {author} {\bibinfo {author} {\bibfnamefont {A.~R.}\ \bibnamefont {Miles}},\ }\bibfield  {title} {\bibinfo {title} {The {{Blast-Wave-Driven Instability}} as a {{Vehicle}} for {{Understanding Supernova Explosion Structure}}},\ }\href@noop {} {\bibfield  {journal} {\bibinfo  {journal} {The Astrophysical Journal}\ }\textbf {\bibinfo {volume} {696}},\ \bibinfo {pages} {498} (\bibinfo {year} {2009})}\BibitemShut {NoStop}%
\bibitem [{\citenamefont {Sedov}(1946)}]{Sedov_1946_PropagationStrongShock}%
  \BibitemOpen
  \bibfield  {author} {\bibinfo {author} {\bibfnamefont {L.~I.}\ \bibnamefont {Sedov}},\ }\bibfield  {title} {\bibinfo {title} {Propagation of strong shock waves},\ }\href@noop {} {\bibfield  {journal} {\bibinfo  {journal} {Journal of Applied Mathematics and Mechanics}\ }\textbf {\bibinfo {volume} {10}},\ \bibinfo {pages} {241} (\bibinfo {year} {1946})}\BibitemShut {NoStop}%
\bibitem [{\citenamefont {Taylor}(1950{\natexlab{a}})}]{Taylor_1950_FormationBlastWave}%
  \BibitemOpen
  \bibfield  {author} {\bibinfo {author} {\bibfnamefont {G.~I.}\ \bibnamefont {Taylor}},\ }\bibfield  {title} {\bibinfo {title} {The formation of a blast wave by a very intense explosion {{I}}. {{Theoretical}} discussion},\ }\href@noop {} {\bibfield  {journal} {\bibinfo  {journal} {Proceedings of the Royal Society of London. Series A. Mathematical and Physical Sciences}\ }\textbf {\bibinfo {volume} {201}},\ \bibinfo {pages} {159} (\bibinfo {year} {1950}{\natexlab{a}})}\BibitemShut {NoStop}%
\bibitem [{\citenamefont {Taylor}(1950{\natexlab{b}})}]{Taylor_1950_FormationBlastWavea}%
  \BibitemOpen
  \bibfield  {author} {\bibinfo {author} {\bibfnamefont {G.~I.}\ \bibnamefont {Taylor}},\ }\bibfield  {title} {\bibinfo {title} {The formation of a blast wave by a very intense explosion. - {{II}}. {{The}} atomic explosion of 1945},\ }\href@noop {} {\bibfield  {journal} {\bibinfo  {journal} {Proceedings of the Royal Society of London. Series A. Mathematical and Physical Sciences}\ }\textbf {\bibinfo {volume} {201}},\ \bibinfo {pages} {175} (\bibinfo {year} {1950}{\natexlab{b}})}\BibitemShut {NoStop}%
\bibitem [{\citenamefont {Goldstine}\ and\ \citenamefont {Neumann}(1955)}]{Goldstine_1955_BlastWaveCalculation}%
  \BibitemOpen
  \bibfield  {author} {\bibinfo {author} {\bibfnamefont {H.~H.}\ \bibnamefont {Goldstine}}\ and\ \bibinfo {author} {\bibfnamefont {J.~V.}\ \bibnamefont {Neumann}},\ }\bibfield  {title} {\bibinfo {title} {Blast wave calculation},\ }\href@noop {} {\bibfield  {journal} {\bibinfo  {journal} {Communications on Pure and Applied Mathematics}\ }\textbf {\bibinfo {volume} {8}},\ \bibinfo {pages} {327} (\bibinfo {year} {1955})}\BibitemShut {NoStop}%
\bibitem [{\citenamefont {Nuckolls}\ \emph {et~al.}(1972)\citenamefont {Nuckolls}, \citenamefont {Wood}, \citenamefont {Thiessen},\ and\ \citenamefont {Zimmerman}}]{Nuckolls_1972_LaserCompressionMatter}%
  \BibitemOpen
  \bibfield  {author} {\bibinfo {author} {\bibfnamefont {J.}~\bibnamefont {Nuckolls}}, \bibinfo {author} {\bibfnamefont {L.}~\bibnamefont {Wood}}, \bibinfo {author} {\bibfnamefont {A.}~\bibnamefont {Thiessen}},\ and\ \bibinfo {author} {\bibfnamefont {G.}~\bibnamefont {Zimmerman}},\ }\bibfield  {title} {\bibinfo {title} {Laser {{Compression}} of {{Matter}} to {{Super-High Densities}}: {{Thermonuclear}} ({{CTR}}) {{Applications}}},\ }\href@noop {} {\bibfield  {journal} {\bibinfo  {journal} {Nature}\ }\textbf {\bibinfo {volume} {239}},\ \bibinfo {pages} {139} (\bibinfo {year} {1972})}\BibitemShut {NoStop}%
\bibitem [{\citenamefont {Betti}\ and\ \citenamefont {Hurricane}(2016)}]{Betti_2016_InertialconfinementFusionLasers}%
  \BibitemOpen
  \bibfield  {author} {\bibinfo {author} {\bibfnamefont {R.}~\bibnamefont {Betti}}\ and\ \bibinfo {author} {\bibfnamefont {O.~A.}\ \bibnamefont {Hurricane}},\ }\bibfield  {title} {\bibinfo {title} {Inertial-confinement fusion with lasers},\ }\href@noop {} {\bibfield  {journal} {\bibinfo  {journal} {Nature Physics}\ }\textbf {\bibinfo {volume} {12}},\ \bibinfo {pages} {435} (\bibinfo {year} {2016})}\BibitemShut {NoStop}%
\bibitem [{\citenamefont {Lindl}\ \emph {et~al.}(2004)\citenamefont {Lindl}, \citenamefont {Amendt}, \citenamefont {Berger}, \citenamefont {Glendinning}, \citenamefont {Glenzer}, \citenamefont {Haan}, \citenamefont {Kauffman}, \citenamefont {Landen},\ and\ \citenamefont {Suter}}]{Lindl_2004_PhysicsBasisIgnition}%
  \BibitemOpen
  \bibfield  {author} {\bibinfo {author} {\bibfnamefont {J.~D.}\ \bibnamefont {Lindl}}, \bibinfo {author} {\bibfnamefont {P.}~\bibnamefont {Amendt}}, \bibinfo {author} {\bibfnamefont {R.~L.}\ \bibnamefont {Berger}}, \bibinfo {author} {\bibfnamefont {S.~G.}\ \bibnamefont {Glendinning}}, \bibinfo {author} {\bibfnamefont {S.~H.}\ \bibnamefont {Glenzer}}, \bibinfo {author} {\bibfnamefont {S.~W.}\ \bibnamefont {Haan}}, \bibinfo {author} {\bibfnamefont {R.~L.}\ \bibnamefont {Kauffman}}, \bibinfo {author} {\bibfnamefont {O.~L.}\ \bibnamefont {Landen}},\ and\ \bibinfo {author} {\bibfnamefont {L.~J.}\ \bibnamefont {Suter}},\ }\bibfield  {title} {\bibinfo {title} {The physics basis for ignition using indirect-drive targets on the {{National Ignition Facility}}},\ }\href@noop {} {\bibfield  {journal} {\bibinfo  {journal} {Physics of Plasmas}\ }\textbf {\bibinfo {volume} {11}},\ \bibinfo {pages} {339} (\bibinfo {year} {2004})}\BibitemShut {NoStop}%
\bibitem [{\citenamefont {Lindl}\ \emph {et~al.}(2014)\citenamefont {Lindl}, \citenamefont {Landen}, \citenamefont {Edwards}, \citenamefont {Moses},\ and\ \citenamefont {{NIC Team}}}]{Lindl_2014_ReviewNationalIgnition}%
  \BibitemOpen
  \bibfield  {author} {\bibinfo {author} {\bibfnamefont {J.}~\bibnamefont {Lindl}}, \bibinfo {author} {\bibfnamefont {O.}~\bibnamefont {Landen}}, \bibinfo {author} {\bibfnamefont {J.}~\bibnamefont {Edwards}}, \bibinfo {author} {\bibfnamefont {E.}~\bibnamefont {Moses}},\ and\ \bibinfo {author} {\bibnamefont {{NIC Team}}},\ }\bibfield  {title} {\bibinfo {title} {Review of the {{National Ignition Campaign}} 2009-2012},\ }\href@noop {} {\bibfield  {journal} {\bibinfo  {journal} {Physics of Plasmas}\ }\textbf {\bibinfo {volume} {21}},\ \bibinfo {pages} {020501} (\bibinfo {year} {2014})}\BibitemShut {NoStop}%
\bibitem [{\citenamefont {Youngs}\ and\ \citenamefont {Williams}(2008)}]{Youngs_2008_TurbulentMixingSpherical}%
  \BibitemOpen
  \bibfield  {author} {\bibinfo {author} {\bibfnamefont {D.~L.}\ \bibnamefont {Youngs}}\ and\ \bibinfo {author} {\bibfnamefont {R.~J.~R.}\ \bibnamefont {Williams}},\ }\bibfield  {title} {\bibinfo {title} {Turbulent mixing in spherical implosions},\ }\href@noop {} {\bibfield  {journal} {\bibinfo  {journal} {International Journal for Numerical Methods in Fluids}\ }\textbf {\bibinfo {volume} {56}},\ \bibinfo {pages} {1597} (\bibinfo {year} {2008})}\BibitemShut {NoStop}%
\bibitem [{\citenamefont {Flaig}\ \emph {et~al.}(2018)\citenamefont {Flaig}, \citenamefont {Clark}, \citenamefont {Weber}, \citenamefont {Youngs},\ and\ \citenamefont {Thornber}}]{Flaig_2018_SinglemodePerturbationGrowth}%
  \BibitemOpen
  \bibfield  {author} {\bibinfo {author} {\bibfnamefont {M.}~\bibnamefont {Flaig}}, \bibinfo {author} {\bibfnamefont {D.}~\bibnamefont {Clark}}, \bibinfo {author} {\bibfnamefont {C.}~\bibnamefont {Weber}}, \bibinfo {author} {\bibfnamefont {D.~L.}\ \bibnamefont {Youngs}},\ and\ \bibinfo {author} {\bibfnamefont {B.}~\bibnamefont {Thornber}},\ }\bibfield  {title} {\bibinfo {title} {Single-mode perturbation growth in an idealized spherical implosion},\ }\href@noop {} {\bibfield  {journal} {\bibinfo  {journal} {Journal of Computational Physics}\ }\textbf {\bibinfo {volume} {371}},\ \bibinfo {pages} {801} (\bibinfo {year} {2018})}\BibitemShut {NoStop}%
\bibitem [{\citenamefont {Heidt}\ \emph {et~al.}(2021)\citenamefont {Heidt}, \citenamefont {Flaig},\ and\ \citenamefont {Thornber}}]{Heidt_2021_EffectInitialAmplitude}%
  \BibitemOpen
  \bibfield  {author} {\bibinfo {author} {\bibfnamefont {L.}~\bibnamefont {Heidt}}, \bibinfo {author} {\bibfnamefont {M.}~\bibnamefont {Flaig}},\ and\ \bibinfo {author} {\bibfnamefont {B.}~\bibnamefont {Thornber}},\ }\bibfield  {title} {\bibinfo {title} {The effect of initial amplitude and convergence ratio on instability development and deposited fluctuating kinetic energy in the single-mode {{Richtmyer}}--{{Meshkov}} instability in spherical implosions},\ }\href@noop {} {\bibfield  {journal} {\bibinfo  {journal} {Computers \& Fluids}\ }\textbf {\bibinfo {volume} {218}},\ \bibinfo {pages} {104842} (\bibinfo {year} {2021})}\BibitemShut {NoStop}%
\bibitem [{\citenamefont {Joggerst}\ \emph {et~al.}(2014)\citenamefont {Joggerst}, \citenamefont {Nelson}, \citenamefont {Woodward}, \citenamefont {Lovekin}, \citenamefont {Masser}, \citenamefont {Fryer}, \citenamefont {Ramaprabhu}, \citenamefont {Francois},\ and\ \citenamefont {Rockefeller}}]{Joggerst_2014_CrosscodeComparisonsMixing}%
  \BibitemOpen
  \bibfield  {author} {\bibinfo {author} {\bibfnamefont {C.~C.}\ \bibnamefont {Joggerst}}, \bibinfo {author} {\bibfnamefont {A.}~\bibnamefont {Nelson}}, \bibinfo {author} {\bibfnamefont {P.}~\bibnamefont {Woodward}}, \bibinfo {author} {\bibfnamefont {C.}~\bibnamefont {Lovekin}}, \bibinfo {author} {\bibfnamefont {T.}~\bibnamefont {Masser}}, \bibinfo {author} {\bibfnamefont {C.~L.}\ \bibnamefont {Fryer}}, \bibinfo {author} {\bibfnamefont {P.}~\bibnamefont {Ramaprabhu}}, \bibinfo {author} {\bibfnamefont {M.}~\bibnamefont {Francois}},\ and\ \bibinfo {author} {\bibfnamefont {G.}~\bibnamefont {Rockefeller}},\ }\bibfield  {title} {\bibinfo {title} {Cross-code comparisons of mixing during the implosion of dense cylindrical and spherical shells},\ }\href@noop {} {\bibfield  {journal} {\bibinfo  {journal} {Journal of Computational Physics}\ }\textbf {\bibinfo {volume} {275}},\ \bibinfo {pages} {154} (\bibinfo {year} {2014})}\BibitemShut {NoStop}%
\bibitem [{\citenamefont {Boureima}\ \emph {et~al.}(2017)\citenamefont {Boureima}, \citenamefont {Ramaprabhu},\ and\ \citenamefont {Attal}}]{Boureima_2017_PropertiesTurbulentMixing}%
  \BibitemOpen
  \bibfield  {author} {\bibinfo {author} {\bibfnamefont {I.}~\bibnamefont {Boureima}}, \bibinfo {author} {\bibfnamefont {P.}~\bibnamefont {Ramaprabhu}},\ and\ \bibinfo {author} {\bibfnamefont {N.}~\bibnamefont {Attal}},\ }\bibfield  {title} {\bibinfo {title} {Properties of the {{Turbulent Mixing Layer}} in a {{Spherical Implosion}}},\ }\href@noop {} {\bibfield  {journal} {\bibinfo  {journal} {Journal of Fluids Engineering}\ }\textbf {\bibinfo {volume} {140}} (\bibinfo {year} {2017})}\BibitemShut {NoStop}%
\bibitem [{\citenamefont {El~Rafei}\ \emph {et~al.}(2019)\citenamefont {El~Rafei}, \citenamefont {Flaig}, \citenamefont {Youngs},\ and\ \citenamefont {Thornber}}]{ElRafei_2019_ThreedimensionalSimulationsTurbulent}%
  \BibitemOpen
  \bibfield  {author} {\bibinfo {author} {\bibfnamefont {M.}~\bibnamefont {El~Rafei}}, \bibinfo {author} {\bibfnamefont {M.}~\bibnamefont {Flaig}}, \bibinfo {author} {\bibfnamefont {D.~L.}\ \bibnamefont {Youngs}},\ and\ \bibinfo {author} {\bibfnamefont {B.}~\bibnamefont {Thornber}},\ }\bibfield  {title} {\bibinfo {title} {Three-dimensional simulations of turbulent mixing in spherical implosions},\ }\href@noop {} {\bibfield  {journal} {\bibinfo  {journal} {Physics of Fluids}\ }\textbf {\bibinfo {volume} {31}},\ \bibinfo {pages} {114101} (\bibinfo {year} {2019})}\BibitemShut {NoStop}%
\bibitem [{\citenamefont {El~Rafei}\ and\ \citenamefont {Thornber}(2020)}]{ElRafei_2020_NumericalStudyBuoyancy}%
  \BibitemOpen
  \bibfield  {author} {\bibinfo {author} {\bibfnamefont {M.}~\bibnamefont {El~Rafei}}\ and\ \bibinfo {author} {\bibfnamefont {B.}~\bibnamefont {Thornber}},\ }\bibfield  {title} {\bibinfo {title} {Numerical study and buoyancy--drag modeling of bubble and spike distances in three-dimensional spherical implosions},\ }\href@noop {} {\bibfield  {journal} {\bibinfo  {journal} {Physics of Fluids}\ }\textbf {\bibinfo {volume} {32}},\ \bibinfo {pages} {124107} (\bibinfo {year} {2020})}\BibitemShut {NoStop}%
\bibitem [{\citenamefont {El~Rafei}\ and\ \citenamefont {Thornber}(2024)}]{ElRafei_2024_TurbulenceStatisticsTransport}%
  \BibitemOpen
  \bibfield  {author} {\bibinfo {author} {\bibfnamefont {M.}~\bibnamefont {El~Rafei}}\ and\ \bibinfo {author} {\bibfnamefont {B.}~\bibnamefont {Thornber}},\ }\bibfield  {title} {\bibinfo {title} {Turbulence statistics and transport in compressible mixing driven by spherical implosions with narrowband and broadband initial perturbations},\ }\href@noop {} {\bibfield  {journal} {\bibinfo  {journal} {Physical Review Fluids}\ }\textbf {\bibinfo {volume} {9}},\ \bibinfo {pages} {034501} (\bibinfo {year} {2024})}\BibitemShut {NoStop}%
\bibitem [{\citenamefont {Besnard}\ \emph {et~al.}(1992)\citenamefont {Besnard}, \citenamefont {Harlow}, \citenamefont {Rauenzahn},\ and\ \citenamefont {Zemach}}]{Besnard_1992_TurbulenceTransportEquations}%
  \BibitemOpen
  \bibfield  {author} {\bibinfo {author} {\bibfnamefont {D.}~\bibnamefont {Besnard}}, \bibinfo {author} {\bibfnamefont {F.~H.}\ \bibnamefont {Harlow}}, \bibinfo {author} {\bibfnamefont {R.~M.}\ \bibnamefont {Rauenzahn}},\ and\ \bibinfo {author} {\bibfnamefont {C.}~\bibnamefont {Zemach}},\ }\href@noop {} {\emph {\bibinfo {title} {Turbulence Transport Equations for Variable-Density Turbulence and Their Relationship to Two-Field Models}}},\ \bibinfo {type} {Tech. Rep.}\ \bibinfo {number} {LA-12303-MS}\ (\bibinfo  {institution} {Los Alamos National Lab. (LANL), Los Alamos, NM (United States)},\ \bibinfo {year} {1992})\BibitemShut {NoStop}%
\bibitem [{\citenamefont {Jones}\ and\ \citenamefont {Launder}(1972)}]{Jones_1972_PredictionLaminarizationTwoequation}%
  \BibitemOpen
  \bibfield  {author} {\bibinfo {author} {\bibfnamefont {W.~P.}\ \bibnamefont {Jones}}\ and\ \bibinfo {author} {\bibfnamefont {B.~E.}\ \bibnamefont {Launder}},\ }\bibfield  {title} {\bibinfo {title} {The prediction of laminarization with a two-equation model of turbulence},\ }\href@noop {} {\bibfield  {journal} {\bibinfo  {journal} {International Journal of Heat and Mass Transfer}\ }\textbf {\bibinfo {volume} {15}},\ \bibinfo {pages} {301} (\bibinfo {year} {1972})}\BibitemShut {NoStop}%
\bibitem [{\citenamefont {Launder}\ and\ \citenamefont {Spalding}(1974)}]{Launder_1974_NumericalComputationTurbulent}%
  \BibitemOpen
  \bibfield  {author} {\bibinfo {author} {\bibfnamefont {B.~E.}\ \bibnamefont {Launder}}\ and\ \bibinfo {author} {\bibfnamefont {D.~B.}\ \bibnamefont {Spalding}},\ }\bibfield  {title} {\bibinfo {title} {The numerical computation of turbulent flows},\ }\href@noop {} {\bibfield  {journal} {\bibinfo  {journal} {Computer Methods in Applied Mechanics and Engineering}\ }\textbf {\bibinfo {volume} {3}},\ \bibinfo {pages} {269} (\bibinfo {year} {1974})}\BibitemShut {NoStop}%
\bibitem [{\citenamefont {Wilcox}(2006)}]{Wilcox_2006}%
  \BibitemOpen
  \bibfield  {author} {\bibinfo {author} {\bibfnamefont {D.~C.}\ \bibnamefont {Wilcox}},\ }\bibfield  {title} {\bibinfo {title} {Turbulence modeling for {{CFD}}},\ }in\ \href@noop {} {\emph {\bibinfo {booktitle} {Turbulence Modeling for {{CFD}}}}}\ (\bibinfo  {publisher} {DCW Industries},\ \bibinfo {address} {La C{\~a}nada, Calif},\ \bibinfo {year} {2006})\ \bibinfo {edition} {3rd}\ ed.\BibitemShut {Stop}%
\bibitem [{\citenamefont {Gr{\'e}goire}\ \emph {et~al.}(2005)\citenamefont {Gr{\'e}goire}, \citenamefont {Souffland},\ and\ \citenamefont {Gauthier}}]{Gregoire_2005_SecondorderTurbulenceModel}%
  \BibitemOpen
  \bibfield  {author} {\bibinfo {author} {\bibfnamefont {O.}~\bibnamefont {Gr{\'e}goire}}, \bibinfo {author} {\bibfnamefont {D.}~\bibnamefont {Souffland}},\ and\ \bibinfo {author} {\bibfnamefont {S.}~\bibnamefont {Gauthier}},\ }\bibfield  {title} {\bibinfo {title} {A second-order turbulence model for gaseous mixtures induced by {{Richtmyer}}---{{Meshkov}} instability},\ }\href@noop {} {\bibfield  {journal} {\bibinfo  {journal} {Journal of Turbulence}\ }\textbf {\bibinfo {volume} {6}},\ \bibinfo {pages} {N29} (\bibinfo {year} {2005})}\BibitemShut {NoStop}%
\bibitem [{\citenamefont {{Mor{\'a}n-L{\'o}pez}}\ and\ \citenamefont {Schilling}(2014)}]{Moran-Lopez_2014_MulticomponentReynoldsaveragedNavier}%
  \BibitemOpen
  \bibfield  {author} {\bibinfo {author} {\bibfnamefont {J.~T.}\ \bibnamefont {{Mor{\'a}n-L{\'o}pez}}}\ and\ \bibinfo {author} {\bibfnamefont {O.}~\bibnamefont {Schilling}},\ }\bibfield  {title} {\bibinfo {title} {Multi-component {{Reynolds-averaged Navier}}--{{Stokes}} simulations of {{Richtmyer}}--{{Meshkov}} instability and mixing induced by reshock at different times},\ }\href@noop {} {\bibfield  {journal} {\bibinfo  {journal} {Shock Waves}\ }\textbf {\bibinfo {volume} {24}},\ \bibinfo {pages} {325} (\bibinfo {year} {2014})}\BibitemShut {NoStop}%
\bibitem [{\citenamefont {{Mor{\'a}n-L{\'o}pez}}\ \emph {et~al.}(2015)\citenamefont {{Mor{\'a}n-L{\'o}pez}}, \citenamefont {Schilling},\ and\ \citenamefont {Holloway}}]{Moran-Lopez_2015_ReynoldsAveragedNavierStokes}%
  \BibitemOpen
  \bibfield  {author} {\bibinfo {author} {\bibfnamefont {J.~T.}\ \bibnamefont {{Mor{\'a}n-L{\'o}pez}}}, \bibinfo {author} {\bibfnamefont {O.}~\bibnamefont {Schilling}},\ and\ \bibinfo {author} {\bibfnamefont {J.~P.}\ \bibnamefont {Holloway}},\ }\bibfield  {title} {\bibinfo {title} {Reynolds-{{Averaged Navier}}--{{Stokes Modeling}} of {{Reshocked Richtmyer}}--{{Meshkov Instability Experiments}} and {{Simulations}}},\ }in\ \href@noop {} {\emph {\bibinfo {booktitle} {29th {{International Symposium}} on {{Shock Waves}} 2}}},\ \bibinfo {editor} {edited by\ \bibinfo {editor} {\bibfnamefont {R.}~\bibnamefont {Bonazza}}\ and\ \bibinfo {editor} {\bibfnamefont {D.}~\bibnamefont {Ranjan}}}\ (\bibinfo  {publisher} {Springer International Publishing},\ \bibinfo {address} {Cham},\ \bibinfo {year} {2015})\ pp.\ \bibinfo {pages} {1047--1052}\BibitemShut {NoStop}%
\bibitem [{\citenamefont {Dimonte}\ and\ \citenamefont {Tipton}(2006)}]{Dimonte_2006_KLTurbulenceModel}%
  \BibitemOpen
  \bibfield  {author} {\bibinfo {author} {\bibfnamefont {G.}~\bibnamefont {Dimonte}}\ and\ \bibinfo {author} {\bibfnamefont {R.}~\bibnamefont {Tipton}},\ }\bibfield  {title} {\bibinfo {title} {K-{{L}} turbulence model for the self-similar growth of the {{Rayleigh-Taylor}} and {{Richtmyer-Meshkov}} instabilities},\ }\href@noop {} {\bibfield  {journal} {\bibinfo  {journal} {Physics of Fluids}\ }\textbf {\bibinfo {volume} {18}},\ \bibinfo {pages} {085101} (\bibinfo {year} {2006})}\BibitemShut {NoStop}%
\bibitem [{\citenamefont {Morgan}\ and\ \citenamefont {Greenough}(2016)}]{Morgan_2016_LargeeddyUnsteadyRANS}%
  \BibitemOpen
  \bibfield  {author} {\bibinfo {author} {\bibfnamefont {B.~E.}\ \bibnamefont {Morgan}}\ and\ \bibinfo {author} {\bibfnamefont {J.~A.}\ \bibnamefont {Greenough}},\ }\bibfield  {title} {\bibinfo {title} {Large-eddy and unsteady {{RANS}} simulations of a shock-accelerated heavy gas cylinder},\ }\href@noop {} {\bibfield  {journal} {\bibinfo  {journal} {Shock Waves}\ }\textbf {\bibinfo {volume} {26}},\ \bibinfo {pages} {355} (\bibinfo {year} {2016})}\BibitemShut {NoStop}%
\bibitem [{\citenamefont {Xiao}\ \emph {et~al.}(2020{\natexlab{a}})\citenamefont {Xiao}, \citenamefont {Zhang},\ and\ \citenamefont {Tian}}]{Xiao_2020_UnifiedPredictionReshocked}%
  \BibitemOpen
  \bibfield  {author} {\bibinfo {author} {\bibfnamefont {M.}~\bibnamefont {Xiao}}, \bibinfo {author} {\bibfnamefont {Y.}~\bibnamefont {Zhang}},\ and\ \bibinfo {author} {\bibfnamefont {B.}~\bibnamefont {Tian}},\ }\bibfield  {title} {\bibinfo {title} {Unified prediction of reshocked {{Richtmyer}}--{{Meshkov}} mixing with {{K-L}} model},\ }\href@noop {} {\bibfield  {journal} {\bibinfo  {journal} {Physics of Fluids}\ }\textbf {\bibinfo {volume} {32}},\ \bibinfo {pages} {032107} (\bibinfo {year} {2020}{\natexlab{a}})}\BibitemShut {NoStop}%
\bibitem [{\citenamefont {Xiao}\ \emph {et~al.}(2020{\natexlab{b}})\citenamefont {Xiao}, \citenamefont {Zhang},\ and\ \citenamefont {Tian}}]{Xiao_2020_ModelingTurbulentMixing}%
  \BibitemOpen
  \bibfield  {author} {\bibinfo {author} {\bibfnamefont {M.}~\bibnamefont {Xiao}}, \bibinfo {author} {\bibfnamefont {Y.}~\bibnamefont {Zhang}},\ and\ \bibinfo {author} {\bibfnamefont {B.}~\bibnamefont {Tian}},\ }\bibfield  {title} {\bibinfo {title} {Modeling of turbulent mixing with an improved {{K}}--{{L}} model},\ }\href@noop {} {\bibfield  {journal} {\bibinfo  {journal} {Physics of Fluids}\ }\textbf {\bibinfo {volume} {32}},\ \bibinfo {pages} {092104} (\bibinfo {year} {2020}{\natexlab{b}})}\BibitemShut {NoStop}%
\bibitem [{\citenamefont {Zhang}\ \emph {et~al.}(2020)\citenamefont {Zhang}, \citenamefont {He}, \citenamefont {Xie}, \citenamefont {Xiao},\ and\ \citenamefont {Tian}}]{Zhang_2020_MethodologyDeterminingCoefficients}%
  \BibitemOpen
  \bibfield  {author} {\bibinfo {author} {\bibfnamefont {Y.-s.}\ \bibnamefont {Zhang}}, \bibinfo {author} {\bibfnamefont {Z.-w.}\ \bibnamefont {He}}, \bibinfo {author} {\bibfnamefont {H.-s.}\ \bibnamefont {Xie}}, \bibinfo {author} {\bibfnamefont {M.-J.}\ \bibnamefont {Xiao}},\ and\ \bibinfo {author} {\bibfnamefont {B.-l.}\ \bibnamefont {Tian}},\ }\bibfield  {title} {\bibinfo {title} {Methodology for determining coefficients of turbulent mixing model},\ }\href@noop {} {\bibfield  {journal} {\bibinfo  {journal} {Journal of Fluid Mechanics}\ }\textbf {\bibinfo {volume} {905}},\ \bibinfo {pages} {A26} (\bibinfo {year} {2020})}\BibitemShut {NoStop}%
\bibitem [{\citenamefont {Kokkinakis}\ \emph {et~al.}(2015)\citenamefont {Kokkinakis}, \citenamefont {Drikakis}, \citenamefont {Youngs},\ and\ \citenamefont {Williams}}]{Kokkinakis_2015_TwoequationMultifluidTurbulence}%
  \BibitemOpen
  \bibfield  {author} {\bibinfo {author} {\bibfnamefont {I.~W.}\ \bibnamefont {Kokkinakis}}, \bibinfo {author} {\bibfnamefont {D.}~\bibnamefont {Drikakis}}, \bibinfo {author} {\bibfnamefont {D.~L.}\ \bibnamefont {Youngs}},\ and\ \bibinfo {author} {\bibfnamefont {R.~J.~R.}\ \bibnamefont {Williams}},\ }\bibfield  {title} {\bibinfo {title} {Two-equation and multi-fluid turbulence models for {{Rayleigh}}--{{Taylor}} mixing},\ }\href@noop {} {\bibfield  {journal} {\bibinfo  {journal} {International Journal of Heat and Fluid Flow}\ }\textbf {\bibinfo {volume} {56}},\ \bibinfo {pages} {233} (\bibinfo {year} {2015})}\BibitemShut {NoStop}%
\bibitem [{\citenamefont {Kokkinakis}\ \emph {et~al.}(2020)\citenamefont {Kokkinakis}, \citenamefont {Drikakis},\ and\ \citenamefont {Youngs}}]{Kokkinakis_2020_TwoequationMultifluidTurbulence}%
  \BibitemOpen
  \bibfield  {author} {\bibinfo {author} {\bibfnamefont {I.~W.}\ \bibnamefont {Kokkinakis}}, \bibinfo {author} {\bibfnamefont {D.}~\bibnamefont {Drikakis}},\ and\ \bibinfo {author} {\bibfnamefont {D.~L.}\ \bibnamefont {Youngs}},\ }\bibfield  {title} {\bibinfo {title} {Two-equation and multi-fluid turbulence models for {{Richtmyer}}--{{Meshkov}} mixing},\ }\href@noop {} {\bibfield  {journal} {\bibinfo  {journal} {Physics of Fluids}\ }\textbf {\bibinfo {volume} {32}},\ \bibinfo {pages} {074102} (\bibinfo {year} {2020})}\BibitemShut {NoStop}%
\bibitem [{\citenamefont {Xiao}\ \emph {et~al.}(2021)\citenamefont {Xiao}, \citenamefont {Zhang},\ and\ \citenamefont {Tian}}]{Xiao_2021_KLModelImproved}%
  \BibitemOpen
  \bibfield  {author} {\bibinfo {author} {\bibfnamefont {M.}~\bibnamefont {Xiao}}, \bibinfo {author} {\bibfnamefont {Y.}~\bibnamefont {Zhang}},\ and\ \bibinfo {author} {\bibfnamefont {B.}~\bibnamefont {Tian}},\ }\bibfield  {title} {\bibinfo {title} {A {{K}}--{{L}} model with improved realizability for turbulent mixing},\ }\href@noop {} {\bibfield  {journal} {\bibinfo  {journal} {Physics of Fluids}\ }\textbf {\bibinfo {volume} {33}},\ \bibinfo {pages} {022104} (\bibinfo {year} {2021})}\BibitemShut {NoStop}%
\bibitem [{\citenamefont {Morgan}\ and\ \citenamefont {Wickett}(2015)}]{Morgan_2015_ThreeequationModelSelfsimilar}%
  \BibitemOpen
  \bibfield  {author} {\bibinfo {author} {\bibfnamefont {B.~E.}\ \bibnamefont {Morgan}}\ and\ \bibinfo {author} {\bibfnamefont {M.~E.}\ \bibnamefont {Wickett}},\ }\bibfield  {title} {\bibinfo {title} {Three-equation model for the self-similar growth of {{Rayleigh-Taylor}} and {{Richtmyer-Meskov}} instabilities},\ }\href@noop {} {\bibfield  {journal} {\bibinfo  {journal} {Physical Review E}\ }\textbf {\bibinfo {volume} {91}},\ \bibinfo {pages} {043002} (\bibinfo {year} {2015})}\BibitemShut {NoStop}%
\bibitem [{\citenamefont {Morgan}\ \emph {et~al.}(2018{\natexlab{a}})\citenamefont {Morgan}, \citenamefont {Olson}, \citenamefont {Black},\ and\ \citenamefont {McFarland}}]{Morgan_2018_LargeeddySimulationReynoldsaveraged}%
  \BibitemOpen
  \bibfield  {author} {\bibinfo {author} {\bibfnamefont {B.~E.}\ \bibnamefont {Morgan}}, \bibinfo {author} {\bibfnamefont {B.~J.}\ \bibnamefont {Olson}}, \bibinfo {author} {\bibfnamefont {W.~J.}\ \bibnamefont {Black}},\ and\ \bibinfo {author} {\bibfnamefont {J.~A.}\ \bibnamefont {McFarland}},\ }\bibfield  {title} {\bibinfo {title} {Large-eddy simulation and {{Reynolds-averaged Navier-Stokes}} modeling of a reacting {{Rayleigh-Taylor}} mixing layer in a spherical geometry},\ }\href@noop {} {\bibfield  {journal} {\bibinfo  {journal} {Physical Review E}\ }\textbf {\bibinfo {volume} {98}},\ \bibinfo {pages} {033111} (\bibinfo {year} {2018}{\natexlab{a}})}\BibitemShut {NoStop}%
\bibitem [{\citenamefont {Morgan}\ \emph {et~al.}(2018{\natexlab{b}})\citenamefont {Morgan}, \citenamefont {Schilling},\ and\ \citenamefont {Hartland}}]{Morgan_2018_TwolengthscaleTurbulenceModel}%
  \BibitemOpen
  \bibfield  {author} {\bibinfo {author} {\bibfnamefont {B.~E.}\ \bibnamefont {Morgan}}, \bibinfo {author} {\bibfnamefont {O.}~\bibnamefont {Schilling}},\ and\ \bibinfo {author} {\bibfnamefont {T.~A.}\ \bibnamefont {Hartland}},\ }\bibfield  {title} {\bibinfo {title} {Two-length-scale turbulence model for self-similar buoyancy-, shock-, and shear-driven mixing},\ }\href@noop {} {\bibfield  {journal} {\bibinfo  {journal} {Physical Review E}\ }\textbf {\bibinfo {volume} {97}},\ \bibinfo {pages} {013104} (\bibinfo {year} {2018}{\natexlab{b}})}\BibitemShut {NoStop}%
\bibitem [{\citenamefont {Morgan}(2021)}]{Morgan_2021_SelfconsistentHighorderSpatial}%
  \BibitemOpen
  \bibfield  {author} {\bibinfo {author} {\bibfnamefont {B.~E.}\ \bibnamefont {Morgan}},\ }\bibfield  {title} {\bibinfo {title} {Self-consistent, high-order spatial profiles in a model for two-fluid turbulent mixing},\ }\href@noop {} {\bibfield  {journal} {\bibinfo  {journal} {Physical Review E}\ }\textbf {\bibinfo {volume} {104}},\ \bibinfo {pages} {015107} (\bibinfo {year} {2021})}\BibitemShut {NoStop}%
\bibitem [{\citenamefont {Morgan}(2022)}]{Morgan_2022_SimulationReynoldsaveragedNavierStokes}%
  \BibitemOpen
  \bibfield  {author} {\bibinfo {author} {\bibfnamefont {B.~E.}\ \bibnamefont {Morgan}},\ }\bibfield  {title} {\bibinfo {title} {Simulation and {{Reynolds-averaged Navier-Stokes}} modeling of a three-component {{Rayleigh-Taylor}} mixing problem with thermonuclear burn},\ }\href@noop {} {\bibfield  {journal} {\bibinfo  {journal} {Physical Review E}\ }\textbf {\bibinfo {volume} {105}},\ \bibinfo {pages} {045104} (\bibinfo {year} {2022})}\BibitemShut {NoStop}%
\bibitem [{\citenamefont {Morgan}\ \emph {et~al.}(2023)\citenamefont {Morgan}, \citenamefont {Ferguson},\ and\ \citenamefont {Olson}}]{Morgan_2023_TwoSelfsimilarReynoldsstress}%
  \BibitemOpen
  \bibfield  {author} {\bibinfo {author} {\bibfnamefont {B.~E.}\ \bibnamefont {Morgan}}, \bibinfo {author} {\bibfnamefont {K.}~\bibnamefont {Ferguson}},\ and\ \bibinfo {author} {\bibfnamefont {B.~J.}\ \bibnamefont {Olson}},\ }\bibfield  {title} {\bibinfo {title} {Two self-similar {{Reynolds-stress}} transport models with anisotropic eddy viscosity},\ }\href@noop {} {\bibfield  {journal} {\bibinfo  {journal} {Physical Review E}\ }\textbf {\bibinfo {volume} {108}},\ \bibinfo {pages} {055104} (\bibinfo {year} {2023})}\BibitemShut {NoStop}%
\bibitem [{\citenamefont {Banerjee}\ \emph {et~al.}(2010)\citenamefont {Banerjee}, \citenamefont {Gore},\ and\ \citenamefont {Andrews}}]{Banerjee_2010_DevelopmentValidationTurbulentmix}%
  \BibitemOpen
  \bibfield  {author} {\bibinfo {author} {\bibfnamefont {A.}~\bibnamefont {Banerjee}}, \bibinfo {author} {\bibfnamefont {R.~A.}\ \bibnamefont {Gore}},\ and\ \bibinfo {author} {\bibfnamefont {M.~J.}\ \bibnamefont {Andrews}},\ }\bibfield  {title} {\bibinfo {title} {Development and validation of a turbulent-mix model for variable-density and compressible flows},\ }\href@noop {} {\bibfield  {journal} {\bibinfo  {journal} {Physical Review E}\ }\textbf {\bibinfo {volume} {82}},\ \bibinfo {pages} {046309} (\bibinfo {year} {2010})}\BibitemShut {NoStop}%
\bibitem [{\citenamefont {Denissen}\ \emph {et~al.}(2012)\citenamefont {Denissen}, \citenamefont {Fung}, \citenamefont {Reisner},\ and\ \citenamefont {Andrews}}]{Denissen_2012_ImplementationValidationBHR}%
  \BibitemOpen
  \bibfield  {author} {\bibinfo {author} {\bibfnamefont {N.~A.}\ \bibnamefont {Denissen}}, \bibinfo {author} {\bibfnamefont {J.}~\bibnamefont {Fung}}, \bibinfo {author} {\bibfnamefont {J.~M.}\ \bibnamefont {Reisner}},\ and\ \bibinfo {author} {\bibfnamefont {M.~J.}\ \bibnamefont {Andrews}},\ }\href@noop {} {\emph {\bibinfo {title} {Implementation and {{Validation}} of the {{BHR Turbulence Model}} in the {{FLAG Hydrocode}}}}},\ \bibinfo {type} {Tech. Rep.}\ \bibinfo {number} {LA-UR-12-24386}\ (\bibinfo  {institution} {Los Alamos National Lab. (LANL), Los Alamos, NM (United States)},\ \bibinfo {year} {2012})\BibitemShut {NoStop}%
\bibitem [{\citenamefont {Schwarzkopf}\ \emph {et~al.}(2016)\citenamefont {Schwarzkopf}, \citenamefont {Livescu}, \citenamefont {Baltzer}, \citenamefont {Gore},\ and\ \citenamefont {Ristorcelli}}]{Schwarzkopf_2016_TwolengthScaleTurbulence}%
  \BibitemOpen
  \bibfield  {author} {\bibinfo {author} {\bibfnamefont {J.~D.}\ \bibnamefont {Schwarzkopf}}, \bibinfo {author} {\bibfnamefont {D.}~\bibnamefont {Livescu}}, \bibinfo {author} {\bibfnamefont {J.~R.}\ \bibnamefont {Baltzer}}, \bibinfo {author} {\bibfnamefont {R.~A.}\ \bibnamefont {Gore}},\ and\ \bibinfo {author} {\bibfnamefont {J.~R.}\ \bibnamefont {Ristorcelli}},\ }\bibfield  {title} {\bibinfo {title} {A {{Two-length Scale Turbulence Model}} for {{Single-phase Multi-fluid Mixing}}},\ }\href@noop {} {\bibfield  {journal} {\bibinfo  {journal} {Flow, Turbulence and Combustion}\ }\textbf {\bibinfo {volume} {96}},\ \bibinfo {pages} {1} (\bibinfo {year} {2016})}\BibitemShut {NoStop}%
\bibitem [{\citenamefont {Braun}\ and\ \citenamefont {Gore}(2021)}]{Braun_2021_MultispeciesTurbulenceModel}%
  \BibitemOpen
  \bibfield  {author} {\bibinfo {author} {\bibfnamefont {N.~O.}\ \bibnamefont {Braun}}\ and\ \bibinfo {author} {\bibfnamefont {R.~A.}\ \bibnamefont {Gore}},\ }\bibfield  {title} {\bibinfo {title} {A multispecies turbulence model for the mixing and de-mixing of miscible fluids},\ }\href@noop {} {\bibfield  {journal} {\bibinfo  {journal} {Journal of Turbulence}\ }\textbf {\bibinfo {volume} {22}},\ \bibinfo {pages} {784} (\bibinfo {year} {2021})}\BibitemShut {NoStop}%
\bibitem [{\citenamefont {Reynolds}(1980)}]{Reynolds_1980_ModelingFluidMotions}%
  \BibitemOpen
  \bibfield  {author} {\bibinfo {author} {\bibfnamefont {W.~C.}\ \bibnamefont {Reynolds}},\ }\bibfield  {title} {\bibinfo {title} {Modeling of fluid motions in engines---an introductory overview},\ }in\ \href@noop {} {\emph {\bibinfo {booktitle} {Symposium on {{Combustion Modeling}} in {{Reciprocatina Engine}}}}}\ (\bibinfo  {publisher} {Plenum Press},\ \bibinfo {year} {1980})\ pp.\ \bibinfo {pages} {41--66}\BibitemShut {NoStop}%
\bibitem [{\citenamefont {Reynolds}(1987)}]{Reynolds_1987_FundamentalsOfTurbulence}%
  \BibitemOpen
  \bibfield  {author} {\bibinfo {author} {\bibfnamefont {W.}~\bibnamefont {Reynolds}},\ }\href@noop {} {\emph {\bibinfo {title} {Fundamentals of Turbulence for Turbulence Modeling and Simulation}}},\ \bibinfo {type} {Tech. Rep.}\ \bibinfo {number} {Agard Report No. 755}\ (\bibinfo  {institution} {Defense Technical Information Center},\ \bibinfo {year} {1987})\BibitemShut {NoStop}%
\bibitem [{\citenamefont {Morel}\ and\ \citenamefont {Mansour}(1982)}]{Morel_1982_ModelingTurbulenceInternal}%
  \BibitemOpen
  \bibfield  {author} {\bibinfo {author} {\bibfnamefont {T.}~\bibnamefont {Morel}}\ and\ \bibinfo {author} {\bibfnamefont {N.~N.}\ \bibnamefont {Mansour}},\ }\href@noop {} {\emph {\bibinfo {title} {Modeling of {{Turbulence}} in {{Internal Combustion Engines}}}}},\ \bibinfo {type} {{{SAE Technical Paper}}}\ \bibinfo {number} {820040}\ (\bibinfo  {institution} {SAE International},\ \bibinfo {address} {Warrendale, PA},\ \bibinfo {year} {1982})\BibitemShut {NoStop}%
\bibitem [{\citenamefont {Campos}\ and\ \citenamefont {Morgan}(2019)}]{Campos_2019_DirectNumericalSimulation}%
  \BibitemOpen
  \bibfield  {author} {\bibinfo {author} {\bibfnamefont {A.}~\bibnamefont {Campos}}\ and\ \bibinfo {author} {\bibfnamefont {B.~E.}\ \bibnamefont {Morgan}},\ }\bibfield  {title} {\bibinfo {title} {Direct numerical simulation and {{Reynolds-averaged Navier-Stokes}} modeling of the sudden viscous dissipation for multicomponent turbulence},\ }\href@noop {} {\bibfield  {journal} {\bibinfo  {journal} {Physical Review E}\ }\textbf {\bibinfo {volume} {99}},\ \bibinfo {pages} {063103} (\bibinfo {year} {2019})}\BibitemShut {NoStop}%
\bibitem [{\citenamefont {Godunov}\ \emph {et~al.}(1976)\citenamefont {Godunov}, \citenamefont {Zabrodin}, \citenamefont {Ivanov}, \citenamefont {Kraiko},\ and\ \citenamefont {Prokopov}}]{Godunov_1976_NumericalSolutionMultidimensional}%
  \BibitemOpen
  \bibfield  {author} {\bibinfo {author} {\bibfnamefont {S.~K.}\ \bibnamefont {Godunov}}, \bibinfo {author} {\bibfnamefont {A.~V.}\ \bibnamefont {Zabrodin}}, \bibinfo {author} {\bibfnamefont {M.~{\relax Ia}.}\ \bibnamefont {Ivanov}}, \bibinfo {author} {\bibfnamefont {A.~N.}\ \bibnamefont {Kraiko}},\ and\ \bibinfo {author} {\bibfnamefont {G.~P.}\ \bibnamefont {Prokopov}},\ }\href@noop {} {\emph {\bibinfo {title} {Numerical Solution of Multidimensional Problems of Gas Dynamics}}}\ (\bibinfo  {publisher} {Nauka Press},\ \bibinfo {year} {1976})\BibitemShut {NoStop}%
\bibitem [{\citenamefont {Kim}\ and\ \citenamefont {Kim}(2005)}]{Kim_2005_AccurateEfficientMonotonic}%
  \BibitemOpen
  \bibfield  {author} {\bibinfo {author} {\bibfnamefont {K.~H.}\ \bibnamefont {Kim}}\ and\ \bibinfo {author} {\bibfnamefont {C.}~\bibnamefont {Kim}},\ }\bibfield  {title} {\bibinfo {title} {Accurate, efficient and monotonic numerical methods for multi-dimensional compressible flows: {{Part II}}: {{Multi-dimensional}} limiting process},\ }\href@noop {} {\bibfield  {journal} {\bibinfo  {journal} {Journal of Computational Physics}\ }\textbf {\bibinfo {volume} {208}},\ \bibinfo {pages} {570} (\bibinfo {year} {2005})}\BibitemShut {NoStop}%
\bibitem [{\citenamefont {Thornber}\ \emph {et~al.}(2008{\natexlab{a}})\citenamefont {Thornber}, \citenamefont {Drikakis}, \citenamefont {Williams},\ and\ \citenamefont {Youngs}}]{Thornber_2008_EntropyGenerationDissipation}%
  \BibitemOpen
  \bibfield  {author} {\bibinfo {author} {\bibfnamefont {B.}~\bibnamefont {Thornber}}, \bibinfo {author} {\bibfnamefont {D.}~\bibnamefont {Drikakis}}, \bibinfo {author} {\bibfnamefont {R.~J.~R.}\ \bibnamefont {Williams}},\ and\ \bibinfo {author} {\bibfnamefont {D.}~\bibnamefont {Youngs}},\ }\bibfield  {title} {\bibinfo {title} {On entropy generation and dissipation of kinetic energy in high-resolution shock-capturing schemes},\ }\href@noop {} {\bibfield  {journal} {\bibinfo  {journal} {Journal of Computational Physics}\ }\textbf {\bibinfo {volume} {227}},\ \bibinfo {pages} {4853} (\bibinfo {year} {2008}{\natexlab{a}})}\BibitemShut {NoStop}%
\bibitem [{\citenamefont {Thornber}\ \emph {et~al.}(2008{\natexlab{b}})\citenamefont {Thornber}, \citenamefont {Mosedale}, \citenamefont {Drikakis}, \citenamefont {Youngs},\ and\ \citenamefont {Williams}}]{Thornber_2008_ImprovedReconstructionMethod}%
  \BibitemOpen
  \bibfield  {author} {\bibinfo {author} {\bibfnamefont {B.}~\bibnamefont {Thornber}}, \bibinfo {author} {\bibfnamefont {A.}~\bibnamefont {Mosedale}}, \bibinfo {author} {\bibfnamefont {D.}~\bibnamefont {Drikakis}}, \bibinfo {author} {\bibfnamefont {D.}~\bibnamefont {Youngs}},\ and\ \bibinfo {author} {\bibfnamefont {R.~J.~R.}\ \bibnamefont {Williams}},\ }\bibfield  {title} {\bibinfo {title} {An improved reconstruction method for compressible flows with low {{Mach}} number features},\ }\href@noop {} {\bibfield  {journal} {\bibinfo  {journal} {Journal of Computational Physics}\ }\textbf {\bibinfo {volume} {227}},\ \bibinfo {pages} {4873} (\bibinfo {year} {2008}{\natexlab{b}})}\BibitemShut {NoStop}%
\bibitem [{\citenamefont {Toro}\ \emph {et~al.}(1994)\citenamefont {Toro}, \citenamefont {Spruce},\ and\ \citenamefont {Speares}}]{Toro_1994_RestorationContactSurface}%
  \BibitemOpen
  \bibfield  {author} {\bibinfo {author} {\bibfnamefont {E.~F.}\ \bibnamefont {Toro}}, \bibinfo {author} {\bibfnamefont {M.}~\bibnamefont {Spruce}},\ and\ \bibinfo {author} {\bibfnamefont {W.}~\bibnamefont {Speares}},\ }\bibfield  {title} {\bibinfo {title} {Restoration of the contact surface in the {{HLL-Riemann}} solver},\ }\href@noop {} {\bibfield  {journal} {\bibinfo  {journal} {Shock Waves}\ }\textbf {\bibinfo {volume} {4}},\ \bibinfo {pages} {25} (\bibinfo {year} {1994})}\BibitemShut {NoStop}%
\bibitem [{\citenamefont {Spiteri}\ and\ \citenamefont {Ruuth}(2002)}]{Spiteri_2002_NewClassOptimal}%
  \BibitemOpen
  \bibfield  {author} {\bibinfo {author} {\bibfnamefont {R.~J.}\ \bibnamefont {Spiteri}}\ and\ \bibinfo {author} {\bibfnamefont {S.~J.}\ \bibnamefont {Ruuth}},\ }\bibfield  {title} {\bibinfo {title} {A {{New Class}} of {{Optimal High-Order Strong-Stability-Preserving Time Discretization Methods}}},\ }\href@noop {} {\bibfield  {journal} {\bibinfo  {journal} {SIAM Journal on Numerical Analysis}\ }\textbf {\bibinfo {volume} {40}},\ \bibinfo {pages} {469} (\bibinfo {year} {2002})}\BibitemShut {NoStop}%
\bibitem [{\citenamefont {Luo}\ \emph {et~al.}(2004)\citenamefont {Luo}, \citenamefont {Baum},\ and\ \citenamefont {L{\"o}hner}}]{Luo_2004_ComputationMultimaterialFlows}%
  \BibitemOpen
  \bibfield  {author} {\bibinfo {author} {\bibfnamefont {H.}~\bibnamefont {Luo}}, \bibinfo {author} {\bibfnamefont {J.~D.}\ \bibnamefont {Baum}},\ and\ \bibinfo {author} {\bibfnamefont {R.}~\bibnamefont {L{\"o}hner}},\ }\bibfield  {title} {\bibinfo {title} {On the computation of multi-material flows using {{ALE}} formulation},\ }\href@noop {} {\bibfield  {journal} {\bibinfo  {journal} {Journal of Computational Physics}\ }\textbf {\bibinfo {volume} {194}},\ \bibinfo {pages} {304} (\bibinfo {year} {2004})}\BibitemShut {NoStop}%
\bibitem [{\citenamefont {Thornber}\ \emph {et~al.}(2017)\citenamefont {Thornber}, \citenamefont {Griffond}, \citenamefont {Poujade}, \citenamefont {Attal}, \citenamefont {Varshochi}, \citenamefont {Bigdelou}, \citenamefont {Ramaprabhu}, \citenamefont {Olson}, \citenamefont {Greenough}, \citenamefont {Zhou}, \citenamefont {Schilling}, \citenamefont {Garside}, \citenamefont {Williams}, \citenamefont {Batha}, \citenamefont {Kuchugov}, \citenamefont {Ladonkina}, \citenamefont {Tishkin}, \citenamefont {Zmitrenko}, \citenamefont {Rozanov},\ and\ \citenamefont {Youngs}}]{ThetaGroup}%
  \BibitemOpen
  \bibfield  {author} {\bibinfo {author} {\bibfnamefont {B.}~\bibnamefont {Thornber}}, \bibinfo {author} {\bibfnamefont {J.}~\bibnamefont {Griffond}}, \bibinfo {author} {\bibfnamefont {O.}~\bibnamefont {Poujade}}, \bibinfo {author} {\bibfnamefont {N.}~\bibnamefont {Attal}}, \bibinfo {author} {\bibfnamefont {H.}~\bibnamefont {Varshochi}}, \bibinfo {author} {\bibfnamefont {P.}~\bibnamefont {Bigdelou}}, \bibinfo {author} {\bibfnamefont {P.}~\bibnamefont {Ramaprabhu}}, \bibinfo {author} {\bibfnamefont {B.}~\bibnamefont {Olson}}, \bibinfo {author} {\bibfnamefont {J.}~\bibnamefont {Greenough}}, \bibinfo {author} {\bibfnamefont {Y.}~\bibnamefont {Zhou}}, \bibinfo {author} {\bibfnamefont {O.}~\bibnamefont {Schilling}}, \bibinfo {author} {\bibfnamefont {K.~A.}\ \bibnamefont {Garside}}, \bibinfo {author} {\bibfnamefont {R.~J.~R.}\ \bibnamefont {Williams}}, \bibinfo {author} {\bibfnamefont {C.~A.}\ \bibnamefont {Batha}}, \bibinfo {author} {\bibfnamefont {P.~A.}\ \bibnamefont {Kuchugov}}, \bibinfo {author} {\bibfnamefont
  {M.~E.}\ \bibnamefont {Ladonkina}}, \bibinfo {author} {\bibfnamefont {V.~F.}\ \bibnamefont {Tishkin}}, \bibinfo {author} {\bibfnamefont {N.~V.}\ \bibnamefont {Zmitrenko}}, \bibinfo {author} {\bibfnamefont {V.~B.}\ \bibnamefont {Rozanov}},\ and\ \bibinfo {author} {\bibfnamefont {D.~L.}\ \bibnamefont {Youngs}},\ }\bibfield  {title} {\bibinfo {title} {Late-time growth rate, mixing, and anisotropy in the multimode narrowband {{Richtmyer}}--{{Meshkov}} instability: {{The}} {$\theta$}-group collaboration},\ }\href@noop {} {\bibfield  {journal} {\bibinfo  {journal} {Physics of Fluids}\ }\textbf {\bibinfo {volume} {29}},\ \bibinfo {pages} {105107} (\bibinfo {year} {2017})}\BibitemShut {NoStop}%
\bibitem [{\citenamefont {Youngs}\ and\ \citenamefont {Thornber}(2020{\natexlab{a}})}]{Youngs_2020_BuoyancyDragModelling}%
  \BibitemOpen
  \bibfield  {author} {\bibinfo {author} {\bibfnamefont {D.~L.}\ \bibnamefont {Youngs}}\ and\ \bibinfo {author} {\bibfnamefont {B.}~\bibnamefont {Thornber}},\ }\bibfield  {title} {\bibinfo {title} {Buoyancy--{{Drag}} modelling of bubble and spike distances for single-shock {{Richtmyer}}--{{Meshkov}} mixing},\ }\href@noop {} {\bibfield  {journal} {\bibinfo  {journal} {Physica D: Nonlinear Phenomena}\ }\textbf {\bibinfo {volume} {410}},\ \bibinfo {pages} {132517} (\bibinfo {year} {2020}{\natexlab{a}})}\BibitemShut {NoStop}%
\bibitem [{\citenamefont {Youngs}\ and\ \citenamefont {Thornber}(2020{\natexlab{b}})}]{Youngs_2020_EarlyTimeModifications}%
  \BibitemOpen
  \bibfield  {author} {\bibinfo {author} {\bibfnamefont {D.~L.}\ \bibnamefont {Youngs}}\ and\ \bibinfo {author} {\bibfnamefont {B.}~\bibnamefont {Thornber}},\ }\bibfield  {title} {\bibinfo {title} {Early {{Time Modifications}} to the {{Buoyancy-Drag Model}} for {{Richtmyer}}--{{Meshkov Mixing}}},\ }\href@noop {} {\bibfield  {journal} {\bibinfo  {journal} {Journal of Fluids Engineering}\ }\textbf {\bibinfo {volume} {142}} (\bibinfo {year} {2020}{\natexlab{b}})}\BibitemShut {NoStop}%
\bibitem [{\citenamefont {Andrews}(1992)}]{Andrews_1992_ExperimentalStudyTurbulent}%
  \BibitemOpen
  \bibfield  {author} {\bibinfo {author} {\bibfnamefont {M.~J.}\ \bibnamefont {Andrews}},\ }\bibfield  {title} {\bibinfo {title} {An {{Experimental Study}} of {{Turbulent Mixing}} by the {{Rayleigh-Taylor Instabilities}} and a {{Two-Fluid Model}} of the {{Mixing Phenomena}}},\ }in\ \href@noop {} {\emph {\bibinfo {booktitle} {Advances in {{Compressible Turbulent Mixing}}}}}\ (\bibinfo  {publisher} {Lawrence Livermore National Laboratory},\ \bibinfo {year} {1992})\ pp.\ \bibinfo {pages} {7--19}\BibitemShut {NoStop}%
\bibitem [{\citenamefont {Youngs}(1994)}]{Youngs_1994_NumericalSimulationMixing}%
  \BibitemOpen
  \bibfield  {author} {\bibinfo {author} {\bibfnamefont {D.~L.}\ \bibnamefont {Youngs}},\ }\bibfield  {title} {\bibinfo {title} {Numerical simulation of mixing by {{Rayleigh}}--{{Taylor}} and {{Richtmyer}}--{{Meshkov}} instabilities},\ }\href@noop {} {\bibfield  {journal} {\bibinfo  {journal} {Laser and Particle Beams}\ }\textbf {\bibinfo {volume} {12}},\ \bibinfo {pages} {725} (\bibinfo {year} {1994})}\BibitemShut {NoStop}%
\bibitem [{\citenamefont {Allaire}\ \emph {et~al.}(2002)\citenamefont {Allaire}, \citenamefont {Clerc},\ and\ \citenamefont {Kokh}}]{Allaire_2002_FiveEquationModelSimulation}%
  \BibitemOpen
  \bibfield  {author} {\bibinfo {author} {\bibfnamefont {G.}~\bibnamefont {Allaire}}, \bibinfo {author} {\bibfnamefont {S.}~\bibnamefont {Clerc}},\ and\ \bibinfo {author} {\bibfnamefont {S.}~\bibnamefont {Kokh}},\ }\bibfield  {title} {\bibinfo {title} {A {{Five-Equation Model}} for the {{Simulation}} of {{Interfaces}} between {{Compressible Fluids}}},\ }\href@noop {} {\bibfield  {journal} {\bibinfo  {journal} {Journal of Computational Physics}\ }\textbf {\bibinfo {volume} {181}},\ \bibinfo {pages} {577} (\bibinfo {year} {2002})}\BibitemShut {NoStop}%
\bibitem [{\citenamefont {Ge}\ \emph {et~al.}(2022)\citenamefont {Ge}, \citenamefont {Li}, \citenamefont {Zhang},\ and\ \citenamefont {Tian}}]{Ge_2022_EvaluatingStretchingCompression}%
  \BibitemOpen
  \bibfield  {author} {\bibinfo {author} {\bibfnamefont {J.}~\bibnamefont {Ge}}, \bibinfo {author} {\bibfnamefont {H.}~\bibnamefont {Li}}, \bibinfo {author} {\bibfnamefont {X.}~\bibnamefont {Zhang}},\ and\ \bibinfo {author} {\bibfnamefont {B.}~\bibnamefont {Tian}},\ }\bibfield  {title} {\bibinfo {title} {Evaluating the stretching/compression effect of {{Richtmyer}}--{{Meshkov}} instability in convergent geometries},\ }\href@noop {} {\bibfield  {journal} {\bibinfo  {journal} {Journal of Fluid Mechanics}\ }\textbf {\bibinfo {volume} {946}},\ \bibinfo {pages} {A18} (\bibinfo {year} {2022})}\BibitemShut {NoStop}%
\bibitem [{\citenamefont {Pope}(2000)}]{Pope_2000_TurbulentFlows}%
  \BibitemOpen
  \bibfield  {author} {\bibinfo {author} {\bibfnamefont {S.~B.}\ \bibnamefont {Pope}},\ }\href@noop {} {\emph {\bibinfo {title} {Turbulent {{Flows}}}}}\ (\bibinfo  {publisher} {Cambridge University Press},\ \bibinfo {year} {2000})\BibitemShut {NoStop}%
\bibitem [{\citenamefont {{Mor{\'a}n-L{\'o}pez}}\ and\ \citenamefont {Schilling}(2013)}]{Moran-Lopez_2013_MulticomponentReynoldsaveragedNavier}%
  \BibitemOpen
  \bibfield  {author} {\bibinfo {author} {\bibfnamefont {J.~T.}\ \bibnamefont {{Mor{\'a}n-L{\'o}pez}}}\ and\ \bibinfo {author} {\bibfnamefont {O.}~\bibnamefont {Schilling}},\ }\bibfield  {title} {\bibinfo {title} {Multicomponent {{Reynolds-averaged Navier}}--{{Stokes}} simulations of reshocked {{Richtmyer}}--{{Meshkov}} instability-induced mixing},\ }\href@noop {} {\bibfield  {journal} {\bibinfo  {journal} {High Energy Density Physics}\ }\textbf {\bibinfo {volume} {9}},\ \bibinfo {pages} {112} (\bibinfo {year} {2013})}\BibitemShut {NoStop}%
\bibitem [{\citenamefont {Gauthier}\ and\ \citenamefont {Bonnet}(1990)}]{Gauthier_1990_KeModelTurbulent}%
  \BibitemOpen
  \bibfield  {author} {\bibinfo {author} {\bibfnamefont {S.}~\bibnamefont {Gauthier}}\ and\ \bibinfo {author} {\bibfnamefont {M.}~\bibnamefont {Bonnet}},\ }\bibfield  {title} {\bibinfo {title} {A k-{$\varepsilon$} model for turbulent mixing in shock-tube flows induced by {{Rayleigh}}--{{Taylor}} instability},\ }\href@noop {} {\bibfield  {journal} {\bibinfo  {journal} {Physics of Fluids A: Fluid Dynamics}\ }\textbf {\bibinfo {volume} {2}},\ \bibinfo {pages} {1685} (\bibinfo {year} {1990})}\BibitemShut {NoStop}%
\bibitem [{\citenamefont {Schilling}(2015)}]{Schilling_2015_ComparativeStudyPredictions}%
  \BibitemOpen
  \bibfield  {author} {\bibinfo {author} {\bibfnamefont {O.}~\bibnamefont {Schilling}},\ }\bibfield  {title} {\bibinfo {title} {Comparative {{Study}} of the {{Predictions}} of {{Four Reynolds-Averaged Navier}}--{{Stokes Turbulence Models Applied}} to a {{Richtmyer}}--{{Meshkov Instability Experiment}}},\ }in\ \href@noop {} {\emph {\bibinfo {booktitle} {29th {{International Symposium}} on {{Shock Waves}} 2}}},\ \bibinfo {editor} {edited by\ \bibinfo {editor} {\bibfnamefont {R.}~\bibnamefont {Bonazza}}\ and\ \bibinfo {editor} {\bibfnamefont {D.}~\bibnamefont {Ranjan}}}\ (\bibinfo  {publisher} {Springer International Publishing},\ \bibinfo {address} {Cham},\ \bibinfo {year} {2015})\ pp.\ \bibinfo {pages} {1041--1046}\BibitemShut {NoStop}%
\bibitem [{\citenamefont {Schilling}(2021)}]{Schilling_2021_SelfsimilarReynoldsaveragedMechanical}%
  \BibitemOpen
  \bibfield  {author} {\bibinfo {author} {\bibfnamefont {O.}~\bibnamefont {Schilling}},\ }\bibfield  {title} {\bibinfo {title} {Self-similar {{Reynolds-averaged}} mechanical--scalar turbulence models for {{Rayleigh}}--{{Taylor}}, {{Richtmyer}}--{{Meshkov}}, and {{Kelvin}}--{{Helmholtz}} instability-induced mixing in the small {{Atwood}} number limit},\ }\href@noop {} {\bibfield  {journal} {\bibinfo  {journal} {Physics of Fluids}\ }\textbf {\bibinfo {volume} {33}},\ \bibinfo {pages} {085129} (\bibinfo {year} {2021})}\BibitemShut {NoStop}%
\bibitem [{\citenamefont {Schilling}(2024)}]{Schilling_2024_SelfsimilarReynoldsaveragedMechanical}%
  \BibitemOpen
  \bibfield  {author} {\bibinfo {author} {\bibfnamefont {O.}~\bibnamefont {Schilling}},\ }\bibfield  {title} {\bibinfo {title} {Self-similar {{Reynolds-averaged}} mechanical--scalar turbulence models for {{Rayleigh}}--{{Taylor}} mixing induced by power-law accelerations in the small {{Atwood}} number limit},\ }\href@noop {} {\bibfield  {journal} {\bibinfo  {journal} {Physics of Fluids}\ }\textbf {\bibinfo {volume} {36}},\ \bibinfo {pages} {075170} (\bibinfo {year} {2024})}\BibitemShut {NoStop}%
\bibitem [{\citenamefont {Speziale}\ and\ \citenamefont {Sarkar}(1991)}]{Speziale_1991_SecondorderClosureModels}%
  \BibitemOpen
  \bibfield  {author} {\bibinfo {author} {\bibfnamefont {C.}~\bibnamefont {Speziale}}\ and\ \bibinfo {author} {\bibfnamefont {S.}~\bibnamefont {Sarkar}},\ }\bibfield  {title} {\bibinfo {title} {Second-order closure models for supersonic turbulent flows},\ }in\ \href@noop {} {\emph {\bibinfo {booktitle} {29th {{Aerospace Sciences Meeting}}}}},\ \bibinfo {series and number} {Aerospace {{Sciences Meetings}}}\ (\bibinfo  {publisher} {{American Institute of Aeronautics and Astronautics}},\ \bibinfo {year} {1991})\BibitemShut {NoStop}%
\bibitem [{\citenamefont {Xie}\ \emph {et~al.}(2021)\citenamefont {Xie}, \citenamefont {Xiao},\ and\ \citenamefont {Zhang}}]{Xie_2021_PredictingDifferentTurbulent}%
  \BibitemOpen
  \bibfield  {author} {\bibinfo {author} {\bibfnamefont {H.-s.}\ \bibnamefont {Xie}}, \bibinfo {author} {\bibfnamefont {M.-j.}\ \bibnamefont {Xiao}},\ and\ \bibinfo {author} {\bibfnamefont {Y.-s.}\ \bibnamefont {Zhang}},\ }\bibfield  {title} {\bibinfo {title} {Predicting different turbulent mixing problems with the same k--{$\varepsilon$} model and model coefficients},\ }\href@noop {} {\bibfield  {journal} {\bibinfo  {journal} {AIP Advances}\ }\textbf {\bibinfo {volume} {11}},\ \bibinfo {pages} {075213} (\bibinfo {year} {2021})}\BibitemShut {NoStop}%
\bibitem [{\citenamefont {Wilcox}(1998)}]{Wilcox_1998_TurbulenceModelingCFD}%
  \BibitemOpen
  \bibfield  {author} {\bibinfo {author} {\bibfnamefont {D.~C.}\ \bibnamefont {Wilcox}},\ }\bibfield  {title} {\bibinfo {title} {Turbulence modeling for {{CFD}}},\ }in\ \href@noop {} {\emph {\bibinfo {booktitle} {Turbulence Modeling for {{CFD}}}}}\ (\bibinfo  {publisher} {DCW Industries},\ \bibinfo {address} {La C{\~a}nada, Calif},\ \bibinfo {year} {1998})\ \bibinfo {edition} {2nd}\ ed.\BibitemShut {Stop}%
\bibitem [{\citenamefont {Vuong}\ and\ \citenamefont {Coakley}(1987)}]{Vuong_1987_ModelingTurbulenceHypersonic}%
  \BibitemOpen
  \bibfield  {author} {\bibinfo {author} {\bibfnamefont {S.~T.}\ \bibnamefont {Vuong}}\ and\ \bibinfo {author} {\bibfnamefont {T.~J.}\ \bibnamefont {Coakley}},\ }\bibfield  {title} {\bibinfo {title} {Modeling of {{Turbulence}} for {{Hypersonic Flows}} with and without {{Separation}}},\ }in\ \href@noop {} {\emph {\bibinfo {booktitle} {25th {{AIAA Aerospace Sciences Meeting}}}}}\ (\bibinfo  {publisher} {{American Institute of Aeronautics and Astronautics}},\ \bibinfo {address} {Reno, Nevada USA},\ \bibinfo {year} {1987})\ p.\ \bibinfo {pages} {287}\BibitemShut {NoStop}%
\bibitem [{\citenamefont {Coakley}\ and\ \citenamefont {Huang}(1992)}]{Coakley_1992_TurbulenceModelingHigh}%
  \BibitemOpen
  \bibfield  {author} {\bibinfo {author} {\bibfnamefont {T.}~\bibnamefont {Coakley}}\ and\ \bibinfo {author} {\bibfnamefont {P.}~\bibnamefont {Huang}},\ }\bibfield  {title} {\bibinfo {title} {Turbulence modeling for high speed flows},\ }in\ \href@noop {} {\emph {\bibinfo {booktitle} {30th {{Aerospace Sciences Meeting}} and {{Exhibit}}}}},\ \bibinfo {series and number} {Aerospace {{Sciences Meetings}}}\ (\bibinfo  {publisher} {{American Institute of Aeronautics and Astronautics}},\ \bibinfo {year} {1992})\BibitemShut {NoStop}%
\bibitem [{\citenamefont {Pope}(1978)}]{Pope_1978_ExplanationTurbulentJet}%
  \BibitemOpen
  \bibfield  {author} {\bibinfo {author} {\bibfnamefont {S.~B.}\ \bibnamefont {Pope}},\ }\bibfield  {title} {\bibinfo {title} {An explanation of the turbulent round-jet/plane-jet anomaly},\ }\href@noop {} {\bibfield  {journal} {\bibinfo  {journal} {AIAA Journal}\ }\textbf {\bibinfo {volume} {16}},\ \bibinfo {pages} {279} (\bibinfo {year} {1978})}\BibitemShut {NoStop}%
\bibitem [{\citenamefont {Read}(1984)}]{Read_1984_ExperimentalInvestigationTurbulent}%
  \BibitemOpen
  \bibfield  {author} {\bibinfo {author} {\bibfnamefont {K.~I.}\ \bibnamefont {Read}},\ }\bibfield  {title} {\bibinfo {title} {Experimental investigation of turbulent mixing by {{Rayleigh-Taylor}} instability},\ }\href@noop {} {\bibfield  {journal} {\bibinfo  {journal} {Physica D: Nonlinear Phenomena}\ }\textbf {\bibinfo {volume} {12}},\ \bibinfo {pages} {45} (\bibinfo {year} {1984})}\BibitemShut {NoStop}%
\bibitem [{\citenamefont {Youngs}(1989)}]{Youngs_1989_ModellingTurbulentMixing}%
  \BibitemOpen
  \bibfield  {author} {\bibinfo {author} {\bibfnamefont {D.~L.}\ \bibnamefont {Youngs}},\ }\bibfield  {title} {\bibinfo {title} {Modelling turbulent mixing by {{Rayleigh-Taylor}} instability},\ }\href@noop {} {\bibfield  {journal} {\bibinfo  {journal} {Physica D: Nonlinear Phenomena}\ }\textbf {\bibinfo {volume} {37}},\ \bibinfo {pages} {270} (\bibinfo {year} {1989})}\BibitemShut {NoStop}%
\bibitem [{\citenamefont {Andrews}\ \emph {et~al.}(2014)\citenamefont {Andrews}, \citenamefont {Youngs}, \citenamefont {Livescu},\ and\ \citenamefont {Wei}}]{Andrews_2014_ComputationalStudiesTwoDimensional}%
  \BibitemOpen
  \bibfield  {author} {\bibinfo {author} {\bibfnamefont {M.~J.}\ \bibnamefont {Andrews}}, \bibinfo {author} {\bibfnamefont {D.~L.}\ \bibnamefont {Youngs}}, \bibinfo {author} {\bibfnamefont {D.}~\bibnamefont {Livescu}},\ and\ \bibinfo {author} {\bibfnamefont {T.}~\bibnamefont {Wei}},\ }\bibfield  {title} {\bibinfo {title} {Computational {{Studies}} of {{Two-Dimensional Rayleigh-Taylor Driven Mixing}} for a {{Tilted-Rig}}},\ }\href@noop {} {\bibfield  {journal} {\bibinfo  {journal} {Journal of Fluids Engineering}\ }\textbf {\bibinfo {volume} {136}} (\bibinfo {year} {2014})}\BibitemShut {NoStop}%
\bibitem [{\citenamefont {Ferguson}\ \emph {et~al.}(2023)\citenamefont {Ferguson}, \citenamefont {Wang},\ and\ \citenamefont {Morgan}}]{Ferguson_2023_MassMomentumTransport}%
  \BibitemOpen
  \bibfield  {author} {\bibinfo {author} {\bibfnamefont {K.}~\bibnamefont {Ferguson}}, \bibinfo {author} {\bibfnamefont {K.~M.}\ \bibnamefont {Wang}},\ and\ \bibinfo {author} {\bibfnamefont {B.~E.}\ \bibnamefont {Morgan}},\ }\bibfield  {title} {\bibinfo {title} {Mass and momentum transport in the {{Tilted Rocket Rig}} experiment},\ }\href@noop {} {\bibfield  {journal} {\bibinfo  {journal} {Physical Review Fluids}\ }\textbf {\bibinfo {volume} {8}},\ \bibinfo {pages} {094502} (\bibinfo {year} {2023})}\BibitemShut {NoStop}%
\bibitem [{\citenamefont {Denissen}\ \emph {et~al.}(2014)\citenamefont {Denissen}, \citenamefont {Rollin}, \citenamefont {Reisner},\ and\ \citenamefont {Andrews}}]{Denissen_2014_TiltedRocketRig}%
  \BibitemOpen
  \bibfield  {author} {\bibinfo {author} {\bibfnamefont {N.~A.}\ \bibnamefont {Denissen}}, \bibinfo {author} {\bibfnamefont {B.}~\bibnamefont {Rollin}}, \bibinfo {author} {\bibfnamefont {J.~M.}\ \bibnamefont {Reisner}},\ and\ \bibinfo {author} {\bibfnamefont {M.~J.}\ \bibnamefont {Andrews}},\ }\bibfield  {title} {\bibinfo {title} {The {{Tilted Rocket Rig}}: {{A Rayleigh}}--{{Taylor Test Case}} for {{RANS Models1}}},\ }\href@noop {} {\bibfield  {journal} {\bibinfo  {journal} {Journal of Fluids Engineering}\ }\textbf {\bibinfo {volume} {136}} (\bibinfo {year} {2014})}\BibitemShut {NoStop}%
\end{thebibliography}%

\end{document}